%% file: HHMP_RR.tex
\RequirePackage{rotating}

\documentclass[12pt,a4paper,fleqn]{article}
\usepackage[utf8]{inputenc}
\usepackage{authblk}
\usepackage{geometry}
\usepackage{xcolor}
\usepackage[round]{natbib}
\usepackage{graphicx}
\usepackage{pdflscape}  
\usepackage{appendix}
\usepackage{threeparttable}
\usepackage{booktabs}
\usepackage{subcaption}
\usepackage{ulem} 

\usepackage{minitoc}

\usepackage{amssymb}
\usepackage{amsbsy}
\usepackage{bm}
\usepackage{amsmath}
\usepackage{amsthm}
\usepackage{graphicx}
\usepackage{mathtools}
\usepackage{setspace}
\usepackage{color, colortbl}
\definecolor{ashgrey}{rgb}{0.7, 0.75, 0.71}
\definecolor{columbiablue}{rgb}{0.61, 0.87, 1.0}
\definecolor{coral}{rgb}{1.0, 0.5, 0.31}

\definecolor{colBVAR}{HTML}{bababa}
\definecolor{colBART}{HTML}{d7191c}
\definecolor{colmixBART}{HTML}{fdae61}
\definecolor{colerrorBART}{HTML}{abd9e9}
\definecolor{colfullBART}{HTML}{2c7bb6}
\definecolor{plusgreen}{RGB}{26,76,57}

\providecommand{\shadeBench}{\rowcolor[rgb]{0.95, 0.3, 0.3}}

\definecolor{colcons}{HTML}{e31a1c}
\definecolor{colSV}{HTML}{a6cee3}
\definecolor{colhBART}{HTML}{1f78b4}

\usepackage{enumitem}
\newlist{steps}{enumerate}{1}
\setlist[steps,1]{label = Step \arabic*:}

\usepackage{adjustbox}
\usepackage{dcolumn}
\newcolumntype{d}[1]{D..{#1}} 

\usepackage{caption}
\definecolor{nblue}{HTML}{000660}
\usepackage[colorlinks=true,urlcolor=nblue,linkcolor=nblue,citecolor=nblue]{hyperref}

\newcommand*{\myeqref}[2][Eq.~]{%
  \hyperref[{#2}]{#1(\ref*{#2})}%
}
\def\equationautorefname#1#2\null{%
  Eq.#1(#2\null)%
}

\begin{document}
\title{\Large{\textbf{Gaussian Process Vector Autoregressions and Macroeconomic Uncertainty}}\thanks{\scriptsize
\textit{Corresponding author}: Massimiliano Marcellino. Bocconi University. \textit{Address}: Via Roentgen 1, I-20136 Milano. \textit{Email}: \href{mailto:massimiliano.marcellino@unibocconi.it}{massimiliano.marcellino@unibocconi.it}. We would like to thank for useful comments on a previous draft  Roberto Casarin, Karim Chalak, Joshua Chan, Todd Clark, Sharada Nia Davidson, Jean-Marie Dufour, Raffaella Giacomini, William McClausland, {Serena Ng}, Luca Onorante, Michael Pfarrhofer, Barbara Rossi, Samad Sarferaz, Frank Schorfheide, Anna Stelzer, Dalibor Stevanovic, Tommaso Tornese, {an associate editor as well as two referees} and the participants of the $8$th Annual Conference of the International Association for Applied Econometrics (IAAE), the $12$th European Seminar on Bayesian Econometrics (ESOBE), the EABCN and Bundesbank conference on challenges in empirical macroeconomics since 2020, the Decision Making under Uncertainty (DeMUr) 2022 workshop and the University of Montreal econometrics seminar. Hauzenberger gratefully acknowledges financial support from the Jubiläumsfonds of the Oesterreichische Nationalbank (OeNB, grant no. 18304 and 18718), and Huber also acknowledges financial support from the Austrian Science Fund (FWF, grant no. ZK 35) and the Jubiläumsfonds of the OeNB (grant no. 18304).}}

\author[a]{\normalsize Niko \textsc{Hauzenberger}}
\author[a]{\normalsize Florian \textsc{Huber}}
\author[b]{\normalsize Massimiliano \textsc{Marcellino}}
\author[a]{\normalsize Nico \textsc{Petz}}
\affil[a]{\normalsize \textit{University of Salzburg}}
\affil[b]{\normalsize \textit{Bocconi University, IGIER, CEPR, Baffi-Carefin and BIDSA}}
\date{\normalsize \today}
\maketitle\thispagestyle{empty}\normalsize\vspace*{-2em}\small\linespread{1.5}
\vspace{7pt}
\begin{center}
\begin{minipage}{0.85\textwidth}
\noindent\small \textbf{Abstract.} We develop a non-parametric multivariate time series model that remains agnostic on the precise relationship between a (possibly) large set of macroeconomic time series and their lagged values. The main building block of our model is a Gaussian process prior on the functional relationship that determines the conditional mean of the model, hence the name of Gaussian process vector autoregression (GP-VAR). A flexible stochastic volatility specification is used to provide additional flexibility and control for heteroskedasticity. Markov chain Monte Carlo (MCMC) estimation is carried out through an efficient and scalable algorithm which can handle large models. The GP-VAR is illustrated by means of simulated data  and in a forecasting exercise with US data. Moreover, we use the GP-VAR to analyze the effects of macroeconomic uncertainty, with a particular emphasis on time variation and asymmetries in the transmission mechanisms.

\end{minipage}
\end{center}

\doparttoc 
\faketableofcontents 
\part{} 

\normalsize\newpage

\section{Introduction}
Economic relations can change over time for a variety of reasons, such as technological progress, institutional changes, major policy interventions, but also  wars, terrorist attacks, stock market crashes and pandemics. Standard econometric models, such as linear single and multivariate regressions assume instead stability of the parameters characterizing the conditional first and second moments of the dependent variables. When stability is formally tested, it is often rejected \citep[see, e.g.,][]{stock1996evidence}. This has led to the development of a variety of methods to handle structural change in econometric models. 

Parameter evolution is assumed to be either observable (i.e., driven by the behavior of observable economic variables) or unobservable, and either discrete and abrupt or continuous and smooth. Examples include threshold and smooth transition models \citep[see, e.g.,][]{tong1990non, terasvirta1994specification}, Markov switching models \citep[see, e.g.,][]{hamilton1989new}, and double stochastic models \citep[see, e.g.,][]{nyblom1989testing}. In all these models, a specific and fully parametrized type of parameter evolution is assumed, and then linear or non-linear filters are used for estimation in a classical context or Markov chain Monte Carlo (MCMC) methods in a Bayesian framework. Examples of economic applications of all these methods include \cite{koop2013large}, \cite{accm2017changing}, \cite{aastveit2017economic}, \cite{caggiano2017estimating}, \cite{alessandri2019financial}, and \cite{caggiano2021uncertainty}.\footnote{A special mention is due to \cite{primiceri2005time} who popularized the use of time-varying parameters and stochastic volatility in macroeconometrics.}

Assuming a specific type of parameter evolution increases estimation efficiency but can lead to mis-specification. A more flexible alternative allows for a smooth evolution of parameters without specifying the form of parameter time variation. In a classical context, the evolution can be either deterministic \citep[see, e.g.,][]{robinson1991time, chen2012testing}, or stochastic \citep[see, e.g.,][for the specific case of (possibly large) vector autoregressive models (VARs)]{giraitis2014inference, giraitis2018inference, kapetanios2019large}. Kernel estimators are the main tool used in this literature. Alternative approaches, which can also capture non-linear relationships between the target and explanatory variables,  are, e.g., based on using functional-coefficient regressions \citep{cai2000functional, kowal2017bayesian}, regression trees \citep{chipman2010bart, huber2020nowcasting, coulombe2020macroeconomy}, neural networks \citep{hornik1989multilayer,gu2021autoencoder, coulombe2022neural} {or infinite mixtures \citep{hirano2002semiparametric, bassetti2014beta, kalli2018bayesian, billio2019bayesian, jin2022infinite}. Most of these approaches, however, are fairly different from the VAR models that are the workhorse of modern time series econometrics, making interpretation of the estimation results and computation of  quantities  such as impulse response functions difficult. In addition, they typically focus on extending the specification of the conditional mean, while assuming a constant conditional variance, which can be restrictive for macroeconomic and financial data. Finally, some of the methods do not scale well into high dimensions and are thus not particularly suited for large datasets nowadays used in macroeconomics.}

In this paper, we propose a new model that belongs to the non-parametric class and is capable of capturing, in a very flexible way, not only parameter evolution but also general non-linear relationships. Our model can be applied in a large data context while keeping the flexibility and ease of use of VARs{, and allowing for time-varying conditional variances}. Specifically, we combine the statistical literature on Gaussian process (GP) regressions \citep[see, e.g.,][]{crawford2019variable}, with that on VARs to obtain a GP-VAR model.  Borrowing ideas from the literature on Bayesian Minnesota-type VARs, the model assumes, for each endogenous variable, a different non-linear relationship with its own lags and with the lags of all the other variables (and possibly of additional exogenous regressors). Gaussian processes are used to model non-linearities in a flexible but efficient way. They can be viewed as a non-parametric alternative to the adaptive Minnesota-type shrinkage proposed in \cite{chan2021ijof}. {They are also similar to neural networks, in the sense that they are universal approximators based on infinite mixtures of Gaussian distributions.\footnote{{In fact, specific choices of the kernels underlying Gaussian processes can produce a variety of neural network models, see \cite{novak2018bayesian} for details. }}}

We develop an efficient (Bayesian) estimation procedure, based on the structural form of the GP-VAR, which has the additional benefit that its complexity is linear in the number of endogenous variables and does not depend on the number of lags. Hence, estimation can be parallelized and, in addition, the conjugate structure of the model we use allows for pre-computing various matrix multiplications and kernel operations, which further speeds up computation. As a result, estimation is feasible also for very large models. 

As in all Bayesian procedures, an assumption on the distribution of the errors is required. As is common in the Bayesian VAR literature, we assume that the errors are Gaussian. Yet, we also permit the variance of the errors to change over time, adopting a stochastic volatility (SV) specification. Already in linear BVARs, the use of SV permits to have time variation in the conditional distribution of the variables, so much so that BVAR-SV are empirically a good alternative to quantile regressions \citep[see ][]{carriero2022specification}. Moreover, multi-step predictive densities, used to produce forecasts and generalized impulse responses, are non-Gaussian. The use of SV in the GP-VAR adds flexibility, and prevents overfitting in the sense of avoiding that large realizations of the shocks are interpreted as changes in the conditional mean.

We illustrate the GP-VAR with synthetic data generated from a highly non-linear multivariate data generating process (DGP) that have both Gaussian and non-Gaussian shocks. This DGP assumes that some equations feature parameters that exhibit structural breaks while others depend non-linearily on the lags of the endogenous variables. To assess whether the GP-VAR is also capable of recovering linear relations, one equation is a standard linear regression model. In all these cases, our approach works reasonably well in terms of detecting the nature of non-linearities. 

Our GP-based model has a vast range of applicability, for both reduced form and structural analysis. This mirrors the possible applications of standard VARs but allows for much more general dynamic relationships across the variables. Besides the evaluation with synthetic data, we consider a forecasting application and a more structural economic analysis. In the forecasting application, we compare the performance of the GP-VAR with other linear and non-linear competitors that allow for parameter change, and with the BVAR with SV. Predicting US output, inflation and interest rates, we show that the GP-VAR improves upon all competing models, with gains that are particularly pronounced at the four-quarter-ahead horizon. 

As an example of a structural economic analysis, and to gather  new insights on a topic that has recently attracted considerable attention \citep[see, e.g.,][]{bloom2014fluctuations}, we use the GP-VAR to investigate the effects of exogenous uncertainty shocks on US macroeconomic and financial time series. Comparing the responses of the GP-VAR with the ones of a standard linear BVAR reveals that our model produces sensible responses for real activity and stock markets. Differences between the responses relate to the shape and magnitudes, with the GP-VAR producing stronger reactions of uncertainty, real GDP growth, and stock markets returns while yielding similar reactions of employment growth. Considering models of differing sizes shows that the impulse responses do not differ markedly across model sizes.

Our proposed framework naturally allows for analyzing potential asymmetries in transmission channels. The responses to a positive uncertainty shock (higher unexpected uncertainty) are typically much stronger than the ones to a negative shock. Interestingly, the shape of the IRFs also differ, with positive shocks leading to responses that peak later. In addition, our findings suggest that the relationship between real activity and uncertainty becomes proportionally slightly smaller for large shocks, while financial markets react relatively more strongly to larger increases in uncertainty. Finally, our framework also allows us to investigate whether transmission mechanisms have changed over time. Doing so reveals that the effects of uncertainty have been smaller in the great inflation period ($1970$Q$1$ to $1984$Q$4$), more pronounced through the great moderation ($1985$Q$1$ to $2006$Q$4$) before again turning more muted during the post great moderation phase ($2007$Q$1$ to $2019$Q$4$).

The paper is structured as follows. Section \ref{sec: intro_GP} provides an introduction to Gaussian process regression. In Section \ref{sec: econometrics}  we develop the GP-VAR model. This section also provides necessary details on the prior setup and posterior computation. We then analyze model performance using synthetic data in Section \ref{sec:synth}. Sections \ref{sec:macrofor} and \ref{sec:empircal} include our empirical work. In the former section we briefly discuss the dataset and provide some in- and out-of-sample model evidence. In the latter, we focus on the macroeconomic implications of an uncertainty shock.  The final section briefly summarizes and concludes the paper. The Online Appendix contains additional empirical results and technical details on the specification, estimation and use of the GP-VAR model.

\section{A brief introduction to Gaussian processes} \label{sec: intro_GP}
In this section we briefly discuss  Gaussian process (GP) regressions with a focus on time series data.\footnote{For a  textbook treatment, see \cite{williams2006gaussian}.} GP regressions are a non-parametric technique to establish a flexible relationship between a scalar time series $y_t$ and a set of $K$ predictors $\bm x_t$ in period $t$. The key advantage of this approach is that it  does not rely on  parametric assumptions on the precise functional relationship between $y_t$ and $\bm x_t$. 

In general, a non-parametric  regression is given by:
\begin{equation*}
  y_t = f(\bm x_t) + \varepsilon_t, \quad \varepsilon_t \sim \mathcal{N}(0, \sigma^2),
\end{equation*}
with $f$ being some unknown regression function $f: \mathbb{R}^K \to \mathbb R$ and $\varepsilon_t$ denoting an independent Gaussian shock with zero mean and constant variance $\sigma^2$. We relax this assumption in Sub-section \ref{sec: GP_VAR} to allow for heteroskedastic shocks. An assumption on the error distribution is needed in a Bayesian context and Gaussianity is the most common one, though different distributions can be easily accommodated by exploiting a scale-location mixture of Gaussians representation \citep[see, e.g.,][]{escobar1995bayesian}.

In standard regression models, the function $f$ is assumed to be linear with $f(\bm x_t) = \bm \beta' \bm x_t$ where $\bm \beta$ is a $K \times 1$ vector of linear coefficients. If mean relations are non-linear, this assumption might be too restrictive. To gain more flexibility one can  embed the covariates in $\bm x_t$ into a higher dimensional space such as the space of powers $\bm x_t \to \psi(\bm x_t) =(\bm x'_t, (\bm x^2_t)', \dots, (\bm x^R_t)')'$, with $\bm x^2_t = (\bm x_t \odot \bm x_t)$ and higher orders defined recursively.  Conditional on choosing a sufficiently large integer $R$, this would provide substantial flexibility to approximate any smooth function $f$. However, adequately selecting $R$ is key and the mapping, moreover, is ad-hoc in the sense that there exist infinitely many non-linear mappings $\psi$.

Standard Bayesian methods place a prior on the coefficients associated with the covariates (and possible non-linear transformations thereof) and thus control for uncertainty with respect to these basis functions but at the cost of remaining within a class of functions (such as linear, polynomial or trigonometric functions). By contrast, in GP regressions we treat the function $f$ as an unknown quantity and let the data decide about the appropriate form (and degree) of non-linearities. 

\subsection{Estimating unknown functions: the function space view}
The key inferential goal in GP regression is to infer the function $f$ from the data under relatively mild assumptions. This is achieved by specifying a prior on $f(\bm x_t)$. A typical assumption is to assume that $f(\bm x_t)$ follows a Gaussian process prior:
\begin{equation*}
f(\bm x_t) \sim \mathcal{GP}\left(\mu(\bm x_t), {k}_{\bm \vartheta}(\bm x_t, \bm x_t)\right),
\end{equation*}
with $\mu(\bm x_t)= \mathbb{E}[f(\bm x_t)]$ being the mean function and
\begin{equation*}
 k_{\bm \vartheta}(\bm x_t, \bm x_\tau) =   \mathbb{E}[(f(\bm x_t) - \mu(\bm x_t)) (f(\bm x_\tau) - \mu(\bm x_\tau))]
\end{equation*}
denoting a kernel (or covariance) function that determines the relationship between $f(\bm x_t)$ and $f(\bm x_\tau)$ for periods $t$ and $\tau$. The kernel is typically parameterized  by a low dimensional vector of hyperparameters $\bm \vartheta$ and controls the behavior of the function $f$. This kernel needs to be positive semidefinite and symmetric.

In what follows, we will set the function $\mu(\bm x_t) = 0$ for all $t$. This is without loss of generality, since any explicit basis function for $\mu(\bm x_t)$ can be used to model the mean process. If the focus is on modeling stationary data,  $\mu(\bm x_t) = 0$ implies that a priori the process is centered around a white noise process. In case one would like to model persistent or non-stationary data it would be straightforward to implement a prior that forces the system towards a set of random walk processes. This can be achieved by setting $\mu(y_{t-1}) = \rho y_{t-1}$, where $\rho$ denotes a persistence parameter with prior mean $\mathbb{E}[\rho] = 1$. Alternatively, one could specify the prior on $f$ to imply persistence in $y_t$. This possibility is discussed in much more detail in Section \ref{sec:persistencel} of the Online Appendix. 

A common choice in GP regressions is the Gaussian (or squared exponential) kernel function:
\begin{equation*}
k_{\bm \vartheta}(\bm x_t, \bm x_\tau) = \xi \times \exp\left( -\frac{\kappa}{2} \lVert \bm x_t - \bm x_\tau \rVert^2 \right), 
\end{equation*}
with $\xi$ denoting a scaling parameter and $\kappa$ the (inverse) length scale and thus $\bm \vartheta = (\xi, \kappa)'$. Larger values of $\kappa$ lead to a GP which displays more high frequency variation whereas lower values imply a slowly varying mean function.  The parameter $\xi$ controls the prior variance of the function $f$. To see this, note that if $\bm x_t = \bm x_\tau$, we obtain $\text{Var}[f(\bm x_t)] =\xi$.  

This specification is quite flexible and fulfills several convenient conditions. For instance, \cite{williams2006gaussian} show that the use of the Gaussian kernel implies that $f(\bm x_t)$ is mean square continuous and differentiable. Moreover, this kernel function represents a positive semidefinite and symmetric covariance function. Furthermore, Mercer's theorem \citep{mercer1909xvi}, under this kernel, states that the GP regression can be written in terms of an infinite number of basis functions. These basis functions are Gaussians with different means and variances. This suggests a connection to the literature on Bayesian non-parametrics  \citep{escobar1995bayesian, neal2000markov,kalli2018bayesian, fruhwirth2019here} that relies on infinite  mixtures of Gaussians to estimate unknown densities. The link between GPs and infinite mixtures of Gaussians shows that the Gaussian assumption on $\varepsilon_t$ is not too restrictive as the model allows to recover non-Gaussian features in the data.

The GP prior represents an infinite dimensional prior over the space of functions. This implies that the estimation problem is infinite dimensional as well. However, since we sample data in a discrete manner, the GP prior becomes a multivariate Gaussian prior on $\bm f = (f(\bm x_1), \dots, f(\bm x_T))'$:
\begin{equation*}
\bm f \sim \mathcal{N}(\bm 0_T, K_{\bm \vartheta}(\bm X, \bm X)),
\end{equation*}
with $\bm 0_T$ being a $T \times 1$ vector of zeros, $K_{\bm \vartheta}(\bm X, \bm X)$ a $T \times T$ kernel matrix with typical element $k_{\bm \vartheta}(\bm x_t, \bm x_\tau)$ and $\bm X = (\bm x_1, \dots, \bm x_T)'$.  This implies that, in terms of full data matrices, the GP regression is given by:
\begin{equation*}
\bm y = \bm f + \bm \varepsilon, \quad \bm f \sim \mathcal{N}(\bm 0_T, K_{\bm \vartheta}(\bm X, \bm X)), \quad \bm \varepsilon \sim \mathcal{N}(\bm 0_T, \sigma^2 \bm I_T),
\end{equation*}
where $\bm I_T$ denotes a $T \times T$ identity matrix. 

Assuming for the moment that $\sigma^2$ is known, the posterior of $\bm f$ follows a multivariate Gaussian distribution:
\begin{equation*}
    \bm f | \bm y \sim \mathcal{N}(\overline{\bm f}, \overline{\bm  V}_{\bm f}),
\end{equation*}
with variance-covariance matrix $\overline{\bm V}_{\bm f}$ and posterior mean vector $\overline{\bm f}$:
\begin{align*}
    \overline{\bm  V}_{\bm f} &= {K}_{\bm \vartheta}(\bm X, \bm X) - {K}_{\bm \vartheta}(\bm X, \bm X) \left({K}_{\bm \vartheta}(\bm X, \bm X) + \sigma^2 \bm I_T\right)^{-1} {K}_{\bm \vartheta}(\bm X, \bm X), \\
    \overline{\bm f} &= {K}_{\bm \vartheta}(\bm X, \bm X) \left({K}_{\bm \vartheta}(\bm X, \bm X) + \sigma^2 \bm I_T\right)^{-1} \bm y.
\end{align*}
The mean function $\overline{\bm f}$ can be interpreted as a weighted average of the values of the endogenous variable:
\begin{equation*}
\overline{\bm f} = \sum_{t=1}^T \alpha_t K_{\bm \vartheta}(\bm X, \bm x_t),
\end{equation*}
where $\bm \alpha = (\alpha_1, \dots, \alpha_T)' = (K_{\bm \vartheta}(\bm X, \bm X) + \sigma^2 \bm I_T)^{-1}\bm y$.   This (finite dimensional) representation shows how one moves from an infinite dimensional problem to a finite dimensional one.

The expression for the variance-covariance matrix $\overline{\bm  V}_{\bm f}$ also has an intuitive interpretation. The first term is the prior variance (i.e., the kernel matrix). The second term measures how much of the variance is expressed through the covariates in $\bm X$ and thus the posterior covariance indicates how much the model learns from $\bm X$.

The predictive distribution of $f(\bm x_{T+h})$ can be easily derived by exploiting basic properties of the multivariate Gaussian: 
\begin{equation}
    f(\bm x_{T+h})|\bm y \sim \mathcal{N}(\overline{f}_{T+h}, \overline{V}_{T+h}), \label{eq: pred_dens}
\end{equation}
whereby
\begin{align*}
    \overline{V}_{T+h} &= k_{\bm \vartheta}(\bm x_{T+h}, \bm x_{T+h})  - K_{\bm \vartheta}(\bm x_{T+h}, \bm X)\left( K_{\bm \vartheta}(\bm X, \bm X) + \sigma^2 \bm I_T\right)^{-1} K_{\bm \vartheta}(\bm X, \bm x_{T+h}), \\
    \overline{f}_{T+h} &= K_{\bm \vartheta}(\bm x_{T+h}, \bm X)\left( K_{\bm \vartheta}(\bm X, \bm X) + \sigma^2 \bm I_T\right)^{-1} \bm y.
\end{align*}
Similar to the posterior mean $\overline{\bm f}$,  the predictive mean $\overline{f}_{T+h}$ is a weighted average of the values of the endogenous variables $\bm y$ with the weights depending on the relationship between $\bm X$ and a realization of the vector of covariates $\bm x_{T+h}$ related to the $h$-step-ahead horizon. The predictive variance $\overline{V}_{T+h}$, again, depends on a term that is purely driven by the prior evaluated at $\bm x_{T+h}$ minus a term that measures the informational content in the covariates.

Before proceeding to the discussion on how to set the kernel it is worth noting that what we have discussed above is often labeled the \textit{function-space view} of the GP. This is because the prior is elicited directly on $f$. Another way of analyzing GPs is based on the \textit{weight-space view}. Under the weight-space view one can rewrite the GP regression as a standard regression model as follows:
\begin{equation*}
    \bm y = \bm W_{\bm \vartheta} \bm \eta + \bm \varepsilon, \quad \bm \eta \sim \mathcal{N}(\bm 0_T, \bm I_T),
\end{equation*}
with $\bm W_{\bm \vartheta}$ denoting the lower Cholesky factor of $K_{\bm \vartheta}(\bm X, \bm X) = \bm W_{\bm \vartheta} \bm W'_{\bm \vartheta}$ and $\bm \eta$ is a Gaussian shock vector with zero mean and unit variance. This is a standard regression model with $T$ regressors, a coefficient vector $\bm \eta$ and a Gaussian prior on $\bm \eta$. Standard textbook formulas for the Bayesian linear regression model \citep[see, e.g.,][Chapter~4]{koop2003bayesian} can be used to carry out posterior inference. 

This also shows that if we set $K_{\bm \vartheta}(\bm X, \bm X) = \bm X \bm V_{\bm \vartheta} \bm X'$, we obtain a linear regression model that features a Gaussian prior with zero mean and a typical prior variance-covariance matrix $\bm V_{\bm \vartheta}$. This kernel implies many more parameters than the parsimony inducing Gaussian kernel. Hence, the resulting fit and forecasts can be expected to have posterior distributions with larger variances than those associated with the Gaussian kernel, though bias would be lower if the true model is linear and features stable parameters.

\subsection{Choosing a kernel and the role of the hyperparameters}
In the previous sub-section  the quantities for the posterior of $\bm f$ and the predictive density for future values of $f(\bm x_{T+h})$ suggest that the kernel and its hyperparameters play an important role.  In this sub-section, we discuss this issue in more detail.

One of the key advantages of GPs is that by constructing suitable kernels, one can determine the space of possible functions. This gives rise to substantial flexibility and allows for capturing a large range of competing models within a single econometric model. For instance, \citet[][Chapter 6]{williams2006gaussian} discuss how kernels can be constructed to mimic the behavior of neural networks, regression splines, polynomial  and  linear regressions. Tree-based techniques such as Bayesian additive regression trees \citep[BART,][]{chipman2010bart} can be cast in this framework by exploiting the ANOVA-representation of the model and then the weight-space view of the GP. In principle, and we will build on this feature later, summing over the corresponding kernels gives rise to another kernel and suitable weights could be constructed to select, in a data-driven way, which model summarizes the data best.

\begin{figure}[!t]
\caption{Effect of different values of $\kappa$ on the prior of $\bm f$ and the posterior $\bm f|\bm y$}
    \begin{minipage}{\textwidth}
    \centering
    Inflation and a linear time trend
    \end{minipage}
    \begin{minipage}{0.49\textwidth}
    \centering
    \scriptsize $\bm f$
    \end{minipage}
    \begin{minipage}{0.49\textwidth}
    \centering
    \scriptsize $\bm f|\bm y$
    \end{minipage}

    \centering
    \includegraphics[width = 1\textwidth]{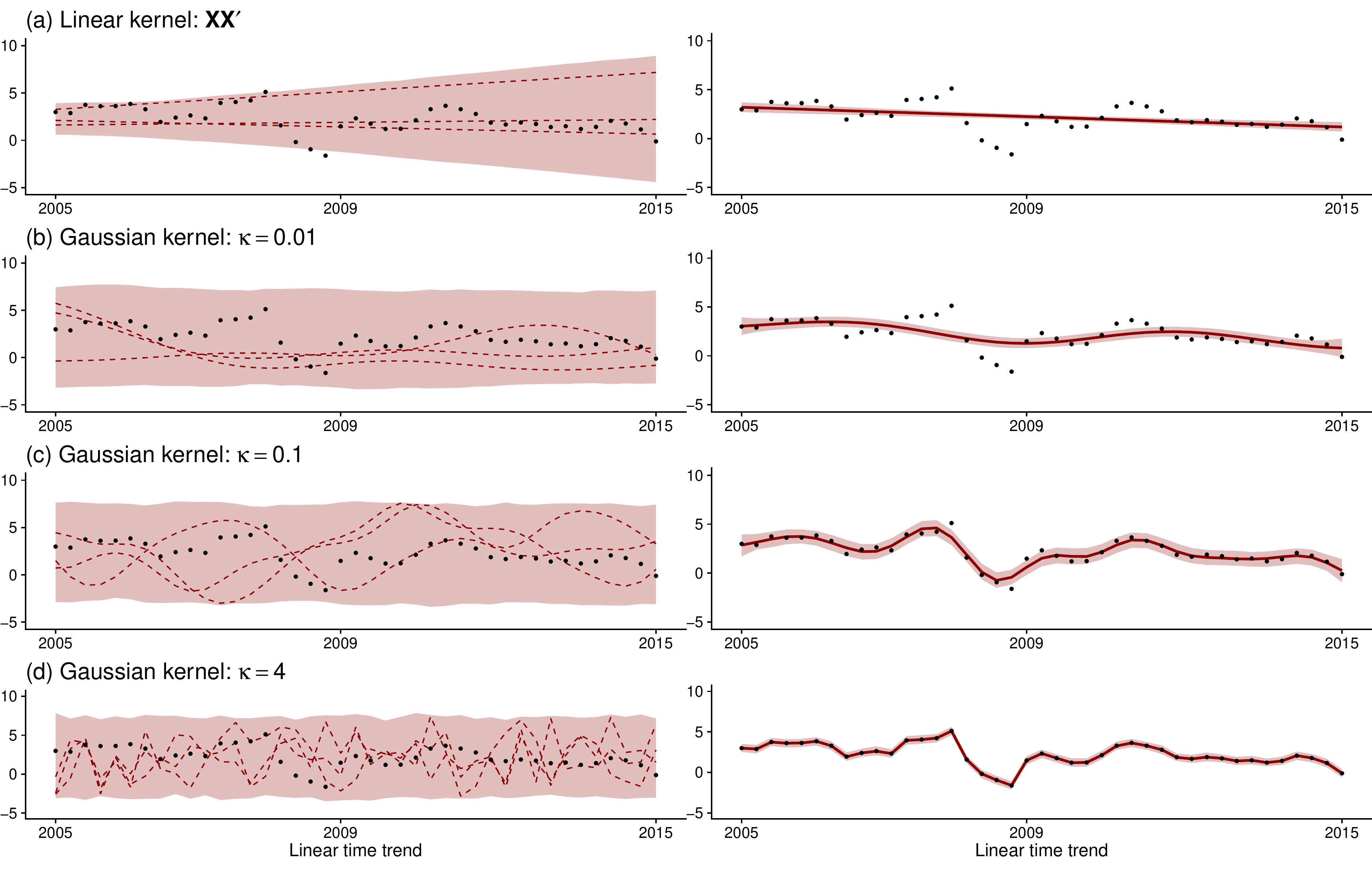}
    \label{fig:choice_kappa_trend}
    \begin{minipage}{\textwidth}
\vspace*{-5pt}
\scriptsize \textit{\textbf{Notes}:} In this figure, we showcase the GP regression with US inflation data using a linear time trend as the only regressor. The left panels report, for different values of $\kappa$, the $5^{th}$ and $95^{th}$ prior percentiles (with the area in between shaded in light red), three draws from the prior (dashed red lines), and the actual values of inflation (black dots). The right panels report the $90\%$ posterior credible sets (shaded in light red), the posterior medians (solid red lines), and actual inflation (black dots).
\end{minipage}
\end{figure}
\begin{figure}[!t]
\caption{Effect of different values of $\kappa$ on the prior of $\bm f$ and the posterior $\bm f|\bm y$}
    \begin{minipage}{\textwidth}
    \centering
    GDP growth and the first lag of macroeconomic uncertainty
    \vspace{10pt}
    \end{minipage}
    \begin{minipage}{0.49\textwidth}
    \centering
    \scriptsize $\bm f$
    \end{minipage}
    \begin{minipage}{0.49\textwidth}
    \centering
    \scriptsize $\bm f|\bm y$
    \end{minipage}
    \centering
    \includegraphics[width = 1\textwidth]{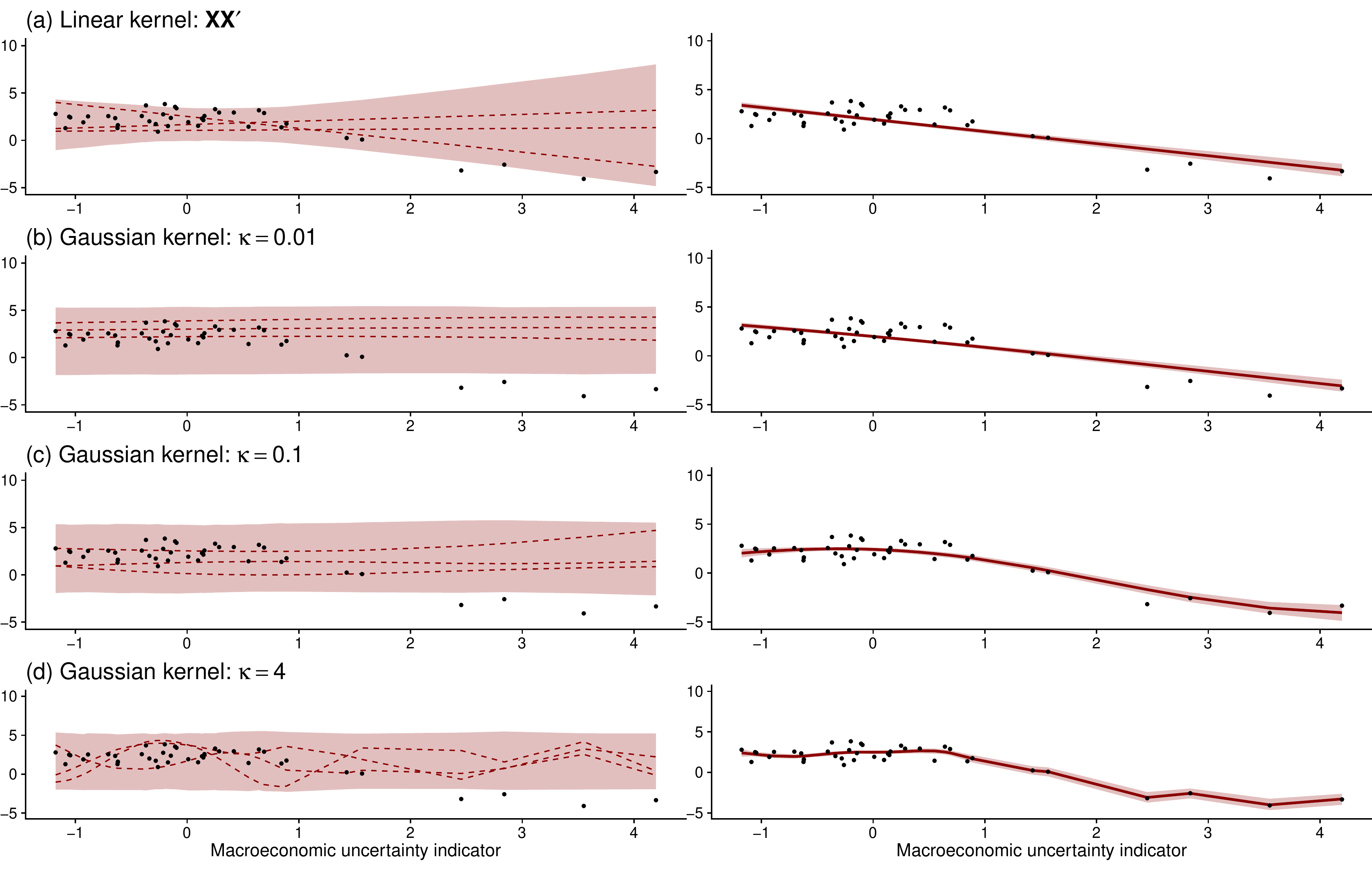}
    \label{fig:choice_kappa_pc}
    \begin{minipage}{\textwidth}
\vspace*{-1pt}
\scriptsize \textit{\textbf{Notes}:} In this figure, we showcase the GP regression with US GDP growth data using the first lag of macroeconomic uncertainty as the only regressor. The left panels report, for different values of $\kappa$, the $5^{th}$ and $95^{th}$ prior percentiles (with the area in between shaded in light red), three draws from the prior (dashed red lines), and the actual values of GDP growth (black dots). The right panels report the $90\%$ posterior credible sets (shaded in light red), the posterior medians (solid red lines), and actual GDP growth (black dots).
\end{minipage}
\end{figure}

As stated in the previous sub-section, our focus will be on the Gaussian kernel due to its excellent empirical properties and analytical tractability. The two hyperparameters $\kappa$ and $\xi$ control the curvature and the marginal variance of the function, respectively.  We illustrate the effect of $\kappa$ on the prior and posterior of $\bm f$ in Figures \ref{fig:choice_kappa_trend} and \ref{fig:choice_kappa_pc} by means of two simple univariate examples.  The first example models quarterly US inflation (in year-on-year terms) and sets $x_t=t$ for periods ranging from $2005$Q$1$ to $2015$Q$4$. The second example models US GDP growth as a function of the first lag of a macroeconomic uncertainty measure for the same sub-sample.\footnote{Throughout the paper, we use the macroeconomic uncertainty measure of \cite{jurado2015measuring} provided (and regularly updated) on the web page of Sydney C. Ludvigson (available online via \href{https://www.sydneyludvigson.com/macro-and-financial-uncertainty-indexes}{sydneyludvigson.com/macro-and-financial-uncertainty-indexes}). Detailed information on this index and the econometric techniques used to obtain this measure can be found in \cite{jurado2015measuring}.} The figures then show (for both the prior and posterior) the value of the function $f(x_t)$ on the y-axis and $x_t$ on the x-axis.  Both figures display in the left (right) panel the $5^{th}$ and $95^{th}$ prior (posterior) percentiles (with the area, the $90\%$ credible set, in between shaded in light red) as well as three random draws from the prior (dashed red lines) in the left panel, and the posterior median (solid red lines) in the right panel.

 \autoref{fig:choice_kappa_trend} reveals that if $y_t$ is a (possibly non-linear) function of time and the inverse length scale parameter is set small, the model generates functions that track the trend in inflation rather well.  This is similar to the unobserved components model of \cite{stock2007has}, which features a persistent stochastic trend in inflation. Once we increase $\kappa$ we observe that the draws from the prior display  more high frequency variation with shorter cycles between peaks and troughs. Once this prior is combined with the data, the estimated mean functions display much more curvature and fit the actual data increasingly well.

Once $\kappa$ is set too high, the functions arising from the prior vary substantially and are likely to capture also very small deviations of inflation from its trend. This translates into a close-to-perfect fit of the posterior mean of the functions and gives rise to serious overfitting concerns. 

To see how a GP regression captures a possibly non-linear relationship between $y_t$ and $x_t$, \autoref{fig:choice_kappa_pc} shows the functional relationship between output growth and lagged macroeconomic uncertainty. If $\kappa = 0.01$, the regression relationship is almost linear and suggests that high levels of (lagged) macroeconomic uncertainty are accompanied by negative output growth rates.   

When we set $\kappa=0.1$ we observe much more curvature (both in the prior and the posterior) in the relationship, indicating that if uncertainty is between $0$ and around $1.7$, GDP growth is between 2.3 and 2.5 percent. However, once a certain threshold in the first lag of uncertainty is reached, the relationship becomes strongly negative until it becomes essentially flat for very high levels of uncertainty. A similar finding, but slightly more pronounced, arises if we set $\kappa = 4$. In this case GDP growth does not change much as long as lagged uncertainty is between 0 and 1.7 and then the relationship becomes, again, strongly negative.

As is clear from these stylized examples, the role of the kernel and its hyperparameters crucially impacts the posterior estimates of the function $f$.  Setting $\kappa$ too small leads to a model which might miss important (higher frequency) information whereas a $\kappa$ set too large translates into an overfitting model which might yield a very strong in-sample fit but poor ouf-of-sample predictions. Setting $\kappa$ is thus of crucial importance and in all our empirical work we will infer it through Bayesian updating.

Another key question is whether the estimated function converges to the true underlying function.  The  literature deals with this question using several assumptions on the error distributions  (mostly setting $\sigma^2 = 0$)  or how the GP regression behaves if the underlying function $f$ differs in terms of smoothness from the GP prior controlled by the kernel \citep[see, e.g.,][]{stone1982optimal, van2008rates, yang2017frequentist, teckentrup2020convergence}. \cite{stone1982optimal}, by focusing on iid data, shows that the optimal rate of estimation of a $\zeta-$smooth function is $T^{- \zeta/(2\zeta + K)}$ and thus decreases in $K$ while it increases in the smoothness of the true function. Building on this finding, \cite{teckentrup2020convergence} analyzes the contraction properties of a Gaussian process regression under a general Mat\'{e}rn kernel function  and provides error bounds that also depend on the relationship of the smoothness of the true and estimated functions. If these agree, one can achieve a convergence rate  of $T^{-\zeta/K}$.  

After having provided the necessary foundations on Gaussian process regression, we will now focus on developing a model that is suitable for macroeconomic analysis.
\section{Large-dimensional Gaussian process VARs}\label{sec: econometrics}
In this section, we first  develop the GP-VAR in Sub-section \ref{sec: GP_VAR}. Next, Sub-sections \ref{sec: functionlearning} to \ref{sec: posterior_comp} are devoted to the development of efficient MCMC schemes to carry out posterior and structural inference. Finally, Sub-section \ref{sec:GIRFs} details how to compute forecasts and (generalized) impulse response functions for the GP-VAR. 

\subsection{The Gaussian process VAR} \label{sec: GP_VAR}

In the following discussion, let $\bm y_t = (y_{1t}, \dots, y_{Mt})'$ denote an $M \times 1$ vector of macroeconomic and financial variables.\footnote{We assume that the elements in $\bm y_t$ are demeaned. In our empirical application we include a constant term with an uninformative prior.} Moreover,  $\bm x_t = (\bm x'_{1t}, \dots, \bm x'_{Mt})'$ denotes an $Mp \times 1$ vector with $\bm x_{jt} = (y_{jt-1}, \dots, y_{jt-p})'$ storing the ``own" lags of the $j^{th}$ endogenous variable and $\bm z_t =(\bm z'_{1t}, \dots, \bm z'_{Mt})'$ an $(M-1) M p \times 1$ vector of ``other" lags. Hence, $\bm z_{jt}=(\bm y'_{-j t-1}, \dots, \bm y'_{-j t-p})'$, where $\bm y_{-j t}$ denotes the vector $\bm y_t$ with the $j^{th}$ element excluded. 

We discriminate between own and other lags of $\bm y_t$ because we assume that lags of other endogenous variables impact a given endogenous variable differently from its own lags. The literature on Bayesian VARs \citep[see][]{banbura2010large, koop2013forecasting, huber2019adaptive, chan2021ijof} has captured this through shrinkage priors that treat coefficients on own and other lags differently. We wish to capture this equation-specific asymmetry by specifying our GP-VAR to depend on two latent processes: one driven by $\bm x_t$ and one  by $\bm z_t$. The structural form of the resulting GP-VAR is then given by:
\begin{equation}\label{eq:gpvar}
        \bm y_t =  F(\bm x_t) +  G(\bm z_t) + \bm Q \bm y_t  + \bm \varepsilon_t, \quad  \bm \varepsilon_t \sim \mathcal{N}(\bm 0_M,\bm H_t), 
\end{equation}
with $F(\bm x_t) = (f_1 (\bm x_{1t}), \dots ,f_M (\bm x_{Mt}) )'$ and $G(\bm z_t) = (g_1 (\bm z_{1t}), \dots ,g_M (\bm z_{Mt}) )'$  and $f_j$ and $g_j$ being equation-specific functions. The function $f_j$ controls how $y_{jt}$ depends on its own lags while $g_j$ encodes the relationship between $y_{jt}$ and the lags of the other endogenous variables. The functions $f_j$ and $g_j$, and hence $F$ and $G$, differ because we construct different kernels with distinct hyperparameters.\footnote{It is worth stressing that one could also think of our decomposition in terms of a new function with a kernel that is given by the sum of the kernels of the functions $f_j$ and $g_j$.} The matrix $\bm Q$ is an $M \times M$ lower triangular matrix with zeros along its main diagonal. This matrix defines the contemporaneous relations across the elements in $\bm y_t$.  

Finally,  $\bm \varepsilon_t$ is an $M \times 1$ vector of Gaussian shocks with zero mean and an $M \times M$ time-varying variance-covariance matrix $\bm H_t = \text{diag}(\omega_{1t},\dots,\omega_{Mt})$. We will assume that $\omega_{jt}$ follows a flexible stochastic volatility (SV) model: 
\begin{equation}\label{eq:svstate}
    h_{jt} = \log \omega_{jt} = \rho_{hj} h_{jt-1} + \nu_{h,jt}, \quad \nu_{h,jt} \sim \mathcal{N}(0, \sigma^2_{hj}), \quad h_{j0} \sim \mathcal{N}\left(0, \frac{\sigma^2_{hj}}{1-\rho^2_{hj}}\right),
\end{equation}
with the logarithm of $h_{jt} = \log \omega_{jt}$ being assumed to evolve according to a stationary AR($1$) state equation. We let $\rho_{hj}$ denote the persistence parameter, $\sigma^2_{hj}$ the error variance, and $h_{j0}$ the initial state of the log-volatility process. 

Allowing for time variation in the shock variances provides additional flexibility and enables us to capture non-Gaussian features in the shocks (not only, but also due to the fact that $h_{jt}$ enters the model non-linearly).\footnote{One could also introduce additional scaling factors that arise from inverse Gamma distributions to obtain a model with t-distributed shocks.} In principle, we could also allow for unknown functional relations between the contemporaneous terms of the preceding $j-1$ equations and the response of equation $j$. However, this would lead to a complicated non-linear covariance structure. Since we are interested in carrying out structural identification based on zero impact restrictions we opt for choosing this simpler approach which implies multivariate Gaussian reduced form shocks, but with a time-varying covariance matrix. Given that the literature on GPs typically assumes the shocks to be Gaussian and homoskedastic, this is already a substantial increase in flexibility.\footnote{A rare exception is \cite{jylanki2011robust}, who propose a GP regression with heavy tailed errors and mainly focus on fast and robust approximate inference of a posterior that is analytically intractable due to a $t$-distributed likelihood.}

The model in \autoref{eq:gpvar} assumes that the shocks in $\bm \varepsilon_t$ are, conditional on $\bm Q \bm y_t$, orthogonal and hence estimation can be carried out equation-by-equation. We will exploit this representation for simplicity and computational tractability. The $j^{th}$ equation, in terms of full-data matrices, is given by:
\begin{equation*}
              \bm Y_j = \bm f_j + \bm g_j + \sum_{k=1}^{j-1} q_{jk}  \bm Y_k  + \bm \epsilon_j, \quad \bm \epsilon_j \sim \mathcal{N}\left(\bm 0_T, \bm \Omega_j\right),
\end{equation*}
with $\bm Y_j =(y_{j1}, \dots, y_{jT})', \bm f_j = (f_j(\bm x_{j1}), \dots, f_j(\bm x_{jT}))', \bm g_j = (g_j(\bm z_{j1}), \dots, g_j(\bm z_{jT}))'$, $\bm \Omega_j = \text{diag}(\omega_{j1}, \dots, \omega_{jT}), \bm \epsilon_j = (\varepsilon_{j1}, \dots, \varepsilon_{jT})'$ and $q_{jk}$ denoting the $(j,k)^{th}$ element of $\bm Q$.  We will use this form to carry out inference about the unknown functions $f_j$ and $g_j$ as well as the remaining parameters and latent states of the model. 


Notice that our estimation strategy is not invariant with respect to reordering the elements in $\bm y_t$, a common problem if this orthogonalization strategy is used.  In Sub-section \ref{app:ordering} of the Online Appendix, we show that different orderings have only a small impact on the estimated impulse responses.


\subsection{Conjugate Gaussian process priors} \label{sec: functionlearning}
In this sub-section, our focus will be on the priors on $f_j$ and $g_j$. The priors on the remaining, linear quantities are standard and thus not discussed in depth. We use a Horseshoe  prior \citep{carvalho2010horseshoe} on the free elements in $\bm Q$, a Beta prior on the (transformed) persistence parameter $(\rho_{h j}+1)/2 \sim \mathcal{B}(25, 5)$, and an inverse Gamma prior on the state innovation variances $\sigma^2_{h j}$. This prior is specified to have mean $0.1$ and variance $0.01$.

For equation-specific functions $f_j$ and $g_j$, we specify two GPs with one conditional on $\bm X_j = (\bm x_{j1}, \dots, \bm x_{jT})'$ and one conditional on $\bm Z_j = (\bm z_{j1}, \dots, \bm z_{jT})'$:
\begin{equation}\label{eq:fg}
\begin{aligned}
\bm f_j  \sim \mathcal{N}\left(\bm 0_T, \sqrt{\bm \Omega_j}  K_{\bm \vartheta_{j1}}(\bm X_j,  \bm X_j) \sqrt{\bm \Omega_j}\right),\quad 
\bm g_j \sim \mathcal{N}\left(\bm 0_T, \sqrt{\bm \Omega_j} K_{\bm \vartheta_{j2}}(\bm Z_j, \bm Z_j) \sqrt{\bm \Omega_j}\right).\quad
\end{aligned}
\end{equation}
We let  $\sqrt{\bm \Omega_j} = \text{diag}(\sqrt{\omega_{j1}}, \dots, \sqrt{\omega_{jT}})$ while $K_{\bm \vartheta_{j1}}(\bm X_j,  \bm X_j)$ and $K_{\bm \vartheta_{j2}}(\bm Z_j, \bm Z_j)$ denote two suitable kernels with typical elements given by: 
\begin{equation*}
\begin{aligned}
k_{\bm \vartheta_{j1}}(\bm x_{jt}, \bm x_{j\tau}) &= \xi_{j1} \times \exp\left( -\frac{\kappa_{j1}}{2}  (\bm x_{jt} - \bm x_{j\tau})' \bm D^{-1}_{\bm X_j} (\bm x_{jt} - \bm x_{j\tau})\right), \quad \bm \vartheta_{j1} = (\xi_{j1}, \kappa_{j1})', \\ k_{\bm \vartheta_{j2}}(\bm z_{jt}, \bm z_{j\tau}) &= \xi_{j2} \times \exp\left( -\frac{\kappa_{j2}}{2} (\bm z_{jt} - \bm z_{j\tau})' \bm D^{-1}_{\bm Z_j} (\bm z_{jt} - \bm z_{j\tau}) \right), \quad \bm \vartheta_{j2} = (\xi_{j2}, \kappa_{j2})'.
\end{aligned}
\end{equation*}
For $j=1, \dots, M$, $\bm \vartheta_{j1}$ and $\bm \vartheta_{j2}$ are equation and kernel-specific hyperparameters and the matrices $\bm D_{\bm X_j}, \bm D_{\bm Z_j}$ are diagonal matrices with typical $i^{th}$ element $\hat{\sigma}^2_{\bm X_j i}, \hat{\sigma}^2_{\bm Z_j i}$. These are set equal to the empirical variances of the $i^{th}$ column of $\bm X_j$ and $\bm Z_j$, respectively. Inclusion of the diagonal scaling matrices  $\bm D_{\bm X_j}$ and $\bm D_{\bm Z_j}$ serves to control for differences in the scaling of the explanatory variables.  Notice that since the hyperparameters are allowed to differ, we essentially treat own and other lags asymmetrically through different functional approximations $f_j$ and $g_j$. 

The kernel is scaled with the error variances in  $\bm \Omega_j$. A typical diagonal element of the corresponding re-scaled kernel is given by $\omega_{jt} \times k_{\bm \vartheta_{j1}}(\bm x_{jt}, \bm x_{jt}) = \omega_{jt} \xi_{j1}$  and $\omega_{jt} \times k_{\bm \vartheta_{j2}}(\bm z_{jt}, \bm z_{jt}) = \omega_{jt} \xi_{j2}$. Typical off-diagonal elements are given by $\sqrt{\omega_{jt}} \sqrt{\omega_{j \tau}} \times k_{\bm \vartheta_{j1}}(\bm x_{jt}, \bm x_{j\tau})$ and $\sqrt{\omega_{jt}} \sqrt{\omega_{j \tau}} \times k_{\bm \vartheta_{j2}}(\bm z_{jt}, \bm z_{j\tau})$. The interaction between the kernel and the error variances gives rise to convenient statistical and computational properties.  

First, note that if $\omega_{jt}$ is large, the corresponding prior on the unknown functions is more spread out. In macroeconomic data, $\omega_{jt}$ is typically large in crisis periods when the $\bm x_{jt}$ and $\bm z_{jt}$ are far away from their previous values. Since the diagonal elements of the kernels are effectively determined by $\bm \xi_j = (\xi_{j1}, \xi_{j2})'$ the presence of $\omega_{jt}$ allows for larger values in the marginal prior variance and thus makes large shifts in the unknown functions more likely. Second, the interaction between $\omega_{jt}$ and $\omega_{jt-1}$ implies that the covariances are scaled down if $\omega_{jt} \gg \omega_{jt-1}$, suggesting that the informational content decreases if increases in uncertainty are substantial (i.e., $\Delta \omega_{jt}$ is large). If $\omega_{jt} \approx \omega_{j\tau}$ and both are large, the corresponding covariance will be scaled upwards. This implies  that our model learns from previous crisis episodes as well. Third, as we will show in Sub-section \ref{sec: posterior_comp}, interacting the kernel with the error variances leads to a conjugate Gaussian process structure which implies that we can factor out the error volatilities and do not need to update several quantities during MCMC sampling. This speeds up computation enormously and allows for estimating large models.

Before discussing how we select the hyperparameters, it is worth highlighting a possible identification problem of our model. In our baseline specification we center $\bm f_j$  and $\bm g_j$ around zero a priori. If we introduce an additional intercept term (or a simpler mean function) no identification issues arise. However, if we believe that $\bm f_j$ and $\bm g_j$ are centered on non-zero values, we can not separately identify them. In our empirical work, we normalize the grand mean of $\bm g_j$ to be equal to zero.\footnote{Notice that this only concerns the posterior distribution since, under the prior, this condition is automatically fulfilled.} 

It is worth stressing, however, that if interest is on predictions or impulse responses, this does not cause any additional issues since the conditional mean function (which is the sum over $\bm f_j$ and $\bm g_j$) is identified. Exploiting basic properties of the Gaussian distribution one can easily show that the sum of $\bm f_j$ and $\bm g_j$ in \autoref{eq:fg}  gives rise to a new latent process $\bm m_j$ which is, again, Gaussian:
\begin{align*}
    \bm m_j \sim \mathcal{N}\left(\bm 0_T, \sqrt{\bm \Omega_j}  \left(K_{\bm \vartheta_{j1}}(\bm X_j,  \bm X_j)+ K_{\bm \vartheta_{j2}}(\bm Z_j,  \bm Z_j) \right) \sqrt{\bm \Omega_j}\right). 
\end{align*}
Hence, our model can be also viewed as a standard Gaussian process that combines information in $\bm X_j$ and $\bm Z_j$ by summing over two different kernels. This immediately implies that if we are interested in sampling from the posterior predictive distribution of $\bm y_{t+h}$ (and related functions such as impulse responses) it is sufficient to estimate $\bm m_j$.

\subsection{Selecting the hyperparameters associated with the kernel}
So far, we always conditioned on the hyperparameters that determine the shape of the Gaussian kernel. A simple way of specifying $\bm \vartheta_{j1}$ and $\bm \vartheta_{j2}$ is the  \textit{median heuristic} approach stipulated in \citet{chaudhuri2017mean}. This choice  works well in a wide range of applications featuring many covariates \citep[see, e.g.,][]{crawford2019variable}.  The median heuristic fixes  $\xi_{j1} = \xi_{j2} = 1$  and defines the inverse of the bandwidth parameter as: \begin{equation*}
\begin{aligned}
\kappa_{j1} &= \bar{\kappa}_{j1} = \text{median}_{t \tau} \left(\frac{1}{\lVert \bm x_{jt} - \bm x_{j\tau} \rVert} \right), \quad \kappa_{j2} = \bar{\kappa}_{j2} = \text{median}_{t \tau} \left(\frac{1}{\lVert \bm z_{jt} - \bm z_{j\tau} \rVert}\right),
\end{aligned}
\end{equation*}
for $j = 1, \dots, M$. This simple approach has the convenient property that it automatically selects a bandwidth which is consistent with the time series behavior of the elements in $\bm y_t$. To illustrate this, suppose that $y_{jt}$ is a highly persistent process (e.g., inflation or short-term interest rates). In this case, for $\tau=t-1$, the Euclidean distance $\lVert \bm x_{jt} - \bm x_{j\tau} \rVert$ will be quite small and, hence, the mean function $\bm f_j$ smoothly adjusts. If $y_{jt}$ is less persistent and displays large fluctuations (e.g., stock market or exchange rate returns), the Euclidean distance $\lVert \bm x_{jt} - \bm x_{j\tau} \rVert$ will be large and, thus, $\bm f_j$ allows for capturing this behavior. The dispersion in $\bm z_{jt}$ might have important implications for $y_{jt}$ if the aim is to model a trend in $y_{jt}$ that depends on other covariates. This could arise in a situation where the prior on $\bm f_j$ is set very tight (i.e., the posterior of $\bm f_j$ will be centered on zero) and information not coming from $\bm x_{jt}$ would then determine the behavior of $y_{jt}$. This discussion highlights how the median heuristic allows for flexibly discriminating between signal and noise and thus acts as a non-linear filter which purges the time series from high frequency variation, if necessary.

Given that we work with potentially large panels of time series, it is questionable that the median heuristic works  equally well for all elements in $\bm y_t$. As a solution, we propose to use the median heuristic to set up a discrete grid for both $\xi_{j1}$ ($\xi_{j2}$) and $\kappa_{j1}$ ($\kappa_{j2}$). For each element in this grid we specify a hyperprior. We use Gamma priors on all elements. For $j = 1, \dots, M$, that is
\begin{equation*}\label{eq:prior_xi}
\xi_{j1} \sim \mathcal{G} \left(\frac{1}{2}, \frac{1}{2c_{\xi1}}\right) \quad \text{and}  \quad \xi_{j2} \sim \mathcal{G} \left(\frac{1}{2}, \frac{1}{2c_{\xi2}}\right), 
\end{equation*}
for the linear shrinkage hyperparameters and 
\begin{equation*}\label{eq:prior_h}
\kappa_{j1} \sim \mathcal{G} \left(\frac{1}{2}, \frac{1}{2c_{\kappa 1}}\right) \quad \text{and} \quad \kappa_{j2} \sim \mathcal{G} \left(\frac{1}{2}, \frac{1}{2c_{\kappa 2}}\right),   
\end{equation*}
for the bandwidth parameters. Here, $c_{\xi1}, c_{\xi2}, c_{\kappa 1}$ and $c_{\kappa 2}$ are scalars that define the tightness of the hyperprior. In the empirical application, we set $c_{\xi1} = c_{\xi2} = c_\xi$ and $c_{\kappa 1} = c_{\kappa 2} = c_\kappa$. These parameters strongly influence the shape of the conditional mean and are crucial modeling choices and we set them through cross-validation. In our empirical application, we find that small values of $c_\kappa$ work reasonably well, yielding an informative prior that forces $\kappa_{j1}$ and $\kappa_{j2}$ towards zero. 

Based on this set of priors we can derive the conditional posterior distribution. Since $\xi_{j1}$ ($\xi_{j2}$) and $\kappa_{j1}$ ($\kappa_{j2}$) are placed on a grid, we can pre-compute several quantities related to the kernel (such as inverses and Cholesky factors) while at the same time infer them from the data with sufficient accuracy, which is crucial for precise inference. In what follows, we center these grids around the median heuristic and additionally take into account the considerations of the informative Gamma priors:  
\begin{equation*}
\kappa_{jk} \in  [0.1 \bar{\kappa}_{jk}, 2 \bar{\kappa}_{jk}] \quad \text{and} \quad \xi_{jk} \in [0.04, 4], \quad \text{for } j = 1, \dots, M \text{ and } k = 1, 2.    
\end{equation*}
Here, the intervals indicate the minimum (maximum) value supported for each hyperparameter. Within this two dimensional range, we define a discrete grid of around $1000$ combinations with equally sized increments along each dimension.\footnote{Implicitly, this two dimensional grid results in a prior view in which any hyperparameter combination not included in the grid has zero support.} The corresponding posterior is discrete and we can use inverse transform sampling to carry out posterior inference. Further details are provided in Sub-section \ref{sec: posterior_comp}.

\subsection{Posterior computation}\label{sec: posterior_comp}
Posterior inference for the GP-VAR is carried out using a novel yet conceptually simple MCMC algorithm which cycles between several steps. In this section we will focus on how to sample from the posterior of $\bm f_j$, $p(\bm f_j | \bullet)$, with $\bullet$ denoting conditioning on everything else, and $\bm \vartheta_{j1}$. Sampling from $p(\bm g_j|\bullet)$ and $p(\bm \vartheta_{j2}|\bullet)$ works analogously with some adjustments. These relate to the fact that we introduce a linear restriction that $(\bm \iota' \bm \iota)^{-1} \bm \iota' \bm g_j = 0$, with $\bm \iota$ denoting a $T \times 1$ vector of ones. The corresponding conditional  posterior distribution is  a hyperplane truncated Gaussian where efficient sampling algorithms are available \citep[see][]{cong2017fast}. Further details can be found in Sub-section \ref{app:restr} of the Online Appendix. It is worth stressing that we sample $\bm f_j$ and $\bm g_j$ separately. This increases the computational burden slightly but allows us to consider both latent processes separately from each other. In case our focus is purely on prediction or impulse response analysis, one can also simulate the process $\bm m_j$ defined in Sub-section \ref{sec: functionlearning} without any additional restriction. Both procedures yield exactly the same results.

{Generalizing the results in Section \ref{sec: intro_GP}, it can be shown that} the posterior of $\bm f_j$ is Gaussian for all $j$:
\begin{equation*}
    \bm f_j | \bullet \sim \mathcal{N}(\overline{\bm f}_j, \overline{\bm V}_{\bm f_j}),
\end{equation*}
with posterior moments given by:
\begin{align*}
    \overline{\bm V}_{\bm f_j} &= \sqrt{\bm \Omega_j} \left(K_{\bm \vartheta_{j1}}(\bm X_j, \bm X_j) - K_{\bm \vartheta_{j1}}(\bm X_j, \bm X_j)\left(K_{\bm \vartheta_{j1}}(\bm X_j, \bm X_j) + \bm I_T \right)^{-1}K_{\bm \vartheta_{j1}}(\bm X_j, \bm X_j)  \right)\sqrt{\bm \Omega_j}, \label{eq: post_covariance} \\
            \overline{\bm f}_j &= \sqrt{\bm \Omega_j} K_{\bm \vartheta_{j1}}(\bm X_j, \bm X_j) \left(K_{\bm \vartheta_{j1}}(\bm X_j, \bm X_j) + \bm I_T    \right)^{-1}  \sqrt{\bm \Omega_j}^{-1} \bm \left(\bm Y_j - \bm g_j - \sum_{k=1}^{j-1} q_{jk} \bm Y_k\right),
\end{align*}
where for $j=1$, the term $\sum_{k=1}^{j-1} q_{jk} \bm Y_k$ is excluded.

In principle, computing the inverse and the Cholesky factor of $\overline{\bm V}_{\bm f_j}^{-1}$ constitutes the main bottleneck when it comes to sampling from $p(\bm f_j|\bullet)$. This is especially so if $T$ is large. But, in common macroeconomic applications which use quarterly US data, $T$ is moderate and thus computation is feasible. In our case, even if interest centers on using monthly data or even higher frequencies, we can exploit the convenient fact that, conditional on the hyperparameters $\xi_{j1}$ and $\kappa_{j1}$, 
\begin{equation*}
\bm B_{\bm f_j} \bm B'_{\bm f_j} = \left(K_{\bm \vartheta_{j1}}(\bm X_j, \bm X_j) - K_{\bm \vartheta_{j1}}(\bm X_j, \bm X_j)\left(K_{\bm \vartheta_{j1}}(\bm X_j, \bm X_j) + \bm I_T \right)^{-1}K_{\bm \vartheta_{j1}}(\bm X_j, \bm X_j)  \right)
\end{equation*}
as well as its Cholesky factor $\bm B_{\bm f_j}$ can be pre-computed. In addition, notice that
\begin{equation*}
    \overline{\bm V}_{\bm f_j} = \bm C_{\bm f_j} \bm C'_{\bm f_j} = \left({\sqrt{\bm \Omega_j}} \bm B_{\bm f_j}\right)\left({\sqrt{\bm \Omega_j}} \bm B_{\bm f_j} \right)'.
\end{equation*}
These practical properties (due to the conjugate structure) substantially speed up computation in terms of sampling from $p(\bm f_j|\bullet)$. 

These results are conditional on the hyperparameters. As outlined in the previous sub-section, we will estimate them by defining a discrete two dimensional grid of $1000$ combinations. For each hyperparameter combination on this grid, we compute the corresponding kernel $K_{\bm \vartheta_{j1}}(\bm X_j, \bm X_j)$ as well as all relevant quantities (i.e., $\bm C_{\bm f_j}$). Based on these values  we jointly evaluate the conditional posterior ordinate by applying Bayes theorem. The exact form of the conditional likelihood is given by: 
\begin{equation*}
\begin{aligned}
p(\bm f_j|\bm \vartheta_{j1}, \bm \Omega_j) =& (2 \pi)^{-\frac{T}{2}} \times \text{det}\left(\sqrt{\bm \Omega_j}  K_{\bm \vartheta_{j1}}(\bm X_j, \bm X_j) \sqrt{\bm \Omega_j}\right)^{-\frac{1}{2}} \\ &\times \exp \left \{- \frac{1}{2}\left(\bm f_j' \left(\sqrt{\bm \Omega_j}  K_{\bm \vartheta_{j1}}(\bm X_j, \bm X_j) \sqrt{\bm \Omega_j}\right)^{-1} \bm f_j \right) \right\}.
\end{aligned}
\end{equation*} 
Note that the shape of $K_{\bm \vartheta_{j1}}(\bm X_j, \bm X_j)$ depends on the hyperparameters $\bm \vartheta_{j1} = (\xi_{j1}, \kappa_{j1})'$, which we want to update. For each pair of values $\bm \vartheta_{j1}^{(s)}$ on our two dimensional grid (with $s$ denoting a specific combination), we compute the corresponding kernel $K_{\bm \vartheta_{j1} = \bm \vartheta_{j1}^{(s)}}(\bm X_j, \bm X_j)$ as well as $\text{det}\left(K_{\bm \vartheta_{j1} = \bm \vartheta_{j1}^{(s)}}(\bm X_j, \bm X_j)\right)$ and $\left(K_{\bm \vartheta_{j1} = \bm \vartheta_{j1}^{(s)}}(\bm X_j, \bm X_j)\right)^{-1}$ prior to MCMC sampling. Hence, within our sampler evaluating the likelihood is straightforward and computationally efficient. All that remains is to multiply the likelihood with the prior. The corresponding posterior ordinates for each $\bm \vartheta_{j1}^{(s)}$ are used to compute probabilities to perform inverse transform sampling to sample from $p(\bm \vartheta_{j1}|\bullet)$. 


Conditional on $\bm f_j$ and $\bm g_j$, the remaining parameters (i.e., the free elements in $\bm Q$, the log-volatilities and the associated coefficients in the corresponding state equations) can be sampled through (mostly) standard steps. One modification relates to how we sample the volatilities in $\bm \Omega_j$.  The main difference stems from the fact that the volatilities in $\bm \Omega_j$ also show up in the prior on $\bm f_j$ and $\bm g_j$.  To circumvent this issue we integrate out the latent processes $\bm f_j$ and $\bm g_j$.  This calls for a minor adjustment of the original sampler by integrating out the latent processes $\bm f_j$ and $\bm g_j$ first and then sampling the log-volatilities using an independent Metropolis Hastings update similar to the one proposed in \cite{chan2017stochastic}. We provide additional details and the full posterior simulator in Section \ref{sec:App C} of the Online Appendix.

\subsection{Forecasts and generalized impulse responses}\label{sec:GIRFs}

In non-parametric models such as the GP regression described in Section \ref{sec: intro_GP}, the effect of the covariates on $\bm y_t$ are typically analyzed through so-called partial dependence plots \citep{friedman2001greedy}. Our large dimensional setting and the fact that we have a VAR-type structure in the conditional mean, imply that partial dependence plots are difficult to compute and visualize since they would require integration over a large number of covariates.  Moreover,  VARs are dynamic models and partial dependence plots are difficult to employ in dynamic settings. To capture non-linear model dynamics and possible relations across variables, impulse responses are used to investigate the effects of structural shocks on $\bm y_t$. In this paper, we will follow this route as well. Because the model is highly non-linear, we need to resort to generalized impulse responses (GIRFs) originally proposed in \cite{koop1996impulse}. {As GIRFs are based on (multi-step-ahead) forecasts, and since forecasting with the GP-VAR is of interest by itself, we start with a discussion on forecast computation.}

We begin by computing the predictive distribution of the one-step-ahead  forecasts $p({\bm y}_{t+1}|\mathcal{I}_t)$, with $\mathcal{I}_t$ denoting all available information up to time $t$. The one-step-ahead predictive distribution is obtained by simulating from the predictive distribution of $\bm m_{t+1}$, $p(\bm m_{t+1}|\mathcal{I}_t)$, and sampling from the marginal distribution of the shocks $\bm \varepsilon_{t+1} \sim \mathcal{N}(\bm 0, \bm H_{t+1})$. The draw from the one-step-ahead predictive density is used to set up $\bm x_{t+2}$ and $\bm z_{t+2}$. Based on these, we can compute the corresponding kernels and obtain a draw $\bm m_{t+2}$ from the  density $p(\bm m_{t+2}|\mathcal{I}_t)$. Again, a draw from $\bm y_{t+2} \sim p(\bm y_{t+2}|\mathcal{I}_t)$ is obtained by sampling from the marginal shock distribution $\bm \varepsilon_{t+2} \sim \mathcal{N}(\bm 0, \bm H_{t+2})$ and adding this draw to $\bm m_{t+2}$. Higher order forecasts are obtained analogously.  The resulting  predictive distribution of $\bm y_{t+h}$ will be highly non-Gaussian and might feature heavy tails and/or asymmetries. This forms the baseline.

The GIRFs are computed as follows.  To analyze the effects of a structural disturbance (such as an uncertainty shock) we assume that the uncertainty indicator is (without loss of generality) in the $j^{th}$ position in $\bm y_t$.  A corresponding shock of size $\varsigma$ in time $t$ to the uncertainty indicator shifts all elements in $\bm y_t$ by $\varsigma ~ \bm q_j$, i.e., the $j^{th}$ column of $(\bm I - \bm Q)^{-1}$ scaled by a scalar that reflects the shock size $\varsigma$. The other shocks are sampled, again, from their marginal distributions, i.e., for all $i \neq j$ we have that $\varepsilon_{it} \sim \mathcal{N}(0, \omega_{it})$. Based on this we draw from  the predictive distribution conditional on the uncertainty shock $\hat{\bm y}_{t+1} \sim p(\bm y_{t+1}|\mathcal{I}_t, \varepsilon_{jt}=\varsigma)$ and use this draw to compute  $\hat{\bm x}_{t+2}$ and $\hat{\bm z}_{t+2}$. For higher order conditional forecasts we proceed as in the case of the unconditional forecast distribution by simulating from the marginal distribution of the structural shocks which are added to the conditional mean forecasts $\hat{\bm m}_{t+h}$.

The corresponding dynamic responses are then obtained by subtracting the mean of the unconditional predictive distribution from the conditional (on the uncertainty shock) predictive density. This yields:
\begin{equation*}
    \bm \delta_{ht} = \mathbb{E}(\bm y_{t+h}|\mathcal{I}_t, \varepsilon_{jt}=\varsigma) - \mathbb{E}(\bm y_{t+h}|\mathcal{I}_t). \label{eq:girf_1}
\end{equation*}
Notice that $\bm \delta_{ht}$ is state-dependent and, due to the non-linear nature of the conditional mean function, allows for asymmetries in how $\bm y_t$ reacts to shocks.\footnote{The full details on computing generalized impulse response functions can be found in Section \ref{app:girfs} of the Online Appendix.} This gives rise to two inferential opportunities. First, one can assess how a given shock has impacted the economy in a given point in time. This allows us to investigate whether transmission mechanisms depend on the underlying state of the economy. Second, the non-linear mean function directly implies that shock transmission can be asymmetric, so that positive shocks might feed through the economy differently than negative shocks, and non-proportional, so that larger shocks can have proportionally different effects than smaller shocks. In all our empirical work we will exploit both dimensions and focus on asymmetries in the sign, and non-proportionality in the size, of the shock as well as explicitly consider state dependencies by computing $\bm \delta_{ht}$ over time. Finally, we also integrate out uncertainty with respect to the state of the economy by averaging over all values of $t$.

\section{Illustration using synthetic data}\label{sec:synth}
In this section, we illustrate the computational merits of our approach and evaluate whether it successfully recovers different features of a highly non-linear DGP. 

To illustrate our methods, we simulate $T=200$ observations from a highly non-linear small-scale VAR with $M=3$ equations. The three equations differ in terms of whether they are linear or non-linear in the parameters but also with respect to the distribution of the shocks. Non-linearities are captured in two ways. First, we assume a break point and second we assume non-linear relations between the response variable and the lags of the other variables. In all these equations, we assume that the functions $f_j$ and $g_j$ differ to assess whether our approach is capable of discriminating between the two. The precise form of our DGP is given by:
\begin{equation*}
\bm y_t =  F(\bm x_t) +  G(\bm z_t) + \bm Q \bm y_{t} + \bm \varepsilon_t, \,\quad  \bm \varepsilon_t \sim \mathcal{N}(\bm 0_M,\bm H_t),
\end{equation*}
with 
\begin{small}
\begin{equation*}
\begin{aligned}
F(\bm x_t) =& \begin{pmatrix}
f_1 (\bm x_{1t}) \\
f_2 (\bm x_{2t}) \\
f_3 (\bm x_{3t})
\end{pmatrix} = 
\begin{pmatrix}
\sum_{k = 1}^{p} \phi_{11,k} y_{1t-k} \\
\sum_{k = 1}^{p} \phi_{22,k} y_{2t-k} \times \mathcal{I}(t \leq 100) + \phi_{22,1} y_{2t-1} \times  \mathcal{I}(t > 100) \\
\frac{1}{12} \sin(\frac{\pi}{2} y_{3t-1} y_{3t-2}) + \frac{1}{3}(y_{3t-3} - 1)^{2} + \frac{1}{12} y_{3t-4} + \frac{1}{12} y_{3t-5} \\
\end{pmatrix},  \\
G(\bm z_t) =& \begin{pmatrix}
g_1 (\bm z_{1t}) \\
g_2 (\bm z_{2t}) \\
g_3 (\bm z_{3t})
\end{pmatrix} = \begin{pmatrix}
0 \\
0 \times \mathcal{I}(t \leq 100) + \sum_{k = 1}^{p} \sum_{j \in \{1,3\}} \phi_{2j,k} y_{jt-k} \times \mathcal{I}(t > 100)  \\
\frac{1}{18} \sin(\frac{\pi}{2} y_{1t-1} y_{2t-1}) + \frac{2}{9}(y_{1t-2} - 1)^{2} + \frac{1}{18} y_{1t-3} + \frac{1}{18} y_{2t-5} \\
\end{pmatrix},
\end{aligned}
\end{equation*}
\end{small}
where $\bm H_t = \text{diag}(\omega_{1t},\omega_{2t},\omega_{3t})$, $\phi_{11,1} = 0.8$, $\phi_{22,1} = 0.65$, $\phi_{ij,k} \sim \mathcal{N}\left(0, \left(\frac{0.3}{k}\right)^2\right)$ for $i \neq j$ and $k = 1, \dots p$, and $\mathcal{I}(\bullet)$ denotes the indicator function that equals one if its argument is true and zero otherwise. The free elements in $\bm Q$, are also simulated from a Gaussian distribution with $q_{jk} \sim \mathcal{N}(0,0.1^2)$. Moreover, we introduce an SV specification with heavy tails for the structural error variances $\omega_{jt} =  \lambda_{jt} \tilde{\omega}_{jt}$ with each $\tilde{h}_{jt} = \log \tilde{\omega}_{jt}$ following an independent random walk law of motion: $\tilde{h}_{jt} = \tilde{h}_{jt-1} + \sigma_{\tilde{h} j} u_{\tilde{h} t}$, with $u_{\tilde{h} t} \sim \mathcal{N}(0, 1)$. For each equation, we set the initial state $\tilde{\omega}_{j0} = \exp \tilde{h}_{j0}  = 0.01$ and the state innovation variance $\sigma_{\tilde{h} j} = 0.01$. We consider (conditionally) $t_3$-distributed errors with three degrees of freedom in the first equation by simulating $\lambda_{1t} \sim \mathcal{G}^{-1}(3/2, 3/2)$ and (conditionally) Gaussian-distributed errors for the second and third equation by setting $\lambda_{2t} = \lambda_{3t} = 1$ for all $t$.

\begin{figure}[!ht]
\caption{Posterior distributions of $\bm f_j$, $\bm g_j$, $\bm m_j (= \bm f_j + \bm g_j)$ and $\bm y_j$ versus actual realizations for each of the three equations in the DGP.}
\label{fig:synth_data_1}
\centering
\hspace*{-0.7em}
\begin{minipage}{\textwidth}
\centering
\includegraphics[width = 0.95\textwidth]{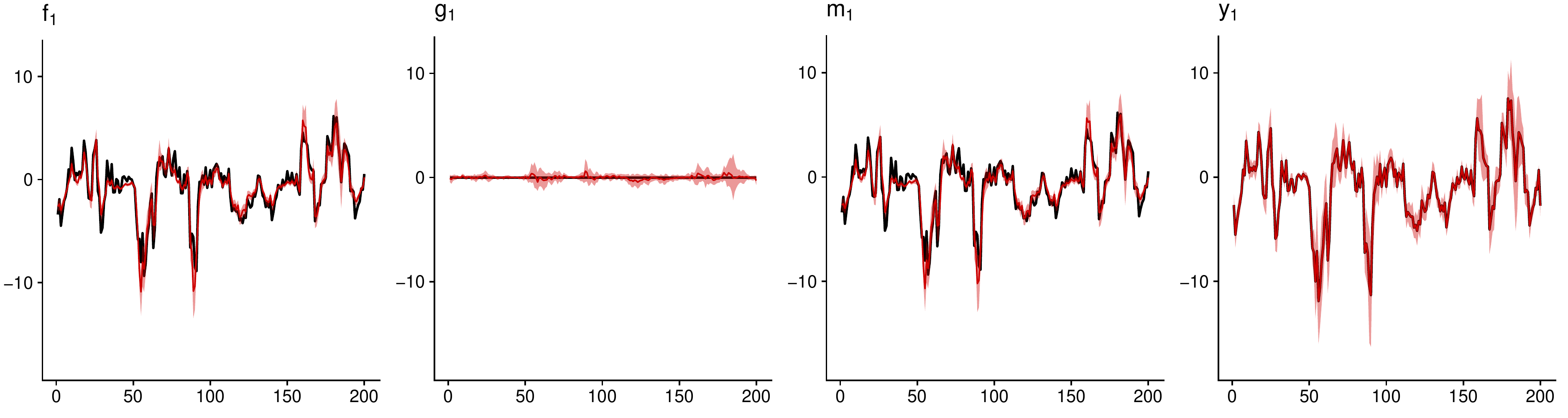}
\end{minipage}
\begin{minipage}{\textwidth}
\centering
\hspace*{-1em}
\includegraphics[width = 0.95\textwidth]{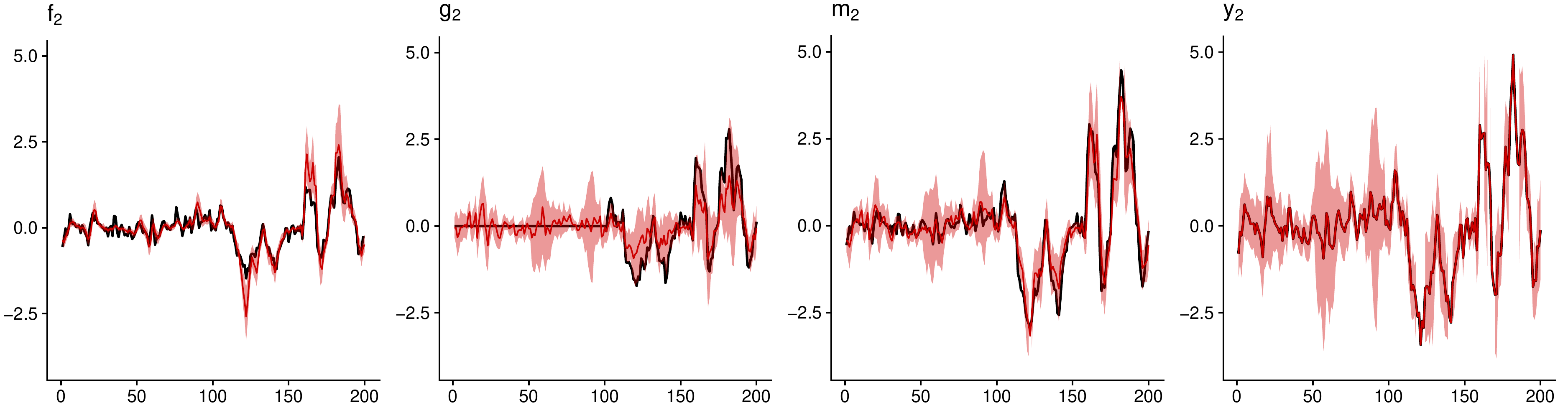}
\end{minipage}
\begin{minipage}{\textwidth}
\centering
\includegraphics[width = 0.95\textwidth]{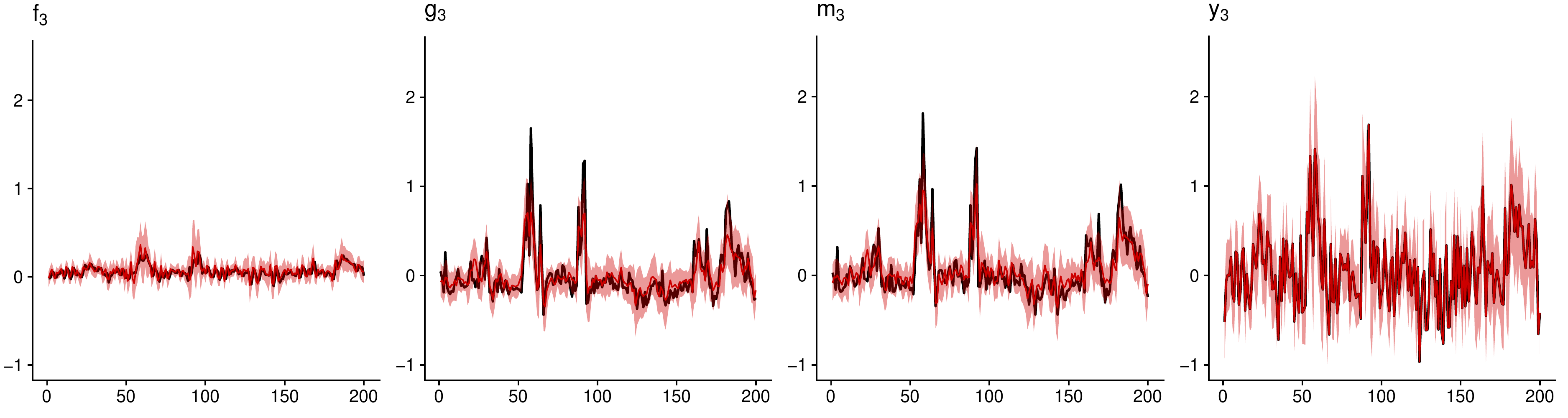}
\end{minipage}
\begin{minipage}{\textwidth}
\centering
\hspace*{15pt}\includegraphics[scale=0.55]{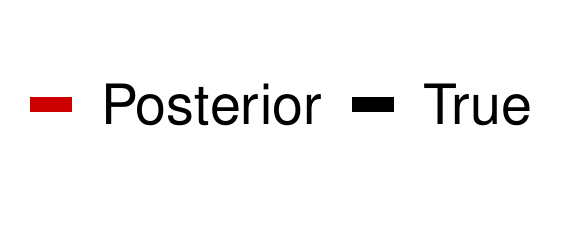}
\end{minipage}
\begin{minipage}{\textwidth}
\scriptsize \textit{Notes:} Results are obtained from simulating a single realization from the proposed DGP. Horizontal panels refer to each equation of the DGP. The solid black lines denote the actual outcomes of the respective functions, while the red solid lines represent the posterior medians and the red shaded areas the $90\%$ posterior credible sets of the respective fitted values.
\end{minipage}
\end{figure}

Figure \ref{fig:synth_data_1} shows results obtained from simulating a single realization from the DGP. Horizontal panels refer to each equation of the DGP while vertical panels show the different components. The red shaded areas represent the $90\%$ posterior credible set of {$\bm f_j, \bm g_j$, $\bm m_j (= \bm f_j + \bm g_j$}) and $\bm y_j$, while the solid black line denotes the actual outcome of these quantities. The figure suggests that our approach is capable of detecting different functional relations between $\bm y_t$, $\bm x_t$ and $\bm z_t$. Across all three equations, we find that the estimated conditional mean function $\bm m_j$ tracks the actual value rather well. It is also worth stressing that if the DGP features non-Gaussian shocks, the model recovers the true mean function particularly well (see the first row of Figure \ref{fig:synth_data_1}).  Zooming into the estimates for the different latent components reveals that most of this strong fit is driven by accurate estimates of $\bm f_j$. Considering the estimates for $\bm g_j$ suggests that, if the actual relationship between $\bm z_t$ and $\bm y_t$ is non-existent, our approach accurately detects this behavior. However, there are some cases where $\bm f_j$ soaks up variation in $\bm g_j$. This, however, does not impact the mean estimate $\bm m_j$. 

Since this discussion has been based on a single draw from the DGP, one might ask whether the strong performance of the GP-VAR is due to a particularly favorable realization from the DGP. To briefly investigate whether this is the case we show, in the first row of Table \ref{tab:Corr}, the average correlations between the true and posterior median of the functions for 100 draws from the DGP. Numerical standard errors (across these $100$ repetitions) are shown in the second row. The first row indicates that correlations are high, reaching $0.95$ for the first, $0.93$ for the second and $0.75$ for the third equation. The fact that mean correlations slightly decrease are mainly driven by the fact that equations two and three feature substantial high frequency movements which, in our framework, is mostly picked up by the stochastic volatility component.

One key advantage of the GP-VAR is that we can remain agnostic on the form of non-linearities the conditional mean function might take. This is confirmed for synthetic data where, irrespective of the non-linear nature of the model, the estimated sum of $\bm f_j$ and $\bm g_j$ (i.e., $\bm m_j$) closely tracks the dynamics of the actual outcome. This finding holds both for the linear case (i.e., the first equation){, even with fat-tailed errors,} and for highly non-linear situations (i.e., the second and third equations).

\begin{table}[!t]
{\scriptsize
\begin{center}
\caption{Correlations between the posterior median of $\bm m_j (= \bm f_j + \bm g_j)$ and the actual realization for each of the three equations in the DGP.\label{tab:Corr}}
\begin{tabular*}{\textwidth}{l @{\extracolsep{\fill}} cccc}
 \toprule
 & $\bm m_1$ & $\bm m_2$ & $\bm m_3$ \\ 
\midrule
\texttt{Avg.} & 0.945 & 0.929 & 0.754 \\ 
\texttt{SD}    & 0.012 & 0.020 & 0.045 \\ 
\bottomrule
\end{tabular*}
\begin{minipage}{\textwidth}
\vspace*{5pt}
\scriptsize 
\noindent \textit{Notes:} Results are obtained from simulating $100$ realizations from the proposed DGP. \texttt{Avg.} refers to the average correlation, while \texttt{SD} to the standard deviation of correlations across realizations. 
\end{minipage}
\end{center}}
\end{table}

\begin{figure}[!t]
\caption{Computation time for 1000 draws from the joint posterior distribution.}
\label{fig:comp_times}
\centering
\includegraphics[width=0.95\textwidth]{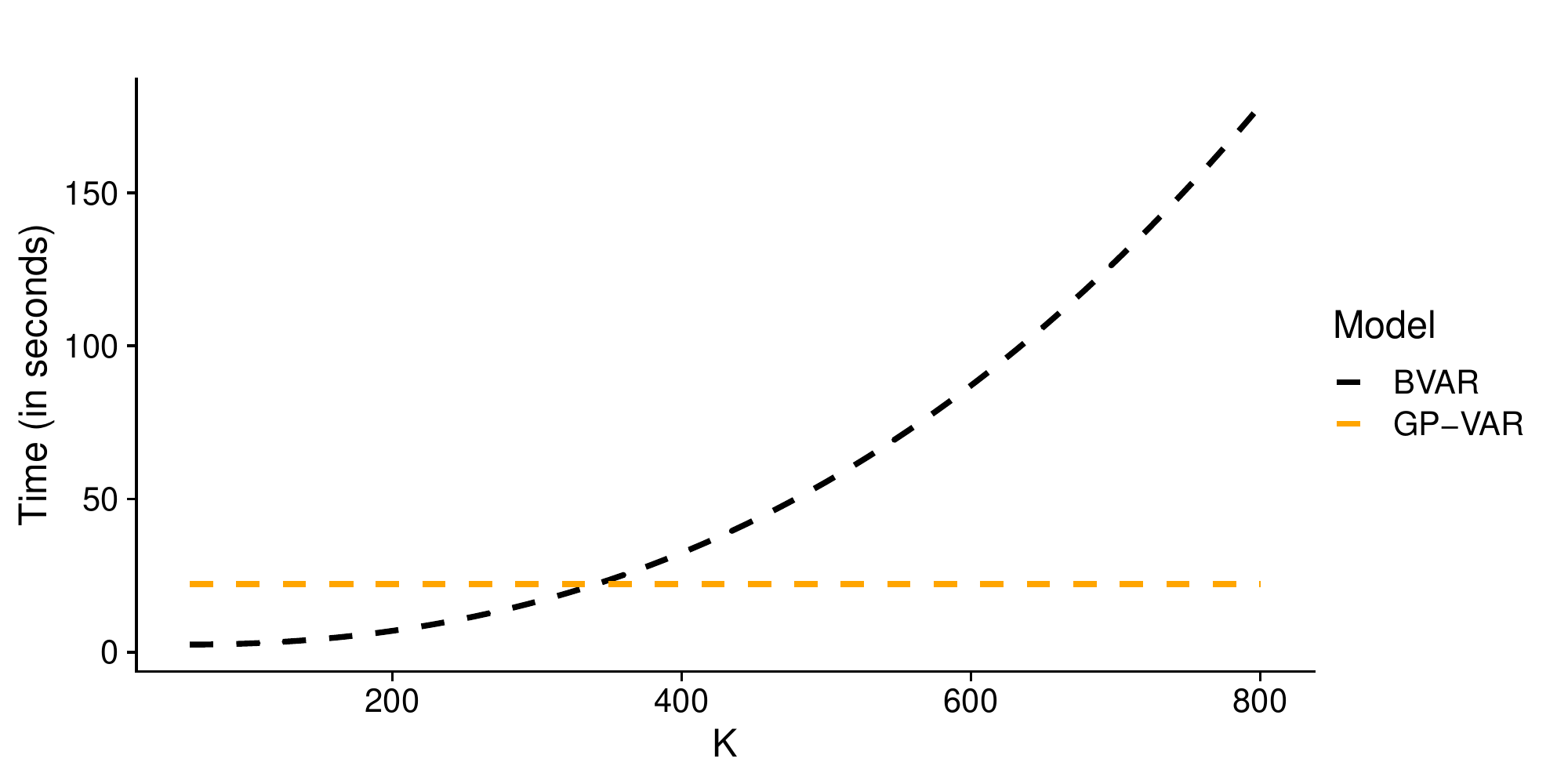}
\begin{minipage}{\textwidth}
\scriptsize \textit{Notes:} Computation time for 1000 draws from the joint posterior distribution of a standard BVAR (black) and the GP-VAR (orange). Since our approach can be parallelized, the actual times for generating 1000 draws from the joint posterior of the full system is approximately $M$ times the runtimes reported in this figure. Computation times are based on a desktop machine with an AMD Ryzen 7 5800X 8-Core processor.
\end{minipage}   
\end{figure}

We have stressed that our approach is computationally efficient and scalable to large datasets. To investigate this claim more carefully, Figure \ref{fig:comp_times} shows the time required to generate 1000 draws from the joint posterior for a given equation across different values of $K$ and for $T=200$. We show the computation times for our GP-VAR and a VAR with SV.  Since our approach is embarrassingly parallel the actual times for generating a draw from the joint posterior of the full system are approximately $M$ times the runtimes reported in the figure.\footnote{Since we need to augment the $j^{th}$ equation with the contemporaneous values of the preceding $j-1$ equations, this statement is only approximately valid.}

The most striking take away from the figure is that the computation time of the GP-VAR does not depend on $K$. This implies that increasing the number of lags and/or endogenous variables does not impact estimation times considerably. By contrast, the time necessary to generate a draw from the joint posterior of the VAR rises rapidly in $K$.  This shows that our approach scales well in high dimensions and, in fact, is much faster than competing approaches to non-linear VAR models such as TVP-VARs or regime-switching VARs. 

To provide a rough gauge on actual estimation times for practitioners, MCMC estimation of the GP-VAR with eight endogenous variables and five lags takes around 30.5 minutes on a standard desktop computer (for 10,000 MCMC draws). These are gross estimation times and thus include pre-computation of matrices used during MCMC simulation and are not based on parallel computation of the individual equations (which is possible due to the structural form). Estimating larger models (such as the 64 variable GP-VAR) takes around four hours. 

\section{Modeling the US economy using GP-VARs}\label{sec:macrofor}
In this section, we apply our GP-VAR to US macroeconomic data. Sub-section \ref{ssec:insmp} provides details on the data and Sub-section \ref{ssec:WAIC} discusses whether the GP-VAR ﬁts the data well, shows some of its key in-sample features,  and investigates its forecast performance.

\subsection{Data overview}\label{ssec:insmp}
We use the quarterly version of the dataset proposed in \cite{mccracken2016fred} and consider time series that range from $1960$Q$1$ to $2019$Q$4$. We exclude the years of the Covid-19 pandemic to make the comparison with a linear VAR similar to that used in \cite{jurado2015measuring} (JLN) sensible. In the following, we consider four different specifications that differ in the number of endogenous variables used. These specifications are:
\begin{itemize}
	\item GP-VAR-8: This dataset is patterned after the original JLN dataset. It includes the JLN macroeconomic uncertainty index (labeled as \texttt{UNC}), real GDP (\texttt{RGDP}), civilian employment (\texttt{EMP}), average weekly working hours in manufacturing (\texttt{AWH}), consumer price index (\texttt{CPI}), average hourly earnings in manufacturing (\texttt{AHE}), Fed funds rate (\texttt{FFR}), and the S\&P 500 (\texttt{SP500}). 
	\item GP-VAR-16: In addition to the variables in the VAR with $M=8$ endogenous variables, we also include the components of GDP (such as real personal consumption and real private fixed investment), additional labour market variables (such as unemployment, initial claims, average weekly working hours and average hourly earnings across all sectors), housing starts, as well as the real M2 money stock.
	\item GP-VAR-32: On top of the variables of the VAR with $M=16$ variables, we include important financial variables, further data on housing and data on loans. 
	\item GP-VAR-64: The largest model we consider features $M=64$ endogenous variables. The set of endogenous variables is obtained by taking the variables with $M=32$ and including additional financial variables and data on manufacturing.
\end{itemize}
All variables are transformed to be approximately stationary and we include five lags of the endogenous variables.
The precise variables included (and transformations applied to each variable) are shown in Table \ref{tab:data} in the Online Appendix.  We consider these different model sizes for several reasons. First, we would like to assess how adding additional information impacts the forecasting performance and the responses of key variables to an uncertainty shock. Second, we are interested in the relationship between non-linearities and the size of the model.

\subsection{Predictive evidence and in-sample features}\label{ssec:WAIC}
In this section, we start by providing some predictive evidence of our GP-VAR and investigate how the role of the parameters associated with the kernel impact predictive accuracy. To this end, we employ a recursive forecasting design. Our initial training period goes from $1960$Q$1$ to $1999$Q$4$. After computing the one- and four-step-ahead predictive distributions, we add an additional observation, recompute all models and simulate from the corresponding predictive densities. This procedure is repeated until the end of the sample ($2019$Q$4$) is reached.  

We consider three ways of specifying  the equation-specific ($j = 1, \dots, M)$ and kernel-specific $(k = \{1, 2\})$ hyperparameters. The first one is a semi-automatic approach that is based on putting the (inverse) length scale and the linear scaling parameter on a two dimensional grid $\kappa_{jk} \in [0.1 \bar{\kappa}_{jk}, 2 \bar{\kappa}_{jk}]$ and $\xi_{jk} \in [0.04, 4]$. Notice that the median heuristic is used to determine the lower and upper bound of the grid. The second approach (labeled ``semi-automatic w/o linear scaling" in Table \ref{tab:lpl3vars}) puts  $\kappa_{jk} \in  [0.1\bar{\kappa}_{jk}, 2 \bar{\kappa}_{jk}]$ on a grid and sets $\xi_{jk} = 1$.  Finally, we also consider a specification that does not rely on the median heuristic (labeled ``naive" in the table).  The grid is $\kappa_{jk} \in [0.1,2]$ and $\xi_{jk}$ is again set equal to $1$.

As competing models, we include the standard Minnesota BVAR with SV, a TVP-VAR-SV  similar to the one used in \cite{primiceri2005time} and the BART-VAR with SV proposed in \cite{huber2022inference}. All models are estimated for different model sizes and benchmarked to a small-scale Minnesota BVAR with SV {for the three variables we focus on: real GDP growth (RGDP), CPI inflation (CPI), and the Fed funds rate (FFR).} For the three focus variables, we compute log predictive Bayes factors (LPBFs) relative to the small-scale BVAR with SV, so that positive numbers indicate that a given model works better than the benchmark while negative values suggest a weaker forecasting performance. The LPBFs do not only take into account how well a given model predicts the realization of a given variable but also factor in forecasting performance for higher order features of the predictive distribution \citep[for a discussion, see][]{geweke2010comparing}. To investigate whether controlling for heteroskedasticity pays off in the GP-VAR, we also consider homoskedastic variants of the GP-VAR in the upper part of the table.

\input{LPSfinal_SVhom}

\autoref{tab:lpl3vars} shows the results of our forecasting exercise across different model sizes.  The columns ``Joint" show the joint LPBF for the three focus variables and thus provide a comparable (across model sizes) metric of overall predictive accuracy. Considering joint LBPFs reveals that GP-VAR SV with $M=16$ and a semi-automatic approach for hyperparameter elicitation improves upon all competing models for both forecast horizons. The smallest model ($M=8$) also yields competitive predictions. Once we further increase the size of the  dataset, predictive accuracy slightly deteriorates. Interestingly, and consistent with findings in  \cite{clark2021tail}, we find that the gains in predictive accuracy increase when we focus on higher forecast horizons. Comparing the models with SV to their homoskedastic counterparts paints a very consistent picture. Models which do not control for time variation in the error variances perform consistently worse than the models that have SV in the error terms. 

To drill deeper into which variables drive the overall forecasting performance, we now focus on the marginal LPBFs for the three focus variables. Starting with one-quarter-ahead predictions of GDP growth, we observe that the GP-VARs with SV are beaten by the BVAR-SV with $M=32$. When we turn off SV, predictive accuracy sometimes increases by small margins. This result, however, changes if we focus on higher order forecasts. For one-year-ahead predictions of GDP growth, the single best performing model is the largest ($M=64$) homoskedastic GP-VAR with the semi-automatic approach that fixes the linear scaling parameter to one. Strikingly, at that horizon and for this specific variable, using a homoskedastic specification is almost uniformly better than the corresponding SV setup. For inflation and the Fed funds rate, the smaller-sized GP-VARs again yield the best density forecasting performance, outperforming all competing models.  Finally, focusing on interest rate forecasts shows that GP-VARs with SV do well (for all model sizes) but once we turn off SV forecasts become highly imprecise. This is driven by the fact that during the zero lower bound, the conditional variance of the interest rate equation approaches zero and a model which assumes homoskedasticity fails to take that into account. 

\begin{figure}[!t]
\caption{Linear shrinkage parameters of equation-specific kernels for the GP-VAR-8.
\label{fig:linearshrink}}
\begin{minipage}{\textwidth}
\centering
\includegraphics[width = \textwidth]{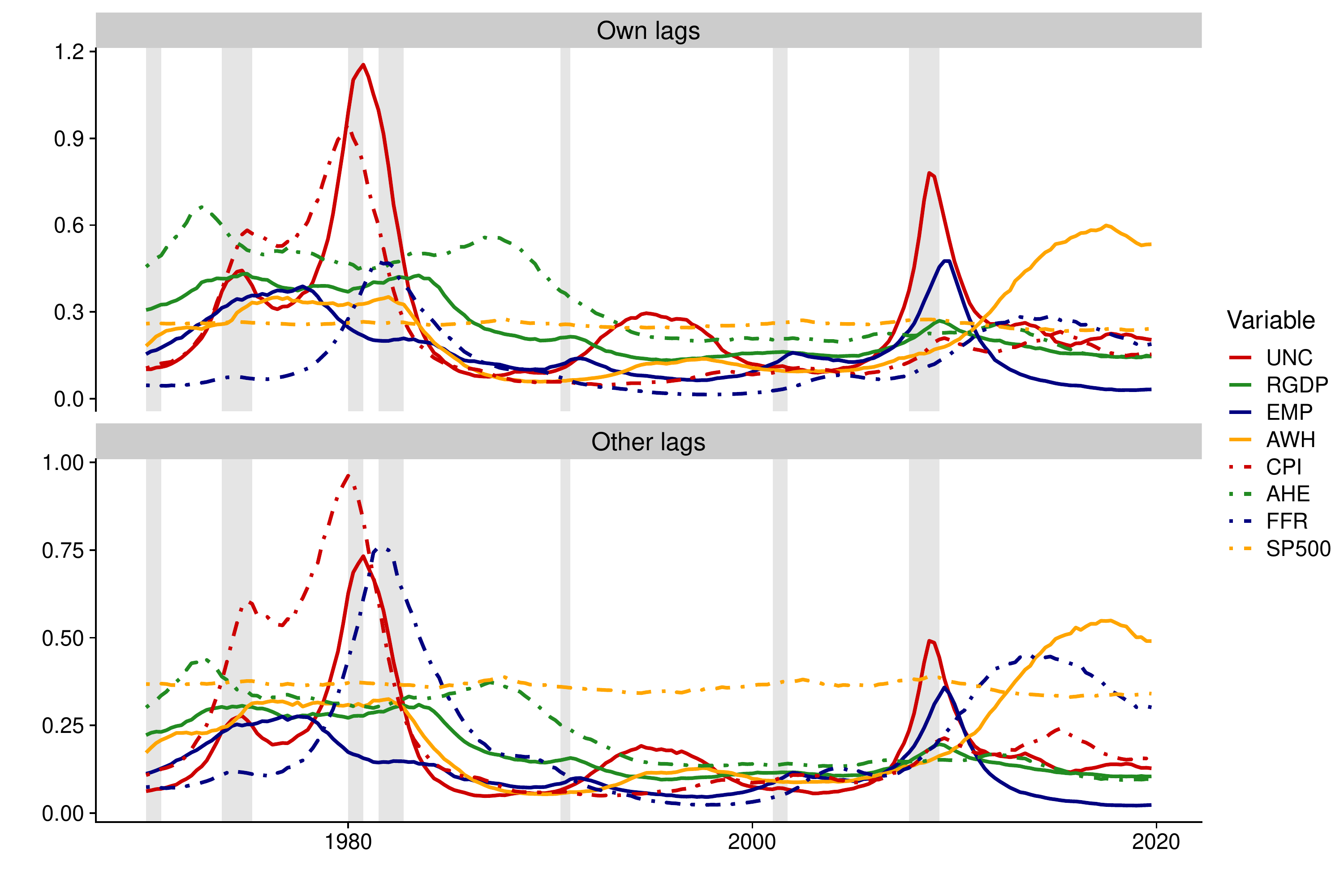}
\end{minipage}
\begin{minipage}{\textwidth}
\vspace*{-5pt}
\scriptsize \textit{Notes:} This figure reports the posterior means of the product of the error variances $\omega_{jt}$ and the linear scaling parameters for own lags ($\xi_{j1}$) and for other lags ($\xi_{j2}$), respectively. These two quantities correspond to the diagonal elements of the re-scaled kernels $\omega_{jt} \times k_{\bm \vartheta_{j1}}(\bm x_t, \bm x_t) = \omega_{jt} \xi_{j1}$ and $\omega_{jt} \times k_{\bm \vartheta_{j2}}(\bm z_t, \bm z_t) = \omega_{jt} \xi_{j2}$.
\end{minipage}
\end{figure}

After having established that the different GP-VARs with SV do well when used to forecast US macroeconomic quantities, we focus on some in-sample features for the GP-VAR with eight endogenous variables. To get an impression on how the linear shrinkage parameter evolves over time, Figure \ref{fig:linearshrink} plots the product of the error variances times the linear shrinkage parameter that determines the kernel of $\bm f_j$ and $\bm g_j$.   Two interesting features emerge. First, less shrinkage (larger parameter values) is applied to the own lags than to the lags of the other variables. This holds for most variables under scrutiny (except for CPI inflation and S\&P 500 returns). Second, less shrinkage is typically applied during recessionary times. In particular, we introduce little shrinkage during the recessions in the early 1980s and the financial crisis. Notice, however, that there are also some exceptions from this pattern (such as stock market returns, hours worked or hours employed).   This finding is, again, in line with results from the forecasting literature showing that more information is particularly useful during problematic times \citep[see, e.g.,][]{koop2013forecasting}.

\begin{figure}[!tbp]
\caption{Inverse length scale parameters of equation-specific kernels for the GP-VAR-8.
\label{fig:shrinkage_eqs}}
\centering
\begin{minipage}{\textwidth}
\centering
\vspace*{-15pt}
\includegraphics[scale=0.5]{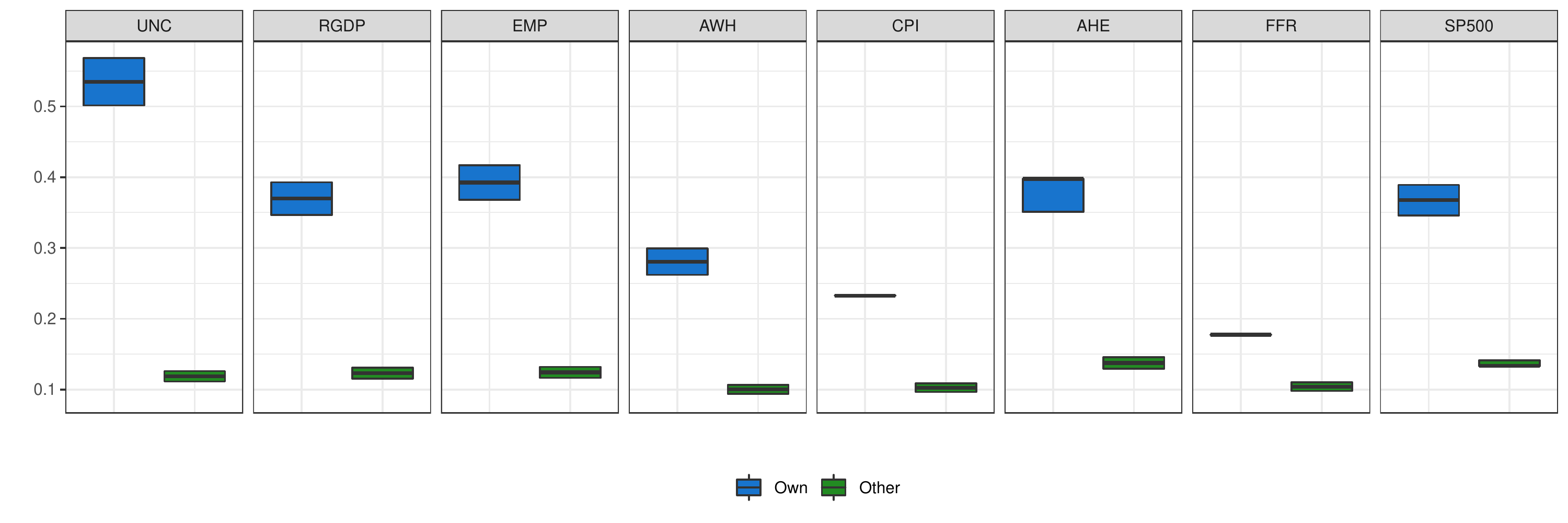}
\end{minipage}
\begin{minipage}{\textwidth}
\scriptsize \textit{Notes:} This figure reports the posterior summaries in the form of simplified boxplots of the inverse length scale parameters for own lags ($\kappa_{j1}$) and other lags ($\kappa_{j2}$), respectively. The solid black lines denote the posterior medians, while the blue (green) shaded areas represent the $50\%$ posterior credible sets (i.e., the posterior interquartile ranges).
\end{minipage}
\end{figure}
Finally, we investigate differences in $\kappa_{j1}$ and $\kappa_{j2}$ across equations ($j = 1, \dots M$) and variable types. Boxplots that show the posterior distribution of the inverse of the length scale parameters for own  and other lags are in \autoref{fig:shrinkage_eqs}. Recall that large values of $\kappa_{jk}~(k=1,2)$  imply more variation in the  latent processes whereas values of $\kappa_{jk}$ close to zero imply less variation in $\bm f_j$ and $\bm g_j$. A general pattern is that for all variables the hyperparameters associated with the kernel on own lags are considerably larger than the ones for the kernel related to the other lags. This indicates that the own lags of a given endogenous variable require more flexibility (i.e., functions that allow for much more variation) whereas the effect of other lags appears to be more linear. For the majority of variables (except for hours worked, CPI inflation and the Fed funds rate), the posterior distribution of the hyperparameter looks similar. For the three exceptions, $\kappa_{j1}$ is much smaller and more precisely estimated.

\section{The macroeconomic effects of uncertainty shocks}\label{sec:empircal}
We now analyze the effects of uncertainty shocks using our GP-VAR and focus on assessing how macroeconomic uncertainty feeds through the economy. In Sub-section \ref{ssec:bench}, as a benchmark exercise, we compare our impulse responses to those obtained from a model similar to that used by JLN.  In Sub-section \ref{ssec:asymmetries} we leverage the non-linear nature of the GP-VAR and analyze how the effects of uncertainty shocks change according to the sign or size of the shocks, and over time.

\subsection{Comparison with standard BVAR analysis}\label{ssec:bench}
We benchmark the IRFs of our GP-VAR-8 to the ones of a BVAR with SV that is closely related to the original JLN specification.\footnote{While they use a classical homoskedastic VAR estimated on monthly data, we work with a quarterly BVAR with SV.} In what follows, our focus will be on the variables discussed in JLN: year-on-year growth rates of output (measured through real GDP) and employment. We also show the responses of the uncertainty indicator and the quarter-on-quarter returns of the S\&P 500 and include the responses of the other variables in Section \ref{sec:App C} of the Online Appendix.
\begin{figure}[!ht]
\caption{Impulse responses of focus variables in the GP-VAR-8 relative to a small-scale BVAR. \label{fig:compJLN}}
\begin{minipage}{\textwidth}
\centering
\vspace*{-10pt}
\includegraphics[width = \textwidth]{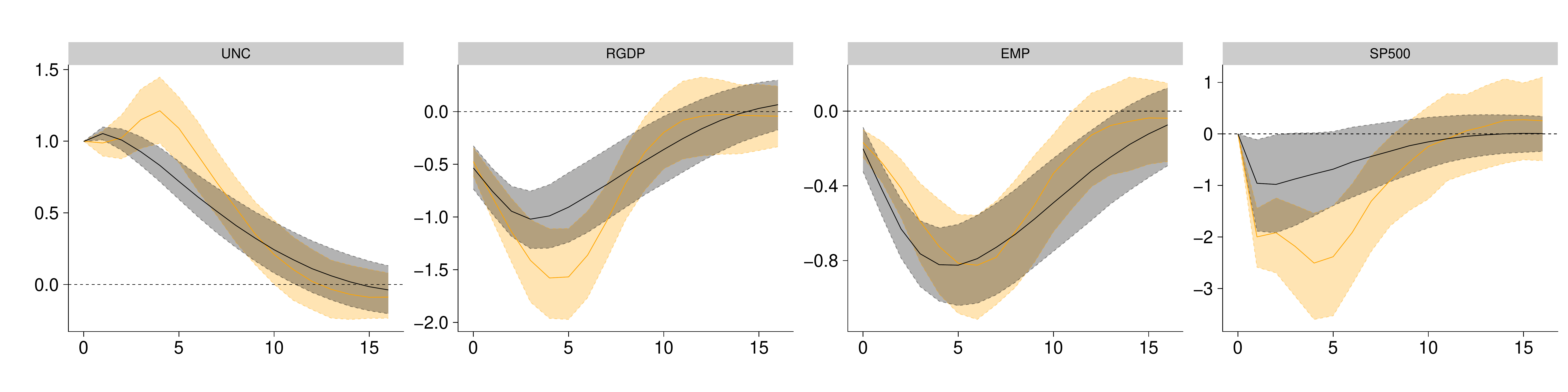}
\end{minipage}
\begin{minipage}{\textwidth}
\centering
\hspace*{50pt}
\includegraphics[scale=0.30]{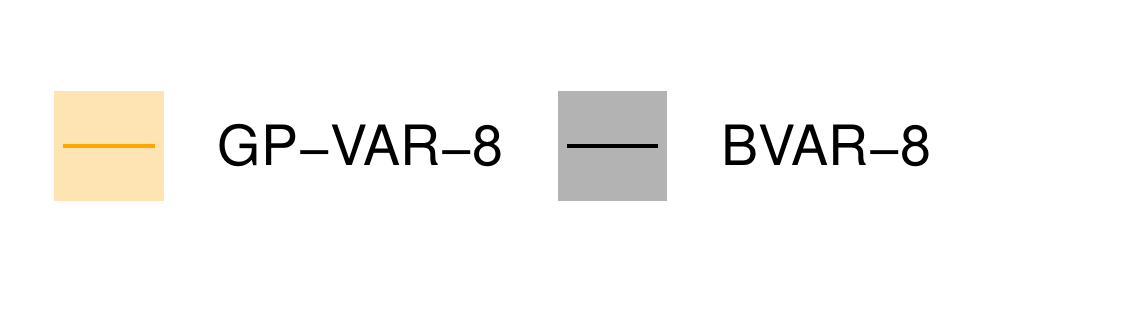}
\end{minipage}
\begin{minipage}{\textwidth}
\vspace*{-5pt}
\scriptsize \textit{Notes:} Average generalized impulse responses (GIRFs, outlined in Sub-section \ref{sec:GIRFs}) to a positive one standard deviation shock in macroeconomic uncertainty. Solid lines denote the posterior medians, while shaded areas correspond to the $68\%$ posterior credible sets. GP-VAR-8 refers to the smallest variant of our non-parametric model and BVAR-8 refers to a small-scale BVAR with SV, which is closely related to the specification used in \citet{jurado2015measuring}.
\end{minipage}
\end{figure}

In \autoref{fig:compJLN} we report the (average over time in the case of the GP-VAR) posterior quantiles ($16^{th}$, $50^{th}$ and $84^{th}$) of the responses to a macroeconomic uncertainty shock in the JLN model (in gray) and in the corresponding GP-VAR (in orange) with eight endogenous variables and uncertainty ordered second after stock market returns. Uncertainty responses to its own shock differ slightly between the GP-VAR and the linear model. These differences relate to responses within the first five quarters after the shock hit the system. The BVAR yields uncertainty reactions that peak after one quarter, declining steadily afterwards.  As opposed to this swift reaction in  uncertainty, the GP-VAR generates endogenous uncertainty reactions which peak after five quarters, declining sharply afterwards. After around eight quarters, both IRFs (almost) coincide. 

This uncertainty reaction has direct implications on how the other variables in the model react. Real GDP growth reacts in an hump-shaped manner under both models. However, driven by the somewhat later peak in uncertainty, the GP-VAR produces much stronger output growth reactions  that  peak slightly later (after around five quarters).  When we focus on employment growth the IRFs differ less. In principle, both models suggest a peak decline of around 0.8 percentage points, with the GP-VAR generating a somewhat slower response, reaching its trough after about six to seven quarters. But in principle, responses between the linear and non-parametric model tell a similar story. Finally, financial market reactions measured through the S\&P500 suggest a much stronger decline in stock prices under the GP-VAR. Interestingly, the shape of the IRFs suggests that the linear model generates the strongest reaction after around two quarters. In the GP-VAR, we find that stock markets react faster and stronger to uncertainty shocks, with substantial reactions within the first year after the shock hit the system. 

\begin{figure}[!ht]
\caption{Impulse responses of focus variables across different information sets. \label{fig:compsize}}
\centering
\begin{minipage}{\textwidth}
\centering
\vspace*{-10pt}
\includegraphics[width=\textwidth]{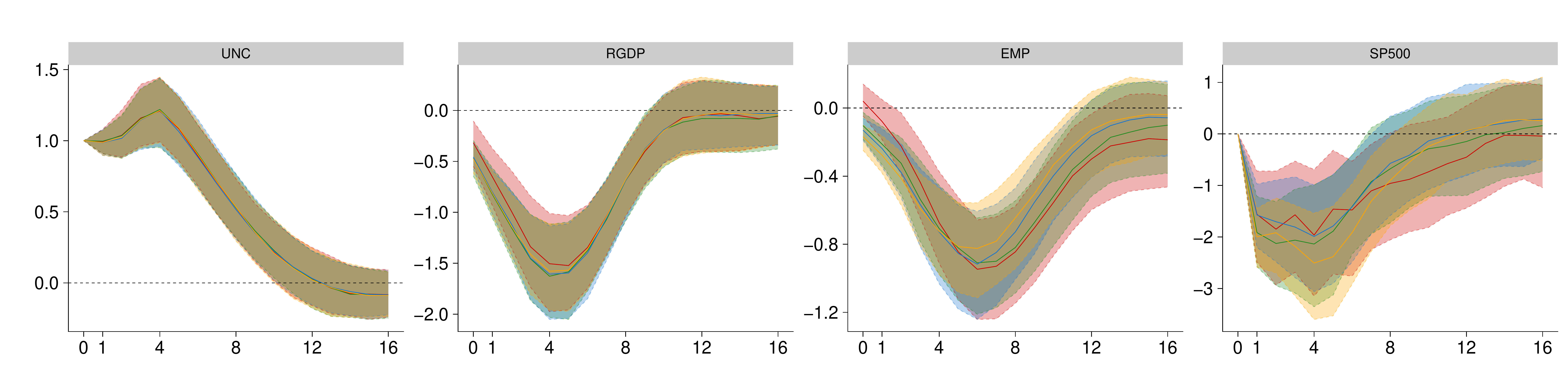}
\end{minipage}
\begin{minipage}{\textwidth}
\centering
\hspace*{50pt}
\includegraphics[scale=0.30]{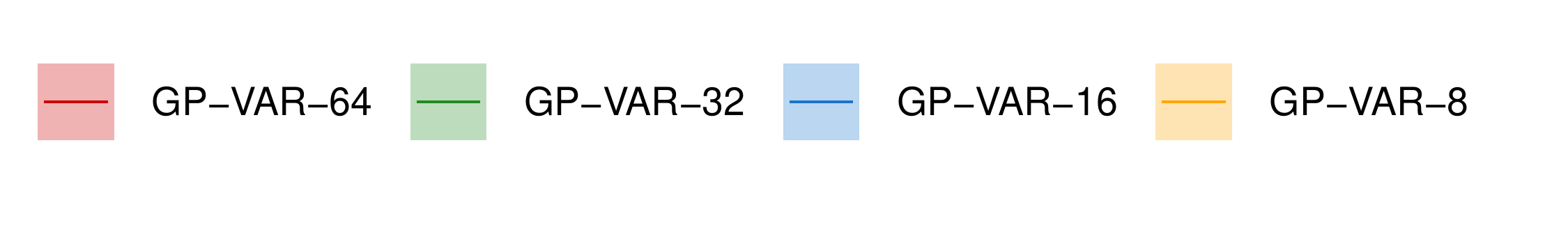}
\end{minipage}
\begin{minipage}{\textwidth}
\vspace*{-5pt}
\scriptsize \textbf{Notes:} Average generalized impulse responses (GIRFs, outlined in Sub-section \ref{sec:GIRFs}) to a positive one standard deviation shock in macroeconomic uncertainty across different information sets. Solid lines denote the posterior medians, while shaded areas correspond to the $68\%$ posterior credible sets. 
\end{minipage}
\end{figure}

To conclude, in \autoref{fig:compsize} we report the GIRFs to the uncertainty shock for the same four variables displayed in  \autoref{fig:compJLN} but obtained from GP-VARs of different dimensions (with 8, 16, 32, and 64 variables). Differences across model sizes are small (or non-existent) for most variables.  Small differences arise for employment growth, with the magnitude of the responses increasing with the model size. Stock market reactions also differ slightly across datasets, with no clear-cut pattern.  Since the GIRFs are very similar across model size and given its excellent forecasting properties, we will focus on asymmetries generated by the GP-VAR-8 model in the following sections. Results for the larger models are provided in the Online Appendix.

\subsection{Asymmetries in the transmission of uncertainty shocks}\label{ssec:asymmetries}
The non-linear and non-parametric nature of our models allows for asymmetries in the impulse response functions. This implies that shocks propagate non-linearily through the model, giving rise to differences in the GIRFs both over time but also for different shock magnitudes or signs.

\subsubsection{Asymmetries with respect to the sign of the shock}
\begin{figure}[!ht]
\caption{Shock sign asymmetries in responses of focus variables for the GP-VAR-8.\label{fig:asym_paper}}
\begin{minipage}{\textwidth}
\centering
\vspace*{-10pt}
\includegraphics[width = \textwidth]{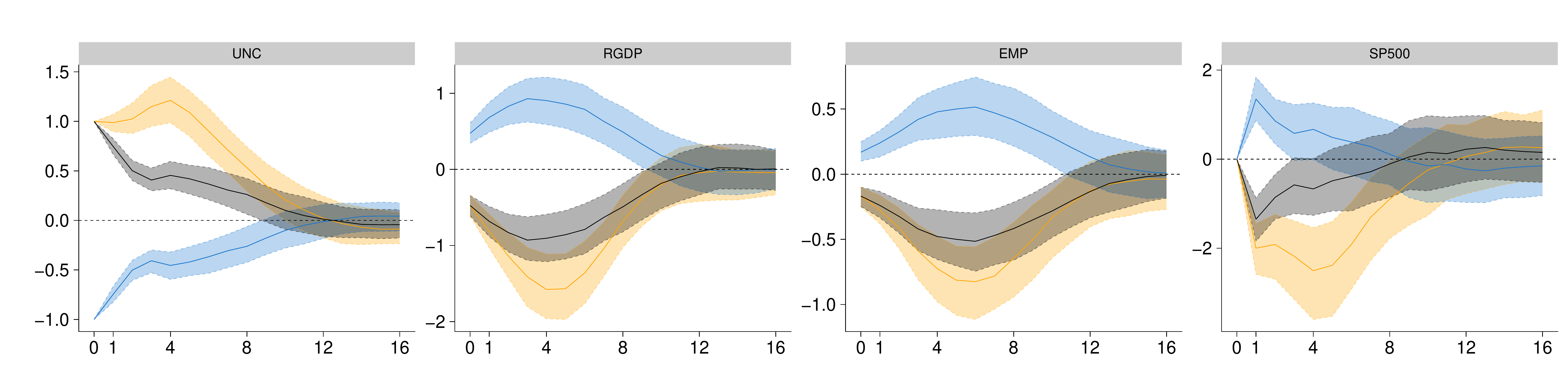}
\end{minipage}
\begin{minipage}{\textwidth}
\centering
\hspace*{50pt}
\includegraphics[scale=0.30]{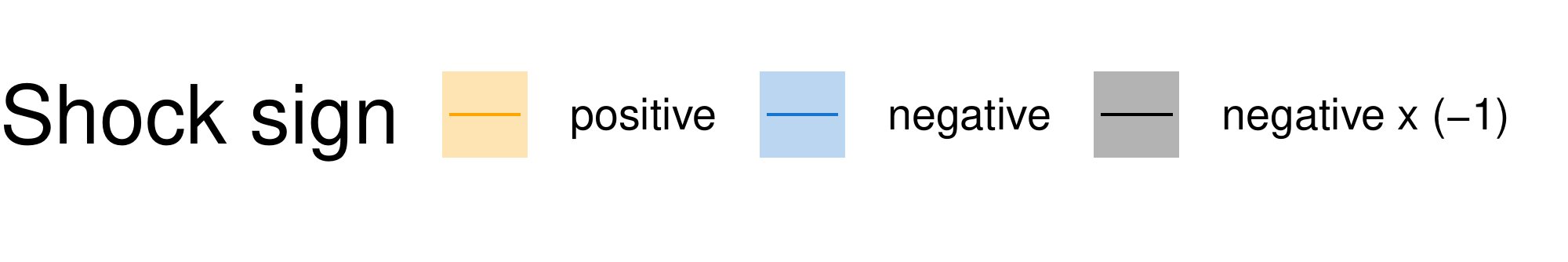}
\end{minipage}
\begin{minipage}{\textwidth}
\vspace*{-5pt}
\scriptsize \textbf{Notes:} Average generalized impulse responses (GIRFs, outlined in Sub-section \ref{sec:GIRFs}) to a negative (positive) one standard deviation shock in macroeconomic uncertainty. Solid lines denote the posterior medians, while shaded areas correspond to the $68\%$ posterior credible sets.  Here, negative  $\times (-1)$ denotes a negative one standard standard deviation shock with the respective responses being mirrored across the x-axis.
\end{minipage}
\end{figure}

The first aspect we consider relates to whether positive and negative uncertainty shocks trigger different responses of the economy.  In \autoref{fig:asym_paper} we report the responses to negative and positive uncertainty shocks from the GP-VAR-8, averaged over time. The figure thus shows GIRFs to a positive (in orange), negative (in blue) and a negative shock multiplied by -1 (in gray, to ease comparison). 
From the figure we observe some differences. These differences mostly relate to peak reactions as well as short-run (i.e. within two years) responses. In general, we find that positive shocks (higher uncertainty) trigger a stronger reaction of uncertainty, which in turn translates into more pronounced reactions of real activity and stock market quantities. 

More specifically, considering the endogenous reaction of the uncertainty indicator shows that responses to a positive uncertainty shock peak after around a year and quickly die out afterwards.  However, if uncertainty unexpectedly declines, the peak happens on impact and is much smaller as opposed to an adverse uncertainty shock. 

Turning to real GDP and employment growth, we find that positive shocks trigger stronger reactions for both variables. Interestingly, the timing of the peak responses is similar for negative and positive shocks but reactions appear much more pronounced for the latter. Stock market reactions also differ markedly across positive and negative shocks. For positive shocks we, again, find that the peak effect happens after one year and that it is more pronounced as compared to the negative shock. Overall, the picture that emerges from the GP-VAR is that higher unexpected uncertainty has stronger effects on the economy than lower uncertainty, a feature that is a priori ruled out in linear VARs.

\subsubsection{Asymmetries with respect to the size of the shock}

Our GP-VAR also allows for analyzing how shocks of different sizes impact the economy. As opposed to a standard VAR which assumes that shocks enter linearly (and thus responses to shocks of different sizes are exactly proportional to each other) our GP-VAR is more flexible and allows for investigating whether shocks of different magnitudes trigger different dynamics in the GIRFs. 

\begin{figure}[!ht]
\caption{Shock size asymmetries in responses of focus variables for the GP-VAR-8. \label{fig:asym_shocks}}
\begin{minipage}{\textwidth}
\centering
\vspace*{-10pt}
\includegraphics[width = \textwidth]{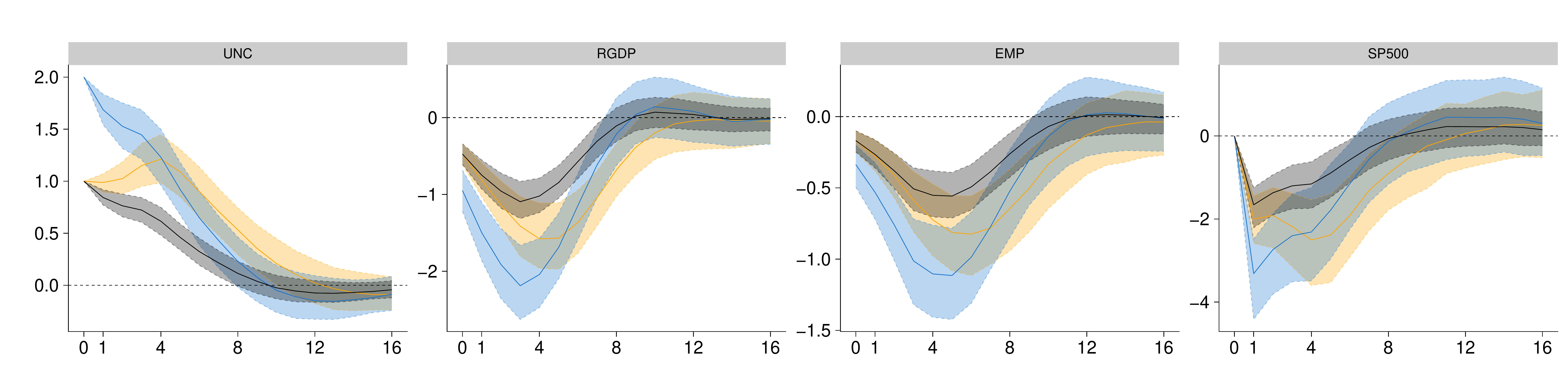}
\end{minipage}
\begin{minipage}{\textwidth}
\centering
\hspace*{50pt}
\includegraphics[scale=0.30]{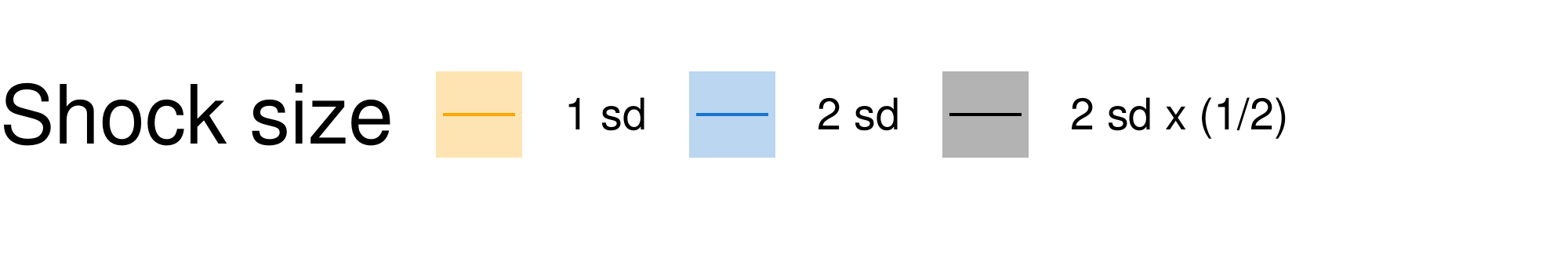}
\end{minipage}
\begin{minipage}{\textwidth}
\vspace*{-5pt}
\scriptsize \textbf{Notes:} Average generalized impulse responses (GIRFs, outlined in Sub-section \ref{sec:GIRFs}) to a positive two (one) standard deviation shock in macroeconomic uncertainty. Solid lines denote the posterior medians, while shaded areas correspond to the $68\%$ posterior credible sets. Here, $1$ sd refers to a one standard deviation shock, $2$ sd indicates a two standard deviation shock, and $2$ sd  $\times (1/2)$ denotes a two standard deviation shock with the respective responses divided by two. 
\end{minipage}
\end{figure}

In \autoref{fig:asym_shocks} we consider two shock sizes: a one standard deviation and a two standard deviation shock. To permit straightforward comparison of the shapes of the responses to differently sized shocks, we also add impulses to a two standard deviation shock which are then re-scaled to match the impact of the one standard deviation shock (the gray shaded area in the figures). 

This figure gives rise to at least two observations. First,  when we compare the shape of the responses to a one standard deviation  to the ones of a two standard deviation shock we find differences in the timing (and more generally in the shape) of the IRFs. A stronger shock triggers a faster peak reaction of GDP and employment growth. Stock market reactions display a somewhat different shape. After a sharp immediate reaction (for both shock sizes) the peak effect happens to be on impact if the size of the shock is large whereas it turns out to materialize after one year if the shock size is smaller. 

Second, in terms of the magnitudes we find that a two standard deviation shock triggers peak responses with magnitudes that are  less than twice the  magnitudes to a one standard deviation shock. This is particularly visible for employment and output reactions. For stock market responses, the impact reactions are (almost) proportional to each other.

\subsubsection{Asymmetries over time}
After showing that the economy reacts asymmetrically with respect to the sign and size of the uncertainty shock, this section asks whether the effect of uncertainty shocks changes over time (see Section 2.5 of \cite{castelnuovo2022jes}, or \cite{mumtaz2018changing} for some evidence using TVP-VARs).  

\begin{figure}[!ht]
\caption{Impulse responses of focus variables in the GP-VAR-8 across different sub-sample periods.}
\centering
\begin{minipage}{\textwidth}
\centering
\includegraphics[width = 0.95\textwidth]{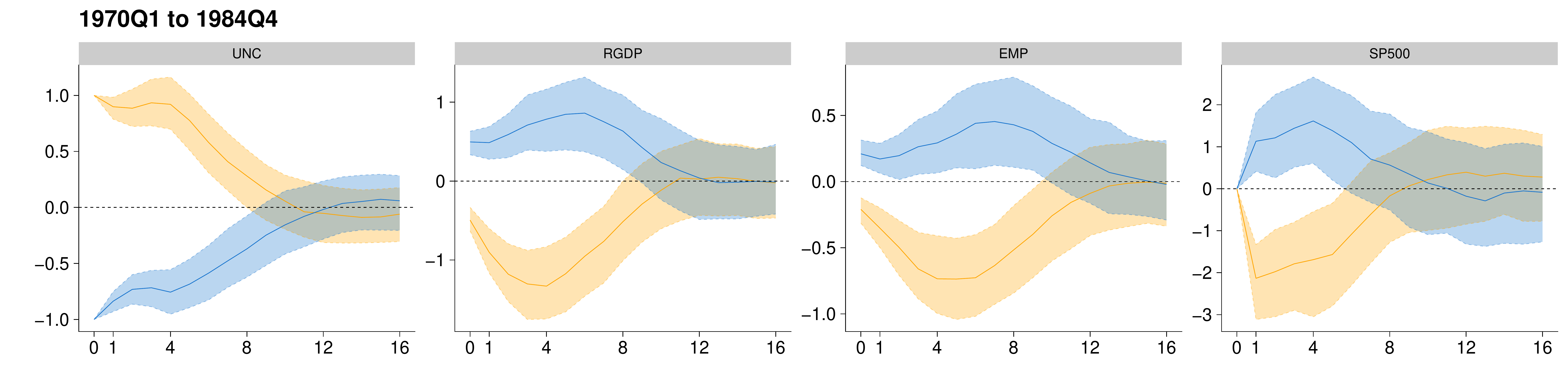}
\end{minipage}
\begin{minipage}{\textwidth}
\centering
\includegraphics[width = 0.95\textwidth]{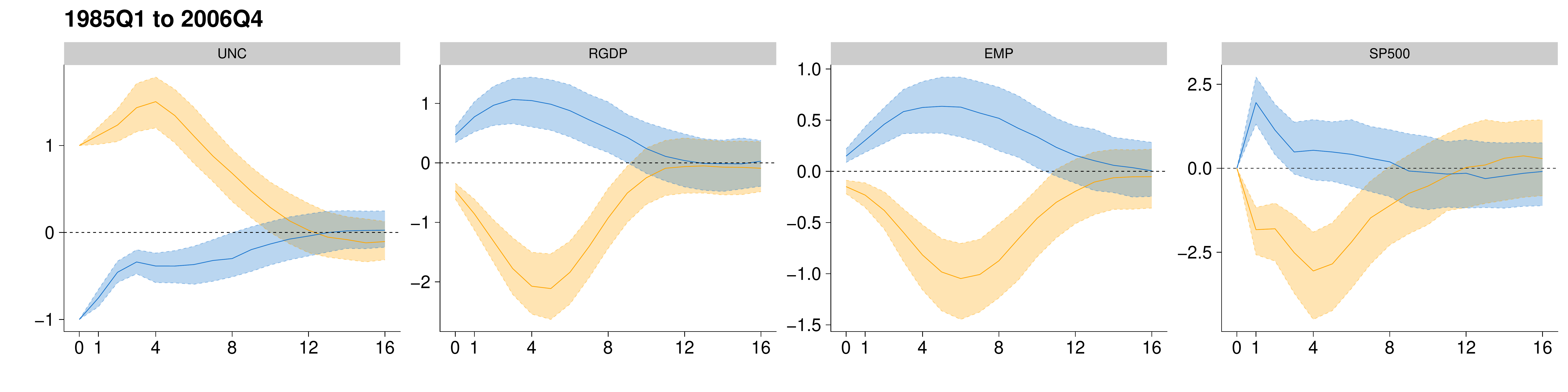}\end{minipage}
\begin{minipage}{\textwidth}
\centering
\includegraphics[width = 0.95\textwidth]{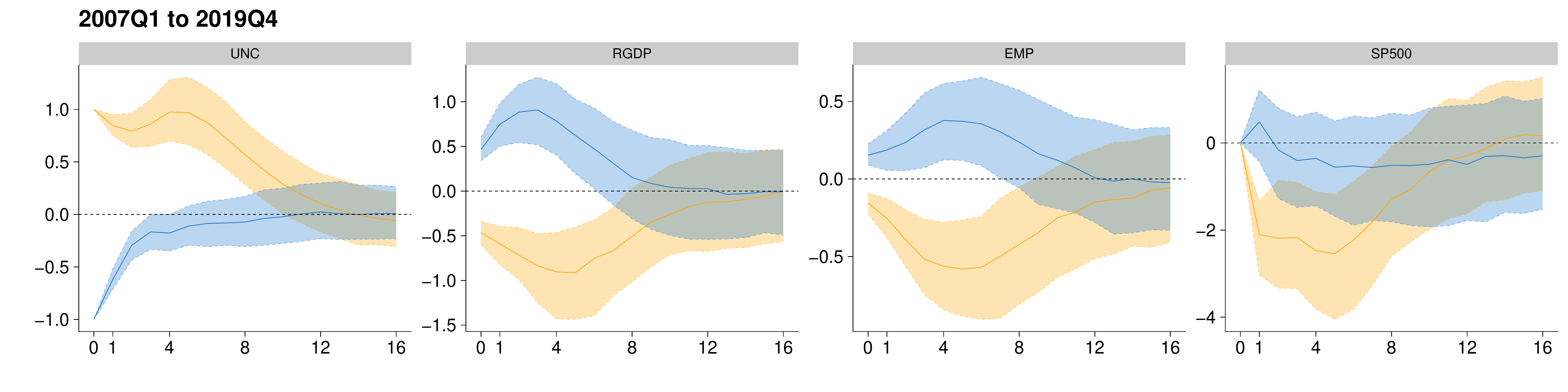}
\end{minipage}
\begin{minipage}{\textwidth}
\centering
\hspace*{15pt}\includegraphics[scale=0.3]{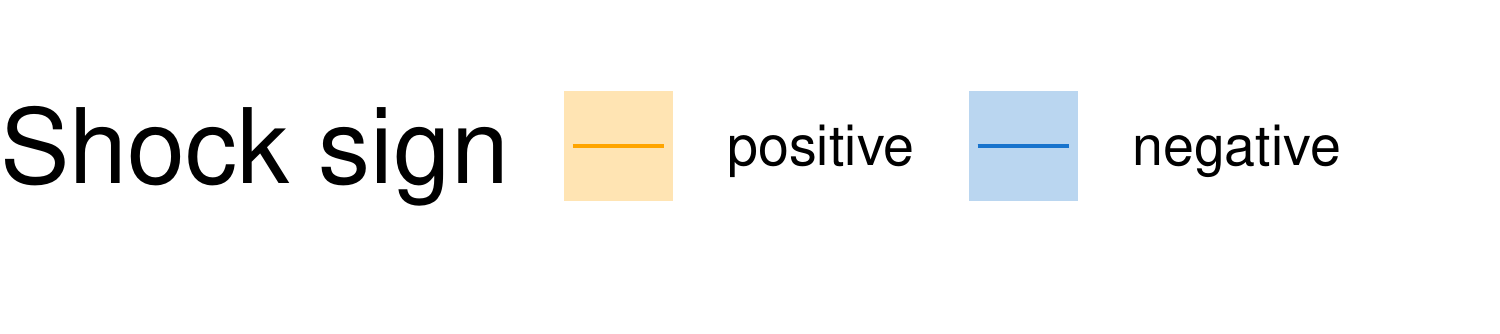}
\end{minipage}
\begin{minipage}{\textwidth}
\scriptsize \textit{Notes:} Period-specific average generalized impulse responses (GIRFs, outlined in Sub-section \ref{sec:GIRFs}) to a positive (negative) one standard deviation shock in macroeconomic uncertainty. Solid lines denote the posterior medians, while shaded areas correspond to the $68\%$ posterior credible sets. 
\end{minipage}
\label{fig:girfovert}
\end{figure}

We start by considering impulse responses averaged over certain sub-periods in \autoref{fig:girfovert}. The classification into sub-periods is mostly taken from \cite{d2012century}, and it is such that the main events in each sub-period include, respectively, the great inflation ($1970$Q$1$ to $1984$Q$4$), the great moderation ($1985$Q$1$ to $2006$Q$4$), and the post great moderation period ($2007$Q$1$ to $2019$Q$4$). To also get a rough feeling about whether asymmetries between positive and negative shocks have changed over time, all figures include the IRFs to positive (in orange) and negative (in blue) shocks.

The main feature emerging from the figure is the different behavior of the response of uncertainty across sub-samples. In the final two sub-samples, uncertainty responses increase up to four quarters after the shock, with peak effects being strongest in the great moderation period and becoming slightly weaker in the final sub-sample. Moreover, sign effects of uncertainty responses increase appreciably in the last two sub-samples. 

These differences in the responses of uncertainty trigger differences in the IRFs of the other quantities which relate not only to the magnitudes but also to the shapes of the responses. We find that GDP growth, employment and the S\&P 500 display the strongest reactions in the great moderation regime, becoming slightly weaker in the post great moderation period. The weaker reaction of real activity over time corroborates findings in \cite{mumtaz2018changing} who also report smaller responses of real activity to uncertainty shocks. As opposed to their findings, we observe that stock market reactions do not change much in magnitude but the shape differs (in accordance with the different shape in the uncertainty reaction described above). We, moreover, observe that asymmetries in terms of the sign of the shocks have decreased over time for GDP and employment growth. Only for stock market reactions these sign asymmetries have increased, with benign uncertainty shocks yielding a much weaker positive reaction of stock markets during the post great moderation regime.

\begin{figure}[!ht]
\caption{Period-specific impulse responses of focus variables in the GP-VAR-8 across different sub-sample periods.}
\centering
\begin{minipage}{\textwidth}
\centering
\includegraphics[width = 0.95\textwidth]{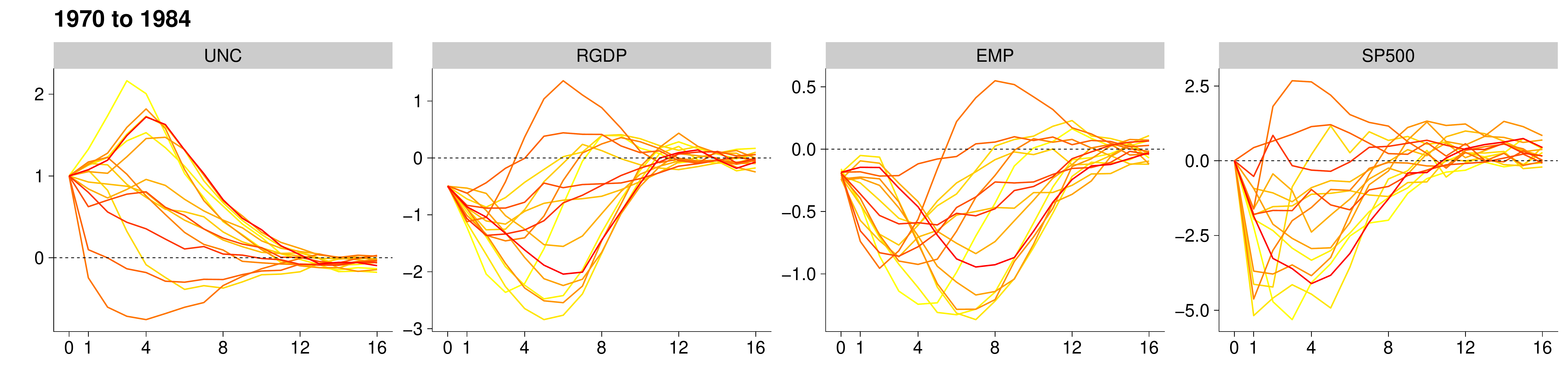}
\end{minipage}
\begin{minipage}{\textwidth}
\centering
\includegraphics[width = 0.95\textwidth]{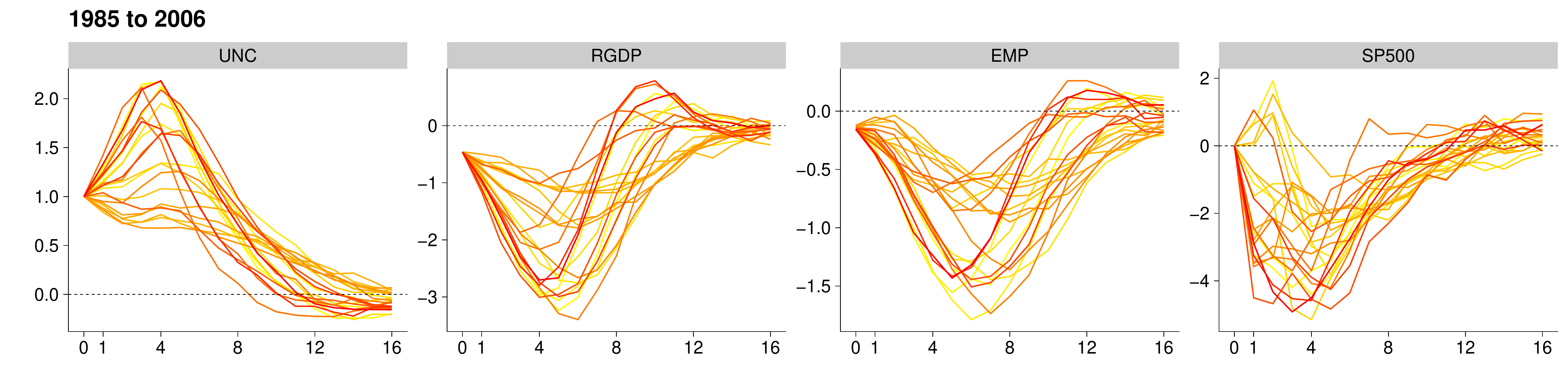}\end{minipage}
\begin{minipage}{\textwidth}
\centering
\includegraphics[width = 0.95\textwidth]{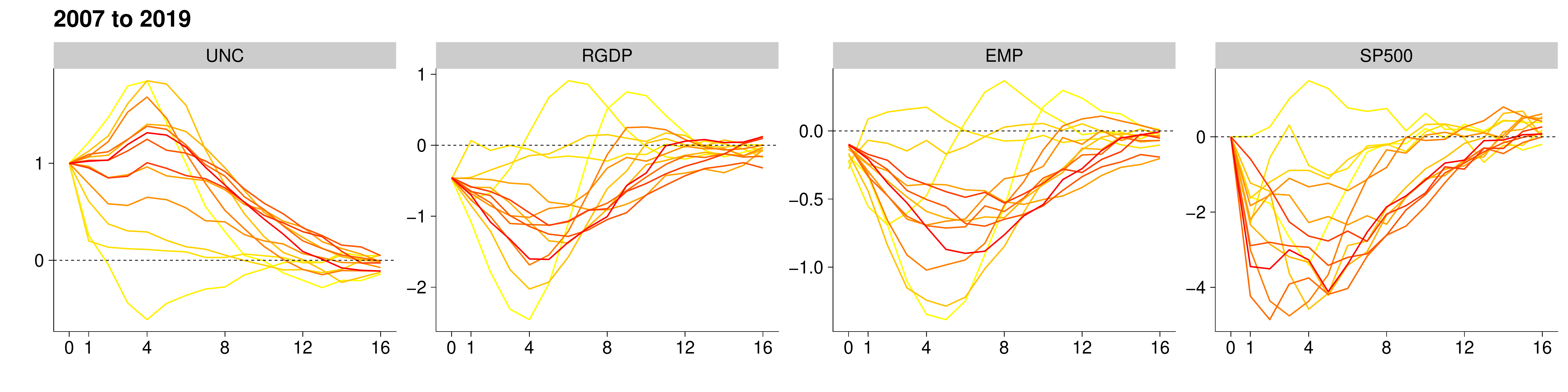}
\end{minipage}
\begin{minipage}{\textwidth}
\centering
\hspace*{15pt}\includegraphics[scale=0.3]{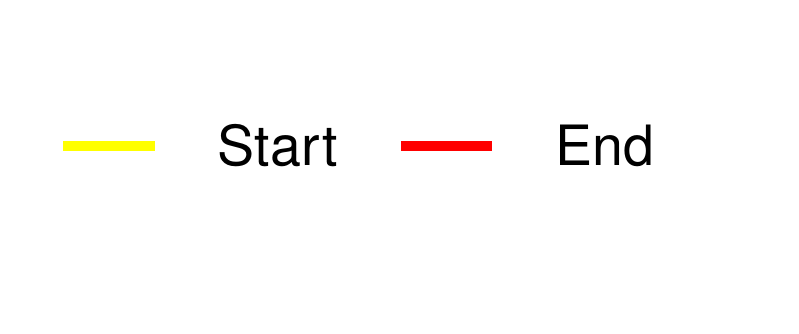}
\end{minipage}
\begin{minipage}{\textwidth}
\scriptsize \textit{Notes:} Impulse response functions to a positive one standard deviation shock in macroeconomic uncertainty in the GP-VAR-8 across sub-sample periods. Solid lines denote the yearly averaged posterior medians, with colors ranging from yellow (start of the sample) to red (end of the sample).
\end{minipage}
\label{fig:girfovert2_1}
\end{figure}

Finally, to conclude this section we raise the issue that considering GIRFs averaged over sub-samples possibly still masks important differences over time within sub-samples. To shed light on whether IRFs change within regimes, \autoref{fig:girfovert2_1} displays yearly averages of posterior medians of the IRFs over time during each of the three periods. Yellow IRFs refer to the beginning of the respective sub-sample and red ones denote IRFs computed towards the end of the sub-sample.  This figure suggests substantial heterogeneity in responses during the great inflation period. Especially towards the end of this sample, reactions of GDP growth and employment point towards a substantial real activity overshoot.  During the great moderation, the intra-period variation of the IRFs becomes much smaller, yielding patterns more consistent with the common wisdom in the uncertainty literature: real activity and stock markets decline in response to increases in economic uncertainty. In the years from 2007 to 2019, we find that IRFs differ especially in the beginning of the sample (from 2007 to 2009). For the remaining years, there is much less variation in responses and these appear to be similar to those observed in the 1985 to 2006 period. 

Overall, we can conclude that the effects of uncertainty change both during sub-samples defined by economic considerations and sometimes also within each sub-sample. This kind of time variation is a priori ruled out in linear VAR models, which can therefore lead to biased estimates of the effects of uncertainty.

\section{Conclusions}\label{sec:concl}
In this paper, we have developed a flexible multivariate model that uses Gaussian processes to model the unknown relationship between a panel of macroeconomic time series and their lagged values. Our GP-VAR is a very flexible model which remains agnostic on the precise relations between the endogenous variables and the predictors. This model can be viewed as a very flexible and general extension of the linear VAR commonly used in empirical macroeconomics. We also control for changes in the error variances by introducing a stochastic volatility speciﬁcation. While a more flexible conditional mean can reduce the need of a time-varying conditional variance, empirically we find heteroskedasticity to be relevant also for GP-VARs.

We develop efficient MCMC estimation algorithms for the GP-VAR, which are scalable to high dimensions, so much so that for large models estimation is even faster than for the corresponding BVAR-SV. Scaling the covariance of the Gaussian process by the latent volatility factors is particularly helpful to achieve computational gains, as it permits to pre-compute several quantities before MCMC sampling. This speeds up computation enormously.

To illustrate the practical working of the GP-VAR, we first test it on simulated data from different linear and non-linear models, finding that it is capable of reproducing a variety of non-linear patterns (but also a linear behavior). Then, we show in a forecasting exercise that our model yields favorable density forecasts of US output, inflation and short-term interest rates with respect to both linear and other non-parametric and time-varying specifications. 

In the main part of our empirical work we re-assess the effects of uncertainty shocks by replicating and extending the analysis carried out by  \cite{jurado2015measuring} based on linear VARs with the GP-VAR. Overall, our empirical results suggest that the measurement of uncertainty and its effects with a simple linear VAR can lead to several incorrect conclusions. Not only the effects of uncertainty can be over-stated, but they can also be treated as stable over time, symmetric for positive and negative shocks, and proportional to the shock size. Instead the GP-VAR model, which is preferred to the linear VAR in terms of fit and forecasting performance, returns time variation in the responses, asymmetry and non-proportionality. Hence, the empirical features we uncover should be also replicated by theoretical models about uncertainty and its effects, which instead at the moment typically assume stability and symmetry \citep[see, e.g., the survey in][]{bloom2014fluctuations}.

\clearpage
\small{\setstretch{0.85}
\addcontentsline{toc}{section}{References}
\bibliographystyle{frbcle}
\bibliography{lit}}\normalsize\clearpage


\newpage \normalsize
\setcounter{page}{0}
\thispagestyle{empty}
\setcounter{footnote}{0}
\begin{appendices}
\begin{center}
\LARGE \textbf{Online Appendix}  \\
 \Large{\textbf{Gaussian Process Vector Autoregressions and Macroeconomic Uncertainty}} \\
 \vspace*{10pt}
\normalsize Niko \textsc{Hauzenberger}$^\text{a}$, Florian \textsc{Huber}$^\text{a}$, Massimiliano \textsc{Marcellino}$^\text{b}$, and Nico \textsc{Petz}$^\text{a}$\\
\vspace*{10pt}
$^\text{a}$\textit{University of Salzburg}  \\[-0.5em]
$^\text{b}$\textit{Bocconi University, IGIER and CEPR} \\
\vspace*{20pt}
\date{}
\end{center}

\vspace{-50pt}
\addcontentsline{toc}{section}{Appendix} 
\part{} 
\parttoc 
\newpage

 \setcounter{equation}{0}
 \setcounter{table}{0}
 \setcounter{figure}{0}
 \renewcommand\theequation{A.\arabic{equation}}
 \renewcommand\thetable{A.\arabic{table}}
 \renewcommand\thefigure{A.\arabic{figure}}
 \renewcommand\thesubsection{A.\arabic{subsection}}

\section{Technical Appendix}\label{sec:App A}
\subsection{Capturing persistence through the kernel}\label{sec:persistencel}
In this sub-section, we discuss how to handle persistent time series with GPs. In principle, appropriately choosing $\kappa$ allows for capturing slowly evolving trends in $y_t$. But one elegant aspect of GPs is that stochastic trends in $y_t$ can also be modeled explicitly through the kernel. This can be achieved as follows.  Let $\bm B$ denote a $T \times T$ lower triangular matrix with $1$'s on the main and off-diagonal elements.  

The corresponding weight-space representation is then given by:
\begin{equation*}
    \bm y = \bm B \bm \eta + \bm \varepsilon, \quad \bm \eta \sim \mathcal{N}(\bm 0_T, r \bm I_T),\quad \bm \varepsilon \sim \mathcal{N}(\bm 0_T, \sigma_\varepsilon^2 \bm I_T),
\end{equation*}
where $r$ is a prior scaling parameter that controls the average jump size of the shocks in $\eta_t$.  Notice that this equation can be represented in component form as follows:
\begin{equation*}
    y_{t} =  \sum_{s=1}^t \eta_{s} + \varepsilon_{t},\quad \eta_t \sim \mathcal{N}(0, r), \quad \varepsilon_t \sim \mathcal{N}(0, \sigma_\varepsilon^2),
\end{equation*}
which implies that $y_{t}$ is driven by a latent random walk factor with the matrix $\bm B$ capturing the state evolution dynamics and $r$ representing the state innovation variance. This is a standard unobserved components model.  Notice that the corresponding kernel matrix can be derived as $r  \bm B \bm B'$.  Adding this to the Gaussian kernel discussed in the main text yields a combination between a kernel that captures unknown relations between $y_t$ and $\bm x_t$ but also possible stochastic trends in $y_{t}$.

\begin{figure}[!t]
\caption{Effect of different values of $r$ on the prior of $\bm f$ and the posterior $\bm f|\bm y$}
    \begin{minipage}{\textwidth}
    \centering
    Inflation
    \vspace{10pt}
    \end{minipage}
    \begin{minipage}{0.49\textwidth}
    \centering
    \scriptsize $\bm f$
    \end{minipage}
    \begin{minipage}{0.49\textwidth}
    \centering
    \scriptsize $\bm f|\bm y$
    \end{minipage}
    \centering
    \includegraphics[width = 1\textwidth]{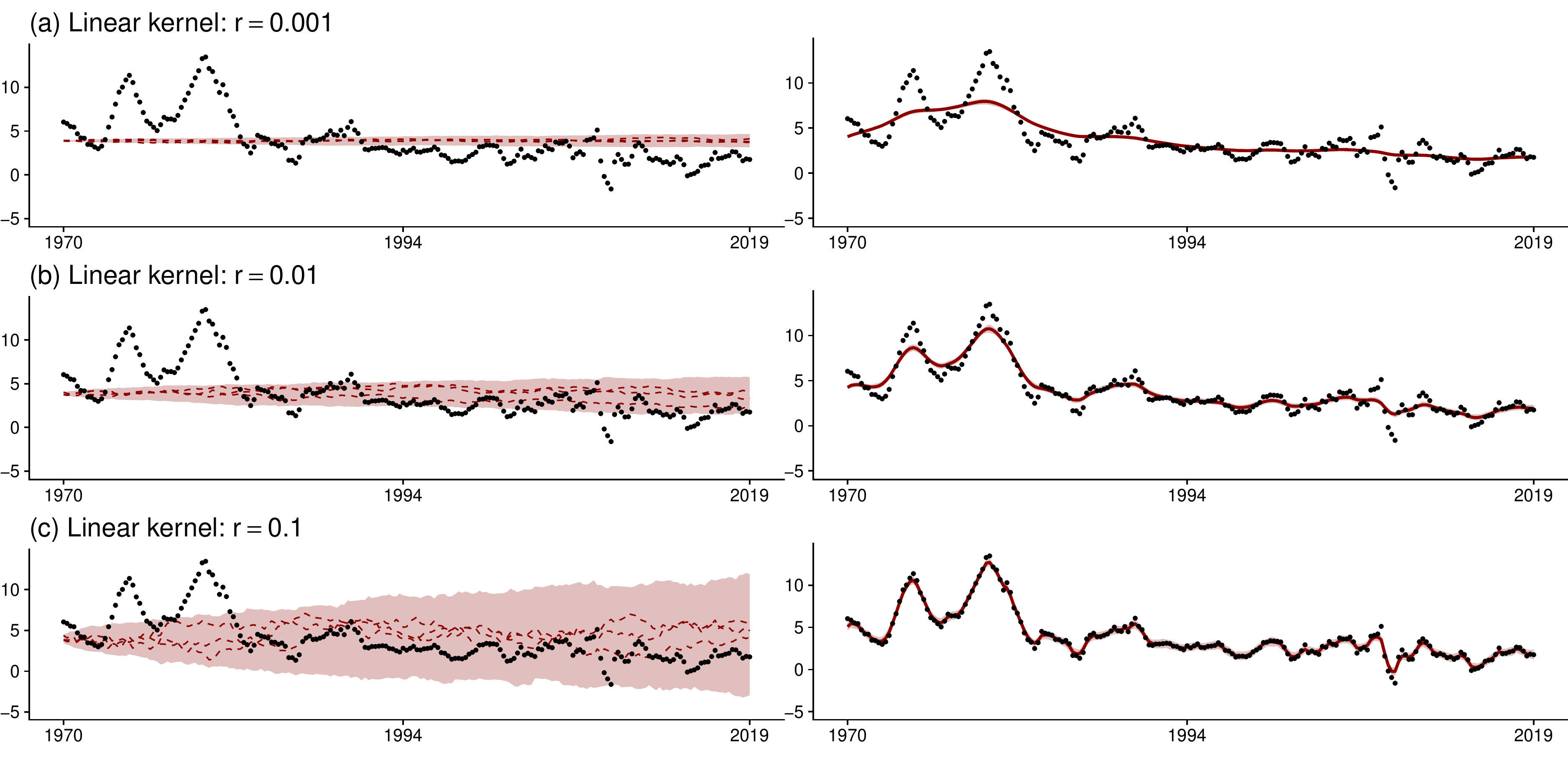}
    \label{fig:choice_k_uc}
    \begin{minipage}{\textwidth}
\vspace*{-1pt}
\scriptsize \textit{\textbf{Notes}:} In this figure we showcase the GP regression with US inflation data, setting the matrix of regressors, $\bm B$, to be a lower triangular matrix with $1$'s as the diagonal and off-diagonal elements. The linear kernel is given by $r \bm B \bm B'$. The left panels report, for different values of $r$, the $5^{th}$ and $95^{th}$ prior percentiles (with the area in between shaded in light red), three draws from the prior (in dashed red), and the actual values of inflation (black dots). The right panels report the $90\%$ posterior credible sets (shaded in light red), the posterior medians (in solid red), and actual inflation (black dots).
\end{minipage}
\end{figure}

The properties of this linear persistence kernel are illustrated in  \autoref{fig:choice_k_uc}. This figure again shows CPI inflation (again in year-on-year terms) but uses a longer sample ($1970$Q$1$ to $2019$Q$4$) and the prior on $\bm f$ (under the linear persistence kernel) and the posterior of $\bm f$. The figure reveals that if we set $r$ close to zero, the corresponding estimate will only capture low frequency movements in $y_t$. The larger $r$ gets, the larger the in-sample fit effectively becomes. If we set $r=0.1$, we observe that the model yields an almost perfect in-sample fit. 

\subsection{Sampling the factors under linear restrictions}\label{app:restr}
As stated in the main text, we identify the conditional mean of the process by introducing the linear restriction that $\bm G \bm g_j = 0$ with $\bm G = (\bm \iota'\bm \iota)^{-1}\bm \iota'$. This can be efficiently achieved through the sampler proposed in \cite{cong2017fast}. 

Suppose our goal is to simulate $\bm g_j$ from $\mathcal{N}_\mathcal{S}(\overline{\bm g}_j, \overline{\bm V}_{\bm g_j})$ with restriction $\mathcal{S}=\{\bm g_j: \bm G \bm g_j = 0 \}$ and moments given by:
    \begin{align*}
    \overline{\bm V}_{\bm g_j} &= \sqrt{\bm \Omega_j} \left(K_{\bm \vartheta_{j2}}(\bm Z_j, \bm Z_j) - K_{\bm \vartheta_{j2}}(\bm Z_j, \bm Z_j)\left(K_{\bm \vartheta_{j2}}(\bm Z_j, \bm Z_j) + \bm I_T \right)^{-1}K_{\bm \vartheta_{j2}}(\bm Z_j, \bm Z_j)  \right)\sqrt{\bm \Omega_j}, \label{eq: post_covariance} \\
            \overline{\bm g}_j &= \sqrt{\bm \Omega_j} K_{\bm \vartheta_{j2}}(\bm Z_j, \bm Z_j) \left(K_{\bm \vartheta_{j2}}(\bm Z_j, \bm Z_j) + \bm I_T    \right)^{-1}  \sqrt{\bm \Omega_j}^{-1} \bm \left(\bm Y_j - \bm f_j - \sum_{k=1}^{j-1} q_{jk} \bm Y_k\right).
\end{align*}
This can be achieved using Algorithm 2 of \cite{cong2017fast}. This algorithm consists of two steps. In the first step we sample $\bm g_j^*$ from the unrestricted distribution $\mathcal{N}(\overline{\bm g}_j, \overline{\bm V}_{\bm g_j})$. In the second step, we obtain a draw from the restricted distribution  by setting:
\begin{equation*}
    \bm g_j = \bm g_j^* - \overline{\bm V}_{\bm g_j} \bm G' (\bm G \overline{\bm V}_{\bm g_j} \bm G')^{-1}  \bm G \bm g_j^*.
\end{equation*}
All involved quantities are trivial to compute and this step does not introduce additional computational hurdles. This normalization essentially subtracts a constant term from $\bm g^*_j$, the unrestricted draw of $\bm g_j$, so that the resulting grand mean will be equal to zero. This leaves the conditional mean function  $\bm m_j$ unchanged.

\subsection{Sampling the log-volatilities}\label{sec: vola_sampler}
We sample the log-volatilities marginally of the factors.  
This is achieved by exploiting the weight-space view of the GP. Integrating out $\bm f_j$ and $\bm g_j$ allows us to rewrite the $j^{th}$ equation as
\begin{equation}\label{eq:svobs}
    \tilde{\bm Y}_j = \left(\bm Y_j - \sum_{k=1}^{j-1} q_{jk} \bm Y_k\right) = \sqrt{\bm \Omega}_j \tilde{\bm W}_j \tilde{\bm \varepsilon}_j, \quad \tilde{\bm \varepsilon}_j \sim \mathcal{N}(\bm 0, \bm I_T),
\end{equation}
where $\tilde{\bm W}_j$ denotes the lower Cholesky factor of $\left( K_{\bm \vartheta_{j1}}(\bm X_j, \bm X_j) + K_{\bm \vartheta_{j2}}(\bm Z_j, \bm Z_j) + \bm I_T \right)$ and $\bm \Omega_j = \text{diag}(\bm \omega_j)$ with $\bm \omega_j = (\omega_{j1}, \dots, \omega_{jT})'$. 
Our goal is to sample the log-volatilities $\bm h_j = \log \bm \omega_j$ from its conditional posterior distribution $p(\bm h_j|\tilde{\bm Y}_j, \tilde{\bm W}_j, \bm \theta_{hj})$, with $\bm \theta_{hj} = (\rho_{hj}, \sigma^2_{hj}, h_{j0})'$ collecting the parameters associated with the state equation of the log-volatility process, which evolves according to a stationary AR($1$). 

The corresponding $T$-dimensional full conditional posterior distribution can be expressed as:
\begin{equation}\label{eq:svpost}
p(\bm h_j|\tilde{\bm Y}_j, \tilde{\bm W}_j, \bm \theta_{hj}) \propto p(\tilde{\bm Y}_j|\bm h_j, \tilde{\bm W}_j) \times p(\bm h_j|\bm \theta_{hj}), 
\end{equation}
where $p(\tilde{\bm Y}_j|\bm h_j, \tilde{\bm W}_j)$ refers to the likelihood and $p(\bm h_j|\bm \theta_{h j})$ to the prior. While the prior is Gaussian and defined by the state equation in Eq. (\ref{eq:svstate}), the likelihood takes a multivariate Gaussian form only when conditioned on the log-volatilities $\bm h_j$ (and thus $\bm \Omega_j$):
\begin{equation}\label{eq:svlik}
\begin{aligned}
p(\tilde{\bm Y}_j|\bm h_j, \tilde{\bm W}_j) =& ~(2 \pi)^{-\frac{T}{2}} \times \text{det}\left(\sqrt{\bm \Omega}_j\tilde{\bm W}_j\tilde{\bm W}_j' \sqrt{\bm \Omega}_j\right)^{-\frac{1}{2}} \\ &\times \exp \left \{- \frac{1}{2}\left(\tilde{\bm Y}_j' \left(\sqrt{\bm \Omega}_j\tilde{\bm W}_j\tilde{\bm W}_j' \sqrt{\bm \Omega}_j\right)^{-1} \tilde{\bm Y}_j \right) \right\},
\end{aligned}
\end{equation}
where $\sqrt{\bm \Omega}_j\tilde{\bm W}_j\tilde{\bm W}_j' \sqrt{\bm \Omega}_j$ refers to the variance-covariance matrix of $\tilde{\bm Y}_j$. At this point it proves convenient to use the fact that $\bm \Omega_j$ is a diagonal matrix. The main challenge  is that $\bm h_j$ enters Eq. (\ref{eq:svlik}) non-linearly and is never observed directly (but evolves according to a latent process), which complicates the evaluation of the likelihood. 

Moreover, the conjugate nature of our GP regression complicates likelihood evaluation further. The algorithm proposed by \cite{kim1998stochastic} based on auxiliary mixture indicators  cannot be used because $\bm \Omega_j$ -- and thus $\bm h_j$ -- are serially correlated and are not independent from each other, as the full symmetric kernel matrices $K_{\bm \vartheta_{j1}}(\bm X_j, \bm X_j)$ and $K_{\bm \vartheta_{j2}}(\bm Z_j, \bm Z_j)$ enter the Cholesky factor $\tilde{\bm W}_j$.\footnote{In particular, this implies that the measurement errors $\log(\tilde{\bm \varepsilon}_j^2)$ of the linearized observation equation (which is obtained by squaring and taking the logarithm of Eq. (\ref{eq:svobs})) are no longer independently $\log(\chi_1^2)$ distributed and cannot readily approximated by a known mixture of Gaussian distribution to render the likelihood Gaussian conditional on the mixture indicators.} 

As a remedy, we follow \cite{chan2017stochastic} who proposes computationally efficient sampling techniques. Our algorithm is a variant of the independence Metropolis-Hastings (MH) which can be readily applied in the present setting.\footnote{For a textbook treatment, see \cite{chan_koop_poirier_tobias_2019}.} As noted by \cite{chan2017stochastic}, this algorithm is fast and exhibits good mixing properties for several reasons. The independent MH step samples the log-volatilities jointly from their full conditional posterior distribution, the proposal distribution is constructed such that the acceptance rate of the MH step is sufficiently high, and in addition, we resort to sparse matrix methods to speed up computation.

In the following, we provide a detailed discussion on our independence MH step.  We use a second-order Taylor approximation with respect  to $\bm h_j$ and then employ the Newton-Raphson optimization algorithm to represent the non-trivial conditional posterior in \autoref{eq:svpost} in the form of a multivariate Gaussian distribution. This approximation is centered on the mode of $p(\bm h_j|\tilde{\bm Y}_j, \tilde{\bm W}_j, \bm \theta_{hj})$ and the variance-covariance  is given by the inverse of the negative Hessian of $\log p(\bm h_j|\tilde{\bm Y}_j, \tilde{\bm W}_j, \bm \theta_{hj})$ evaluated at the obtained mode. This Gaussian distribution provides a sufficiently accurate approximation to the full conditional posterior of $\bm h_j$ and thus ensures a high acceptance rate when used as the proposal density in our MH step, leading to favorable mixing properties.  
 
To construct our Gaussian proposal density thus requires evaluating the gradient and Hessian of the logarithm of the full conditional posterior of $\bm h_j$. According to \autoref{eq:svpost}, the log-posterior distribution is the sum of the log-likelihood and log-prior distribution, implying that the gradient/Hessian of the log-posterior is also the sum of the gradient/Hessian of these two components.

We first focus on the gradient and the Hessian of the log-prior distribution, since these quantities are mostly standard \citep[see, e.g.,][]{chan_koop_poirier_tobias_2019}. For the $t^{th}$ element of $\bm h_j$, the prior is defined by \autoref{eq:svstate} and depends on the additional state equation hyperparameters $\bm \theta_{hj}$. For  $\bm h_j$, \autoref{eq:svstate} can be written more compactly as:
\begin{equation}
\begin{aligned}
\bm D_{hj} \bm h_j =& \tilde{\bm \mu}_{hj} + \tilde{\bm u}_{hj}, \quad \tilde{\bm u}_{hj} \sim \mathcal{N}(\bm 0_{T}, \sigma^2_{hj} \bm I_T),\\
\bm h_j =& \bm \mu_{hj} + \bm u_{hj}, \quad \bm u_{hj} \sim \mathcal{N}(\bm 0_{T}, \sigma^2_{hj} \left(\bm D_{hj}'\bm D_{hj})^{-1}\right),\\
\end{aligned}
\end{equation}
with 
\begin{equation*} 
\bm D_{hj} = 
\begin{pmatrix} 
1          & 0          & 0       & \dots  & 0          & 0 \\
-\rho_{hj} & 1          & 0      & \dots  & 0          & 0 \\
0          & -\rho_{hj} & 1      & \dots  & 0          & 0 \\
\vdots     & \vdots     & \vdots & \ddots & \vdots     & \vdots \\
0          & 0          & 0      & \dots  & -\rho_{hj} & 1
\end{pmatrix} 
\end{equation*}
denoting a $T \times T$- matrix, $\tilde{\bm \mu}_{hj} = (\rho_{hj} h_{j0}, 0, \dots, 0)'$ being $T \times 1$-vector, and $\bm \mu_{hj} = \bm D_{hj}^{-1}\tilde{\bm \mu}_{hj}$. Then the log-prior distribution of $\bm h_j$ is given by: 
\begin{equation}
\log p(\bm h_j|\bm \theta_{hj}) = -\frac{T}{2} \log 2 \pi -\frac{T}{2} \log \sigma^2_{hj} - \frac{1}{2\sigma^2_{hj}} (\bm h_j - \bm \mu_{hj})'\bm D_{hj}'\bm D_{hj}(\bm h_j - \bm \mu_{hj}),\\
\end{equation}
by noting that $\text{det}(\bm D_{hj}'\bm D_{hj}) = 1$.

In the following, let $\bm r_{P}(\tilde{\bm h}_j)$ denote the first derivative (gradient) and $\bm R_{P}(\tilde{\bm h}_j)$ the second derivative (Hessian) of the log-prior distribution with respect to $\bm h_j$ evaluated at $\tilde{\bm h}_j$:
\begin{equation*}
\begin{aligned}
\bm r_{P}(\tilde{\bm h}_j) = \frac{\partial \log p(\bm h_j|\bm \theta_{hj})}{\partial \bm h_{j}} \bigg|_{\bm h_j = \tilde{\bm h}_j} =&  -\frac{1}{2 \sigma^2_{hj}}\bm D_{hj}'\bm D_{hj}(\bm h_j - \bm \mu_{hj}), \\
\bm R_{P}(\tilde{\bm h}_j) = -\frac{\partial^2  \log p(\bm h_j|\bm \theta_{hj})}{\partial \bm h_{j} \partial \bm h_{j}'} \bigg|_{\bm h_j = \tilde{\bm h}_j}= & -\frac{1}{2 \sigma^2_{hj}} \bm D_{hj}'\bm D_{hj}.
\end{aligned}
\end{equation*}
While $\bm r_{P}(\tilde{\bm h}_j)$ is a $T \times 1$ vector, $\bm R_{P}(\tilde{\bm h}_j)$ is a tridiagonal matrix of dimension $T$, with non-zero elements on the main diagonal and the two diagonals below and above the main one. 

Next, we focus on the log-likelihood, which has the following form:
\begin{equation}
\begin{aligned}
\log p(\tilde{\bm Y}_j|\bm h_j, \tilde{\bm W}_j) =& -\frac{T}{2} \log 2 \pi -\frac{1}{2} \left(\sum_{t = 1}^T h_{jt}\right) - \frac{1}{2} \log \left( \text{det}\left(\tilde{\bm W}_j\tilde{\bm W}_j'\right)\right) \\
&- \frac{1}{2}\left(\hat{\bm Y}_j'\left(\tilde{\bm W}_j\tilde{\bm W}_j'\right)^{-1} \hat{\bm Y}_j\right),
\end{aligned}
\end{equation}
where all the quantities involving $\tilde{\bm W}_j$ can be pre-computed and $\hat{\bm Y}_j = \left(\tilde{\bm Y}_j \odot \exp \left\{-\frac{\bm h_j}{2}\right\} \right)$ with $\odot$ denoting the component-wise multiplication. Now, let a $T \times 1$ vector $\bm r_{L}(\tilde{\bm h}_j)$ denote the gradient and a $T \times T$-matrix $\bm R_{L}(\tilde{\bm h}_j)$, the Hessian of the log-likelihood again with respect to $\bm h_j$ evaluated at $\tilde{\bm h}_j$:
\begin{equation*}
\begin{aligned}
\bm r_{L}(\tilde{\bm h}_j) = \frac{\partial \log p(\tilde{\bm Y}_j|\bm h_j, \tilde{\bm W}_j)}{\partial \bm h_{j}} \bigg|_{\bm h_j = \tilde{\bm h}_j} =& -\frac{1}{2}\bm{\iota_T} +\frac{1}{2} \left(\left(\tilde{\bm W}_j\tilde{\bm W}_j'\right)^{-1}\hat{\bm Y}_j\right)\odot \hat{\bm Y}_j, \\
\bm R_{L}(\tilde{\bm h}_j) = -\frac{\partial^2  \log p(\tilde{\bm Y}_j|\bm h_j, \tilde{\bm W}_j)}{\partial \bm h_{j} \partial \bm h_{j}'} \bigg|_{\bm h_j = \tilde{\bm h}_j}= & 
\frac{1}{4} \text{diag}\left(\hat{\bm Y}_j\right)\left(\tilde{\bm W}_j\tilde{\bm W}_j'\right)^{-1}\text{diag}\left(\hat{\bm Y}_j\right)\\
&+\frac{1}{4}\text{diag}\left( \left(\left(\tilde{\bm W}_j\tilde{\bm W}_j'\right)^{-1}\hat{\bm Y}_j\right)\odot \hat{\bm Y}_j \right). \end{aligned}
\end{equation*}
It is easy to see that the presence of $\left(\tilde{\bm W}_j\tilde{\bm W}_j'\right)^{-1}$ might lead to a fairly dense $T \times T$ Hessian, which makes the use of band sparse matrix algorithms impractical and leads to non-negligible computational costs for matrix calculations involving the Hessian.\footnote{A band sparse matrix is symmetric and contains only non-zero entries along the main diagonal band.} 

In what follows, we exploit the special structure of $\left(\tilde{\bm W}_j\tilde{\bm W}_j'\right)^{-1}$. Elements of the two kernels $K_{\bm \vartheta_{j1}}(\bm X_j, \bm X_j)$ and $K_{\bm \vartheta_{j2}}(\bm Z_j, \bm Z_j)$ are defined by a squared exponential function that per construction imposes a view that periods close to each other (usually this is the case for $t-1$, $t$, and $t+1$) feature relatively large covariances, while dissimilar periods have very small covariances. This structure implies that the elements of $\left(\tilde{\bm W}_j\tilde{\bm W}_j'\right)^{-1}$ on the tridiagonal main band carry most of the information, while entries far away from this main band are typically close to zero. Hence, we approximate $\bm R_{L}(\tilde{\bm h}_j)$ by a $T \times T$ matrix $\hat{\bm R}_{L}(\tilde{\bm h}_j)$ that keeps only the elements of $\bm R_{L}(\tilde{\bm h}_j)$ on the tridiagonal main band and sets all other entries to zero. This approximation has the convenient feature that both $\hat{\bm R}_{L}(\tilde{\bm h}_j)$ and $\bm R_{P}(\tilde{\bm h}_j)$ have the same band sparse structure. Consequently, the Hessian of the log-posterior distribution is also a tridiagonal matrix, which greatly facilitates computation through the use of band sparse matrix algorithms. 

Combining the gradient (the approximated Hessian) of the log-likelihood with that of the log-prior distribution, we obtain the gradient (the approximated Hessian) of the log-posterior distribution for $\bm h_j$ evaluated at $\tilde{\bm h}_j$: \begin{equation*}
\begin{aligned}
\bm r(\tilde{\bm h}_j) = \frac{\partial \log p(\bm h_j|\tilde{\bm Y}_j, \tilde{\bm W}_j, \bm \theta_{hj}) }{\partial \bm h_{j}} \bigg|_{\bm h_j = \tilde{\bm h}_j} =& \bm r_{L}(\tilde{\bm h}_j) +  \bm r_{P}(\tilde{\bm h}_j), \\
\bm R(\tilde{\bm h}_j) = -\frac{\partial^2  p(\bm h_j|\tilde{\bm Y}_j, \tilde{\bm W}_j, \bm \theta_{hj})}{\partial \bm h_{j} \partial \bm h_{j}'} \bigg|_{\bm h_j = \tilde{\bm h}_j} =&  \bm R_{L}(\tilde{\bm h}_j) + \bm R_{P}(\tilde{\bm h}_j) \quad \text{and} \\
\bm R(\tilde{\bm h}_j) \approx \hat{\bm R}(\tilde{\bm h}_j)  =& \hat{\bm R}_{L}(\tilde{\bm h}_j) + \bm R_{P}(\tilde{\bm h}_j). 
\end{aligned}
\end{equation*}

Finally, to define the moments of our Gaussian proposal density for our independence MH step, we employ the Newton-Raphson optimization method. To obtain the mode $\hat{\bm h}_j$ of the log-posterior distribution of $\bm h_j$, we initialize the algorithm with a suitable starting value $\bm h_j = \tilde{\bm h}^{(1)}_j$ and iterate through
\begin{equation*} 
\tilde{\bm h}^{(s+1)}_j =  \tilde{\bm h}^{(s)}_j - \left(\hat{\bm R}(\tilde{\bm h}^{(s)}_j)\right)^{-1} \bm r(\tilde{\bm h}^{(s)}_j)  
\end{equation*}
until $\lVert \tilde{\bm h}^{(s+1)}_j - \tilde{\bm h}^{(s)}_j \rVert < \epsilon$, with $\epsilon = 10^{-4}$ being sufficiently close to zero and acting as convergence criterion for the numerical maximization algorithm. While $\hat{\bm h}_j$ is used as the mean of the Gaussian proposal density, the variance-covariance is set to the inverse of the (approximated) negative Hessian evaluated at $\hat{\bm h}_j$. Hence, the proposal is given by: $\mathcal{N}\left(\hat{\bm h}_j, \left(-\hat{\bm R}(\hat{\bm h}_j)\right)^{-1}\right)$.

This proposal has good empirical properties. In all our empirical work it leads to acceptance rates between 35 and 50 percent.

\subsection{A sketch of the posterior simulator}\label{app:sampler}
Our posterior simulator is comprised of several steps that involve the full conditional distributions of the corresponding latent quantities and parameters of the model. Since we use a collapsed sampler the ordering of the steps is crucial for sampling from the correct joint posterior distribution. 

The joint posterior distribution associated with the $j^{th}$ equation is given by:
\begin{equation}
    p(\bm f_j, \bm g_j, q_{j1}, \dots, q_{j j-1}, \bm \Omega_j, \bm \vartheta_{j1}, \bm \vartheta_{j2}, \bm d_j, \bm \nu_j, \bm \varpi_j| Data),
\end{equation}
where $d_j$ denotes the mixture indicators used in the Gaussian approximation to the log-$\chi^2_1$ distribution, $\bm \nu_j$ are the coefficients associated with the SV state equation and $\bm \varpi_j$ are the hyperparameters associated with the Horseshoe prior. Let $\bm q_{j \bullet} =(q_{j1}, \dots, q_{j j-1})'$ denote the $j^{th}$ row of $\bm Q$.

Conditional on adequately chosen starting values, our MCMC algorithm draws from the joint posterior of equation $j^{th}$ coefficients and states by iterating through the following steps.
\begin{enumerate}
    \item[Step 1:] Sample a value of $\bm q_{j \bullet}$ from $p(\bm q_{j \bullet}| \bm f_j, \bm g_j, \bm \Omega_j, \bm \varpi_j, Data) \sim \mathcal{N}(\overline{\bm q}_{j \bullet}, \overline{\bm V}_{q_j})$ if $j >1$.\footnote{If $j=1$, ignore this step and proceed to step 2.}  The moments $\overline{\bm V}_{q_j}$ and $\overline{\bm q}_{j \bullet}$ take a standard form. 
    \item[Step 2:] Simulate the full history of the (log) volatilities from $p(\bm \Omega_j|\bm q_{j \bullet}, \bm \vartheta_{j1}, \bm \vartheta_{j2}, \bm d_j, \bm \nu_j, Data)$ using the independent MH algorithm outlined in Sub-section \ref{sec: vola_sampler}. Notice that this step is marginally of $\bm f_j$ and $\bm g_j$
    \item[Step 3:] The factor $\bm f_j$ is obtained from the multivariate Gaussian conditional posterior $p(\bm f_j|\bm g_j, \bm q_{j \bullet}, \bm \vartheta_{j1}, \bm \Omega_j, Data)$ as described in Sub-section \ref{sec: posterior_comp}.
    \item[Step 4:] The factor $\bm g_j$ is obtained from the multivariate Gaussian conditional posterior $p(\bm g_j|\bm f_j, \bm q_{j \bullet}, \bm \vartheta_{j2}, \bm \Omega_j, Data)$ as described in Sub-section \ref{app:restr}.
    \item[Step 5:] $\bm \vartheta_{j1}$ is obtained from its discrete posterior distribution using multinomial sampling. 
    \item[Step 6:] Similarly, $\bm \vartheta_{j2}$ is also obtained through multinomial sampling. 
    \item[Step 7:] Sample the parameters of state equation associated with the log-volatilities. The corresponding conditional distributions all take standard forms.
    \item[Step 8:] Sample the hyperparameters of the Horseshoe prior in $\bm \varpi$. We use the Gibbs updating step described in \cite{makalic2015simple} which only samples from inverse Gamma distributions.
\end{enumerate}
The precise ordering of the steps in our algorithm is crucial for obtaining draws from the correct stationary distribution since we rely on marginalizing out some of the states/parameters in some steps of the MCMC algorithm.\footnote{\cite{van2008partially} discuss a general approach on how to construct collapsed Gibbs samplers in order to maintain the correct stationary distribution.} 
Notice that $\bm f_j$ and $\bm g_j$ are sampled conditionally on $\bm \Omega_j$ while $\bm \Omega_j$ is sampled marginally from $\bm f_j$ and $\bm g_j$. This implies that we again sample $\bm \Omega_j$ marginally from the factors and the factors conditionally on the volatilities, leading to a joint update from $p(\bm \Omega_j, \bm f_j, \bm g_j|\bullet)$. 


\subsection{Computing generalized impulse response functions}\label{app:girfs}

Computing the GIRFs is achieved as follows. We proceed in an equation-by-equation basis using the structural representation of the GP-VAR in \autoref{eq:gpvar}. In this case, the one-step-ahead  predictive distribution of $m_{1t}$ is:
\begin{equation*}
    m_{1t+1} \sim \mathcal{N}(\overline{m}_{1t+1}, \overline{V}_{1t+1}),
\end{equation*}
with predictive moments given by:
\begin{align*}
\overline{V}_{1t+1} &= \sigma_{1t+1} \left(k_{\bm \vartheta_1}^*(\bm W_{1t+1}, \bm W_{1t+1}) - {K}^*_{\bm \vartheta_1}(\bm W_{1t+1}, \bm W_1) (K^*_{\bm \vartheta_1}(\bm W_1, \bm W_1)+\bm I_T)^{-1}  {K}^*_{\bm \vartheta_1}(\bm W_1 \bm W_{1t+1})\right) ,\\
     \overline{m}_{1t+1} &= \sigma_{1t+1} {K}^*_{\bm \vartheta_1}(\bm W_{1t+1}, \bm W_1) (K^*_{\bm \vartheta_1}(\bm W_1, \bm W_1)+\bm I_T)^{-1} \sqrt{\bm \Omega_1}^{-1}\bm Y_1.
\end{align*}
Here, we let $\bm W_j =(\bm X_j, \bm Z_j)$ with $t^{th}$ row $\bm W_{jt}$  and ${K}^*_{\bm \vartheta_j}(\bm W_j, \bm W_j) = K_{\bm \vartheta_{j1}}(\bm X_j,  \bm X_j)+ K_{\bm \vartheta_{j2}}(\bm Z_j,  \bm Z_j)$.  A draw from the one-step-ahead predictive distribution of $y_{1t+1}$ is obtained from:
\begin{equation*}
    y_{1t+1} \sim \mathcal{N}(m_{1t+1}, \omega_{1t}).
\end{equation*}


Draws from the forecast distribution of $y_{1t+1}$ are used to form the predictive distribution for $m_{2t}$ (and thus $y_{2t+1}$ and so on). In general, the predictive density for the conditional mean of equation $j>2$ is given by:
\begin{equation*}
    m_{jt+1} \sim \mathcal{N}(\overline{m}_{jt+1}, \overline{V}_{jt+1}),
\end{equation*}
with 
\begin{equation*}
    \overline{m}_{jt+1} = \sigma_{jt+1} {K}^*_{\bm \vartheta_j}(\bm W_{jt+1}, \bm W_j) (K^*_{\bm \vartheta_j}(\bm W_j, \bm W_j)+\bm I_T)^{-1} \sqrt{\bm \Omega_j}^{-1}\left(\bm Y_j - \sum_{k=1}^{j-1} q_{jk} ({m}_{kt+1} + \varepsilon_{kt+1})\right),
\end{equation*}
and the predictive variance defined as in the case of $j=1$. A draw from the predictive distribution of $y_{jt+1}$ is obtained by drawing $\varepsilon_{jt+1} \sim \mathcal{N}(0, \omega_{jt+1})$ and adding this to $m_{jt+1}$.

Drawing from the posterior of  the  forecasts for all $M$  elements in $\bm y_t$ yields the one-step-ahead  forecast distribution $p(\bm y_{t+1}|\mathcal{I}_t)$. Higher order forecast distributions are then obtained by using $\bm y_{t+1} \sim p(\bm y_{t+1}|\mathcal{I}_t)$ to construct $\overline{\bm W}_{jt+2}$ and then compute $\overline{m}_{jt+2}$, for $j=1,\dots, M$, and drawing from the marginal distribution of the structural shocks. In general, $h$-step-ahead predictions $\bm y_{t+h}$ are obtained similarly by drawing from $p(\bm y_{t+h}|\mathcal{I}_t)$. 

As described in the text, we focus on an uncertainty shock with the uncertainty index being located in the $j^{th}$ position in $\bm y_t$. Let $\varepsilon_{jt}$ denote the structural innovation  in time $t$ and we assume that  $\varepsilon_{jt}= \varsigma$. This implies that $\bm y_t$ changes by $\bm q_j$. Again, we can compute point forecasts but instead of using $\bm W_{jt+1}$ (which comprises of $\bm y_t$ and its $p-1$ lags), $\hat{\bm W}_{jt+1}$ is constructed based on $\hat{\bm y}_t = \bm y_t + \bm q_1 + \tilde{\bm \varepsilon}_t$. The shock $\tilde{\bm \varepsilon}_t$ is obtained from the marginal distribution of the shocks but its $j^{th}$ element equals zero. Using  $\hat{\bm W}_{jt+1}$, we can compute the predictive distributions of $\bm y_{t+1}$ given a  shock of size $\varsigma$ in $\varepsilon_{jt}$. As an intermediate step, we need to compute the mean forecast based on $\varepsilon_{jt}=\varsigma$:
\begin{equation*}
    \hat{m}_{jt+1} \sim \mathcal{N}(\overline{m}^\dagger_{jt+1}, \overline{V}^\dagger_{jt+1})
\end{equation*}
with mean and variance given by:
\begin{align*}
         \overline{m}^\dagger_{jt+1} &= \sigma_{jt+1} {K}^*_{\bm \vartheta_j}(\hat{\bm W}_{jt+1}, \bm W_j) (K^*_{\bm \vartheta_j}(\bm W_j, \bm W_j)+\bm I_T)^{-1} \sqrt{\bm \Omega_j}^{-1}(\bm Y_j - \sum_{k=1}^{j-1} q_{jk} (\hat{m}_{kt+1} + \varepsilon_{kt+1})),\\
         \overline{V}^\dagger_{jt+1} &= \sigma_{jt+1} \left(k_{\bm \vartheta_j}^*(\hat{\bm W}_{1t+1}, \hat{\bm W}_{1t+1}) - {K}^*_{\bm \vartheta_j}(\hat{\bm W}_{1t+1}, \bm W_j) (K^*_{\bm \vartheta_j}(\bm W_j, \bm W_j)+\bm I_T)^{-1}  {K}^*_{\bm \vartheta_j}(\bm W_j \hat{\bm W}_{1t+1})\right)
\end{align*}
These one-step-ahead forecasts can again be used to construct iterative higher order predictions conditional on a unit structural shock to the first variable. Doing so for each of the $M$  variables yields a  draw from $y_{t+h} \sim p(\bm y_{t+h}|\mathcal{I}_t, \varepsilon_{jt}=\varsigma)$.

\clearpage
\setcounter{equation}{0}
\setcounter{table}{0}
\setcounter{figure}{0}
\renewcommand\theequation{B.\arabic{equation}}
\renewcommand\thetable{B.\arabic{table}}
\renewcommand\thefigure{B.\arabic{figure}}
\renewcommand\thesubsection{B.\arabic{subsection}}
\section{Data Appendix}\label{sec:App B}

\input{data_desc}

\clearpage

\setcounter{equation}{0}
\setcounter{table}{0}
\setcounter{figure}{0}
\renewcommand\theequation{C.\arabic{equation}}
\renewcommand\thetable{C.\arabic{table}}
\renewcommand\thefigure{C.\arabic{figure}}
\renewcommand\thesubsection{C.\arabic{subsection}}

\section{Empirical Appendix}\label{sec:App C}
\subsection{In-sample results for non-focus variables in the GP-VAR-8}
\begin{figure}[!ht]
\caption{Linear shrinkage parameters of equation-specific kernels for the GP-VAR-64.
\label{fig:linearshrinkL}}
\begin{minipage}{\textwidth}
\centering
\includegraphics[scale=0.5]{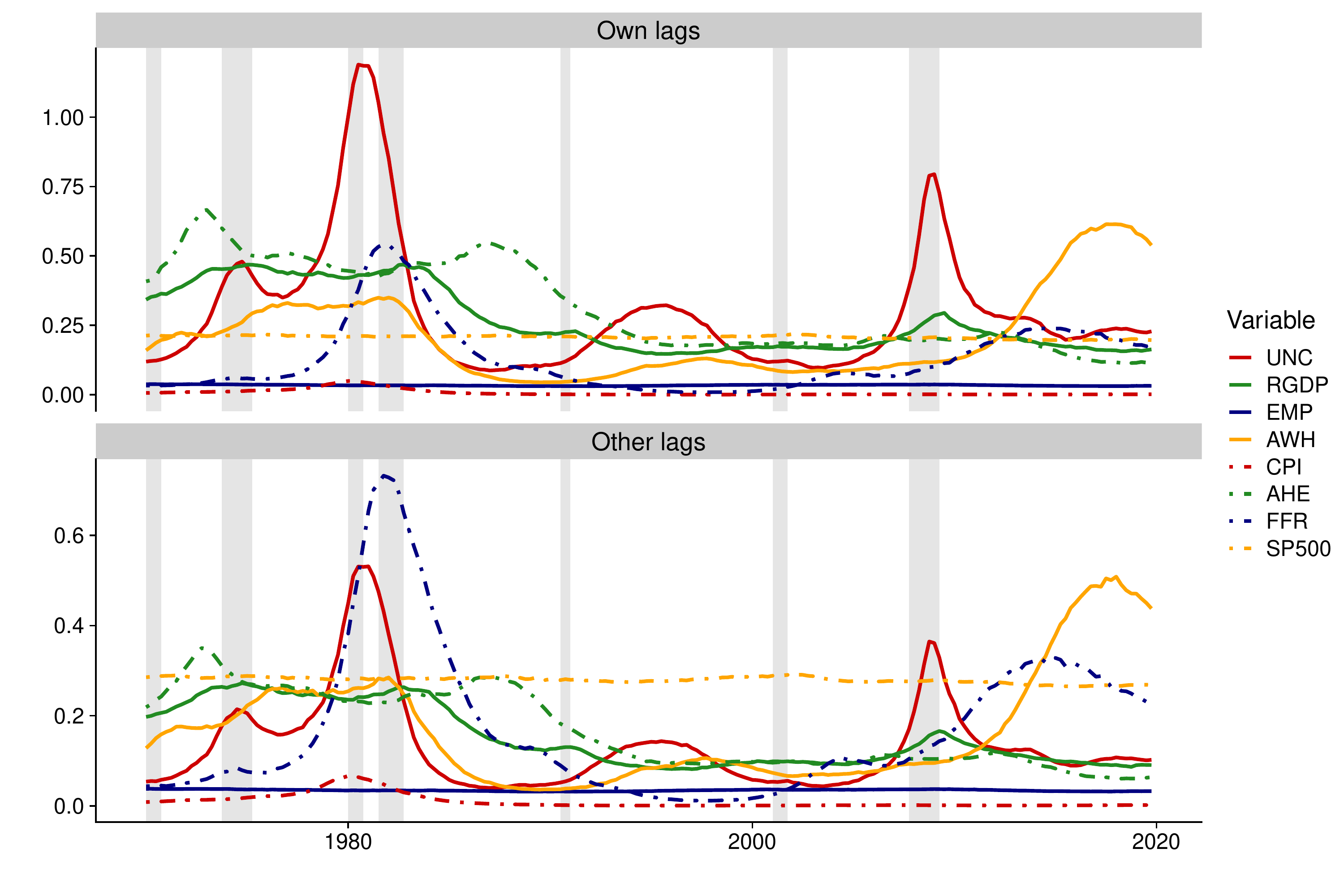}
\end{minipage}
\begin{minipage}{\textwidth}
\vspace*{-5pt}
\scriptsize \textit{Notes:} This figure reports the posterior means of the product of the error variances $\omega_{jt}$ and the linear scaling parameters for own lags ($\xi_{j1}$) and for other lags ($\xi_{j2}$), respectively. These two quantities correspond to the diagonal elements of the re-scaled kernels $\omega_{jt} \times k_{\bm \vartheta_{j1}}(\bm x_t, \bm x_t) = \omega_{jt} \xi_{j1}$ and $\omega_{jt} \times k_{\bm \vartheta_{j2}}(\bm z_t, \bm z_t) = \omega_{jt} \xi_{j2}$.
\end{minipage}
\end{figure}

\begin{figure}[!ht]
\caption{Inverse length scale parameters of equation-specific kernels for the GP-VAR-64.
\label{fig:shrinkage_eqsL}}
\centering
\begin{minipage}{\textwidth}
\centering
\includegraphics[scale=0.5]{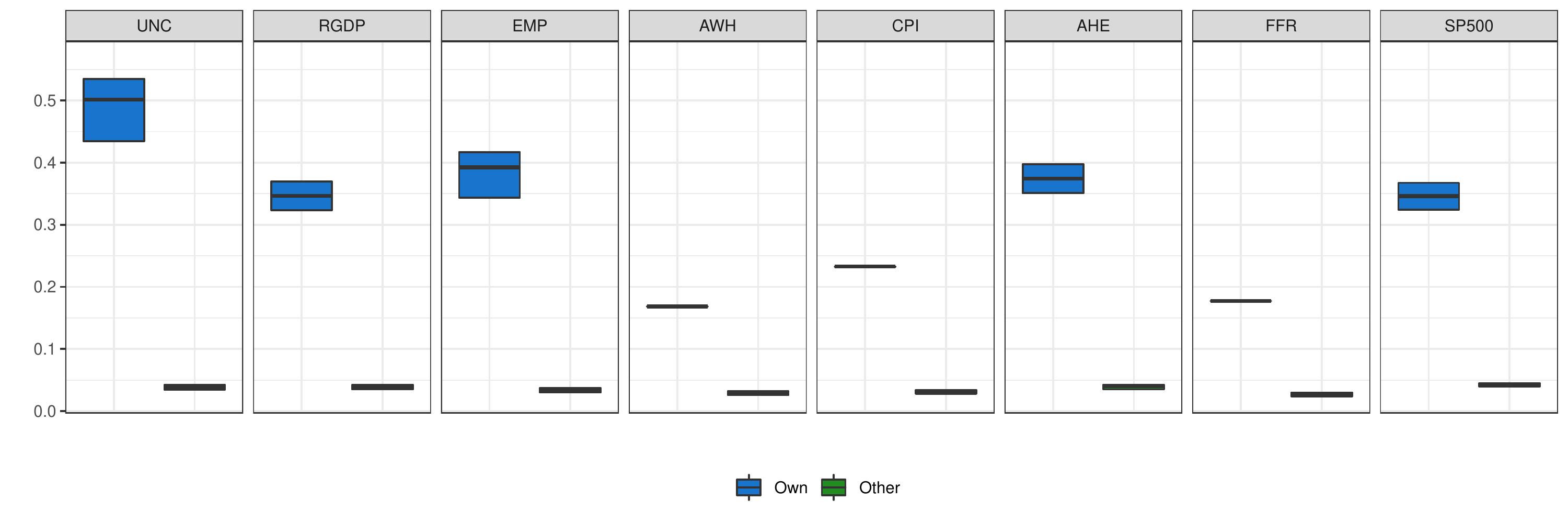}
\end{minipage}
\begin{minipage}{\textwidth}
\scriptsize \textit{Notes:} This figure reports the posterior summaries in the form of simplified boxplots of the inverse length scale parameters for own lags ($\kappa_{j1}$) and other lags ($\kappa_{j2}$), respectively. The solid black lines denote the posterior medians, while the blue (green) shaded areas represent the $50\%$ posterior credible sets (i.e., the posterior interquartile ranges).
\end{minipage}
\end{figure}
\newpage
\subsection{Robustness with respect to different orderings}\label{app:ordering}
This figure shows heatmaps of the correlations between the posterior median of the GIRFs for 25 different variable orderings (chosen at random). To ensure that the identification does not impact the results, we fix the ordering of the first two elements in $\bm y_t$ (the S\&P 500 is ordered first and the uncertainty index second). The heatmaps show that for our focus variables, the correlations are (almost) always above 0.9. This indicates that the ordering does not play a particular role for the estimation of the IRFs. The only responses that display slightly stronger changes are the ones of the S\&P 500. In this case, the differences in IRFs are, however, mostly related to higher order IRFs which are insignificant. Short-term reactions are almost identical and thus do not change much if we alter the orderings.
\begin{figure}[!ht]
\caption{Correlation of GIRFs across different orderings in the estimation of the GP-VAR-8. }
\label{fig:perm_corrmat}
\begin{minipage}{\textwidth}
\centering
\vspace*{-10pt}
\includegraphics[width = \textwidth]{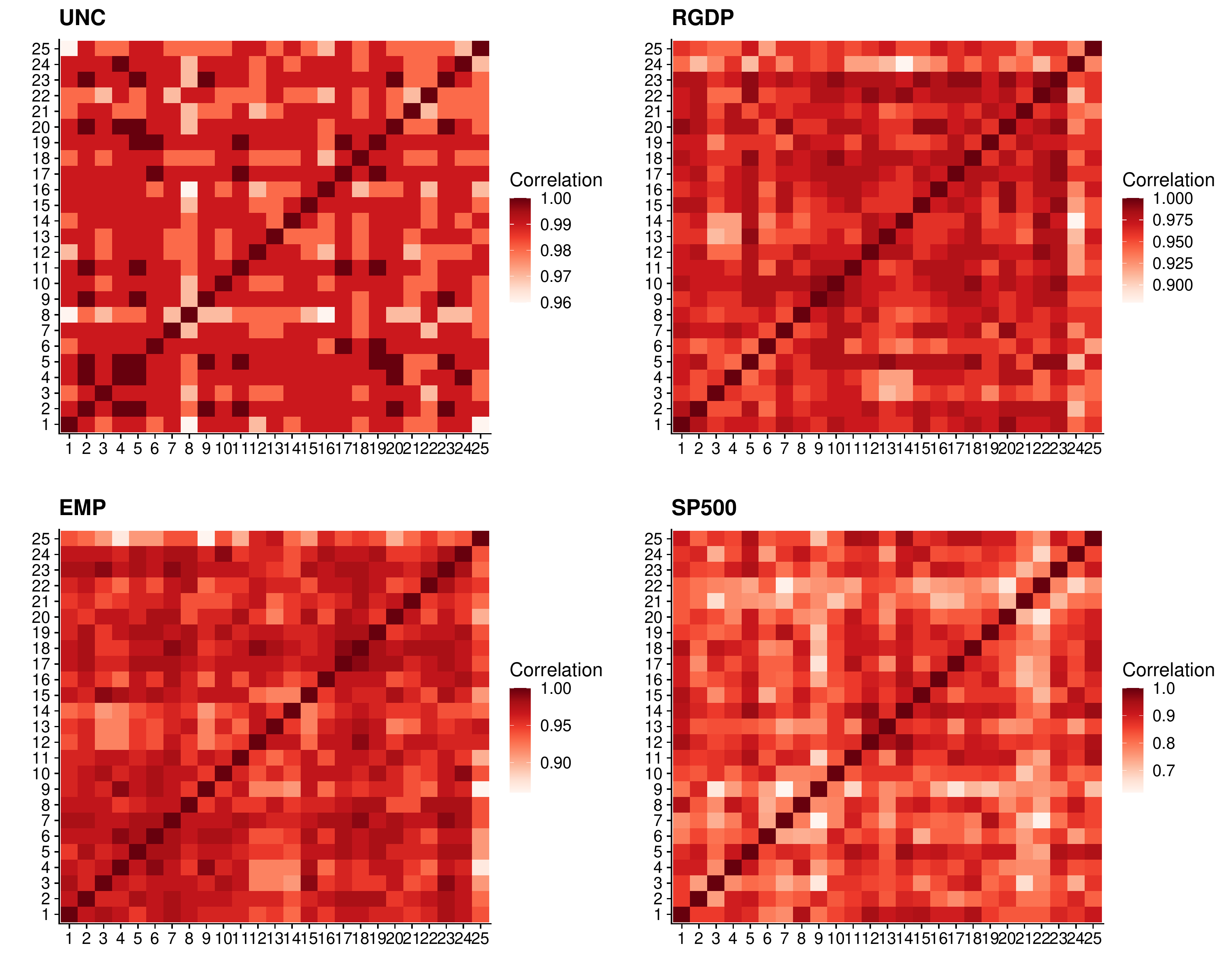}
\end{minipage}
\begin{minipage}{\textwidth}
\vspace*{-5pt}
\scriptsize \textit{Notes:} The figure shows, for each focus variable, the correlation of the median of the average GIRF for 25 different variable orderings in the estimation. The macroeconomic uncertainty index is always ordered second and the S\&P 500 is always ordered first. The reference model specification used is the GP-VAR-8.
\end{minipage}
\end{figure}
\newpage
\subsection{Impulse responses of non-focus variables in the GP-VAR-8}
\begin{figure}[!ht]
\caption{Impulse responses of non-focus variables in the GP-VAR-8 relative to a small-scale BVAR. \label{fig:Sothers_BVAR}}
\begin{minipage}{\textwidth}
\centering
\vspace*{-10pt}
\includegraphics[width = \textwidth]{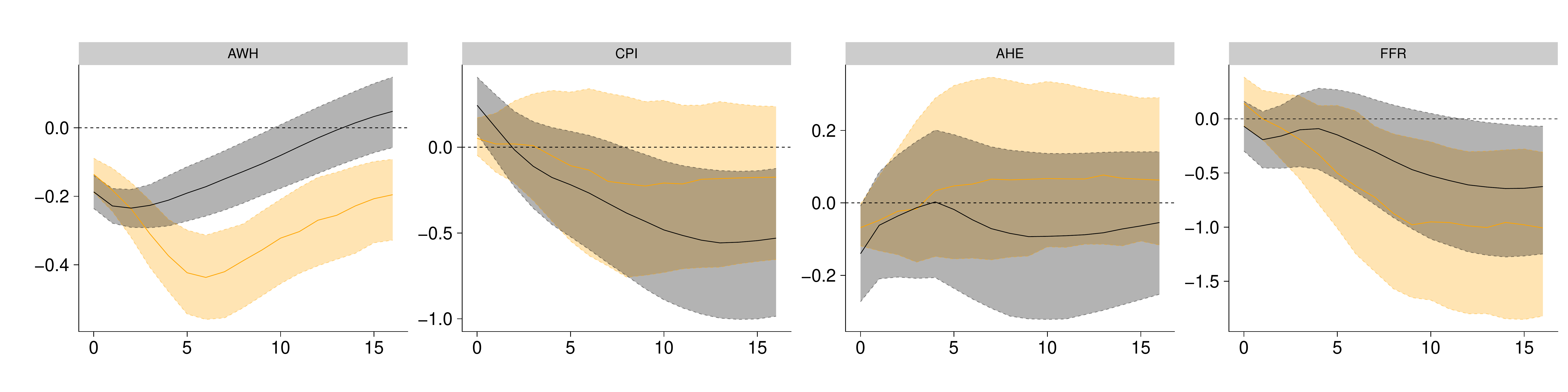}
\end{minipage}
\begin{minipage}{\textwidth}
\centering
\hspace*{50pt}
\includegraphics[scale=0.30]{key_avg_BVAR_GPS_legend.pdf}
\end{minipage}

\begin{minipage}{\textwidth}
\vspace*{-5pt}
\scriptsize \textit{Notes:} Average generalized impulse responses (GIRFs, outlined in Sub-section \ref{sec:GIRFs}) to a positive one standard deviation shock in macroeconomic uncertainty. Solid lines denote the posterior medians, while shaded areas correspond to the $68\%$ posterior credible sets. GP-VAR-8 refers to the smallest variant of our non-parametric model and BVAR-8 refers to a small-scale BVAR with SV, which is closely related to the specification used in \citet{jurado2015measuring}.
\end{minipage}
\end{figure}

\begin{figure}[!ht]
\caption{Impulse responses of non-focus variables in the GP-VAR-8 for three different sub-sample periods.}
\centering
\begin{minipage}{\textwidth}
\centering
\includegraphics[width = 0.95\textwidth]{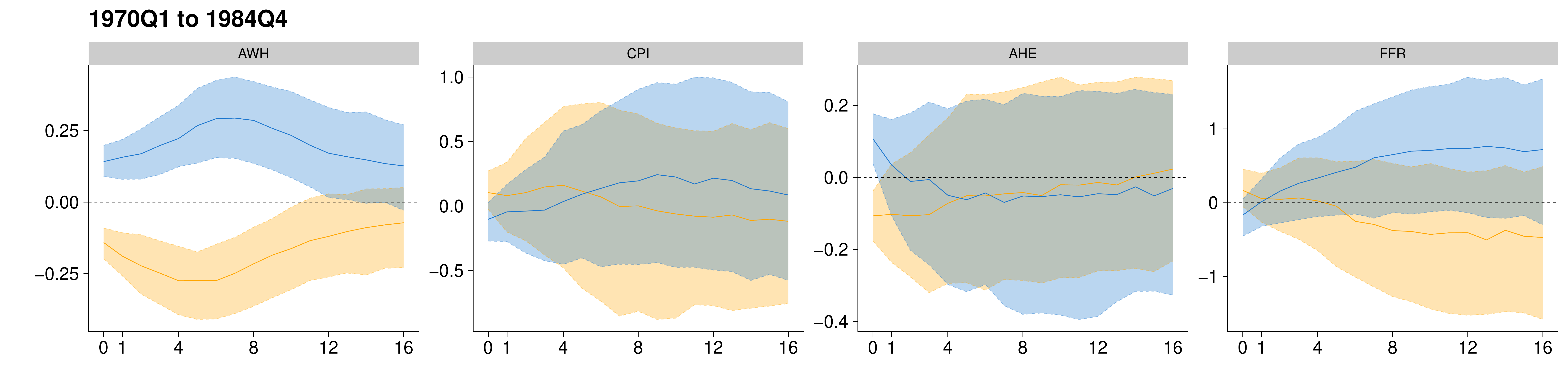}
\end{minipage}
\begin{minipage}{\textwidth}
\centering
\includegraphics[width = 0.95\textwidth]{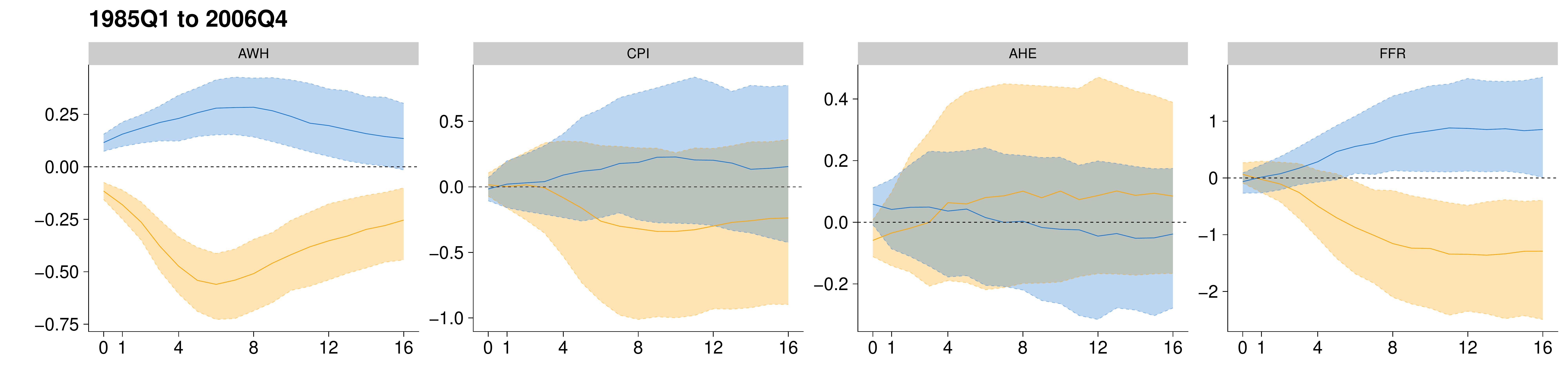}\end{minipage}
\begin{minipage}{\textwidth}
\centering
\includegraphics[width = 0.95\textwidth]{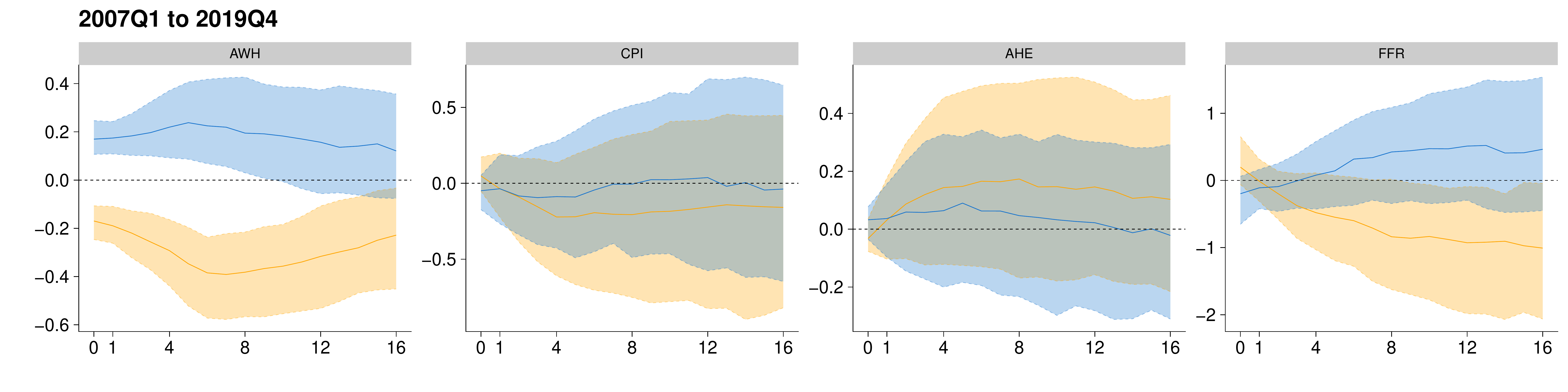}
\end{minipage}
\begin{minipage}{\textwidth}
\centering
\hspace*{15pt}\includegraphics[scale=0.3]{avg_subsample_legend.pdf}
\end{minipage}
\begin{minipage}{\textwidth}
\scriptsize \textit{Notes:} Period-specific average generalized impulse responses (GIRFs, outlined in Sub-section \ref{sec:GIRFs}) to a positive (negative) one standard deviation shock in macroeconomic uncertainty. Solid lines denote the posterior medians, while shaded areas correspond to the $68\%$ posterior credible sets. 
\end{minipage}
\end{figure}

\begin{figure}[!ht]
\caption{Impulse responses of focus variables in the GP-VAR-8 in recessions and expansions.}
\label{fig:synth_data}
\centering
\begin{minipage}{\textwidth}
\centering
\includegraphics[width = 0.95\textwidth]{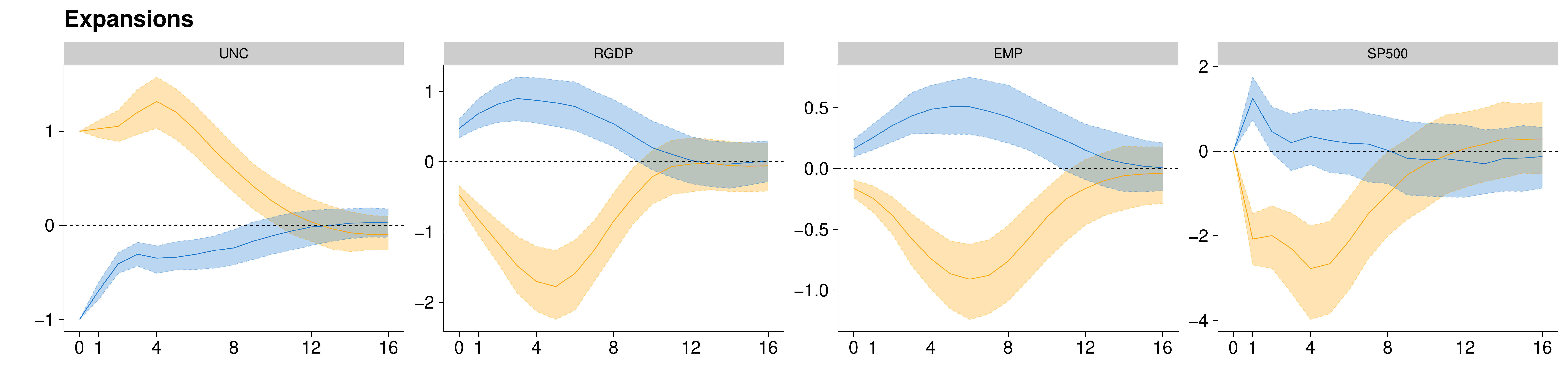}
\end{minipage}
\begin{minipage}{\textwidth}
\centering
\includegraphics[width = 0.95\textwidth]{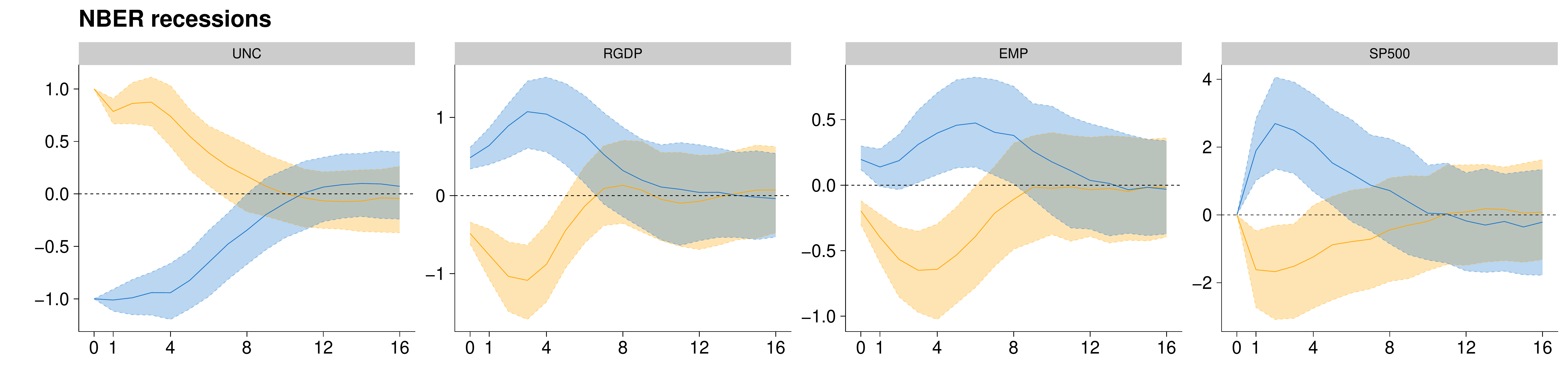}\end{minipage}
\begin{minipage}{\textwidth}
\centering
\hspace*{15pt}\includegraphics[scale=0.3]{avg_subsample_legend.pdf}
\end{minipage}
\begin{minipage}{\textwidth}
\scriptsize \textit{Notes:} Period-specific generalized impulse responses (GIRFs, outlined in Sub-section \ref{sec:GIRFs}) to a positive (orange) and negative (blue) one standard deviation shock in macroeconomic uncertainty. Solid lines denote the posterior medians, while shaded areas correspond to the $68\%$ posterior credible sets. We consider economic recessions and expansions according to the NBER Business Cycle Dating Committee.
\end{minipage}
\label{fig:rec_exp_app}
\end{figure}


\begin{figure}[!ht]
\caption{Period-specfic impulse responses of non-focus variables in the GP-VAR-8 across different sub-sample periods.}
\centering
\begin{minipage}{\textwidth}
\centering
\includegraphics[width = 0.95\textwidth]{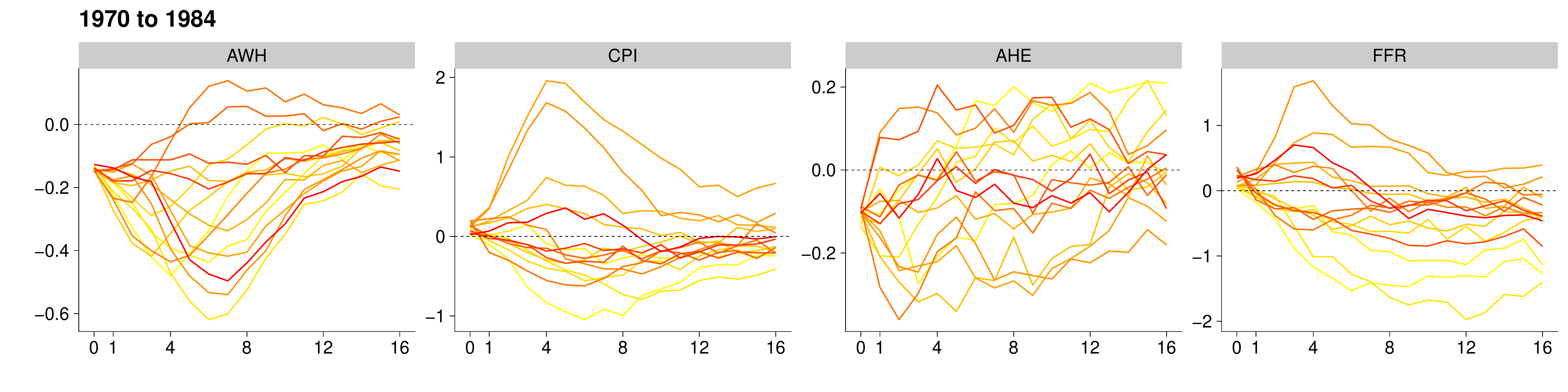}
\end{minipage}
\begin{minipage}{\textwidth}
\centering
\includegraphics[width = 0.95\textwidth]{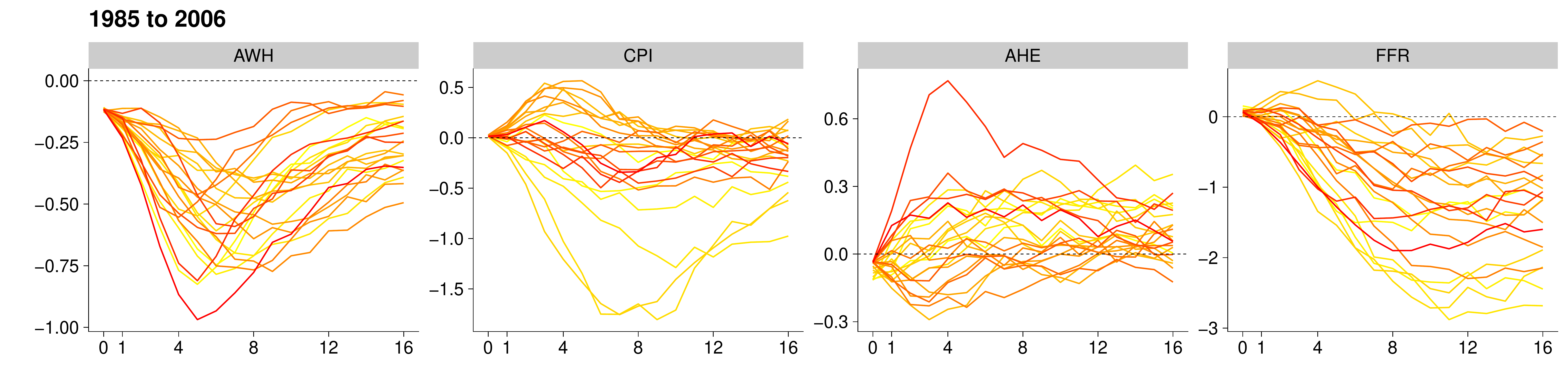}\end{minipage}
\begin{minipage}{\textwidth}
\centering
\includegraphics[width = 0.95\textwidth]{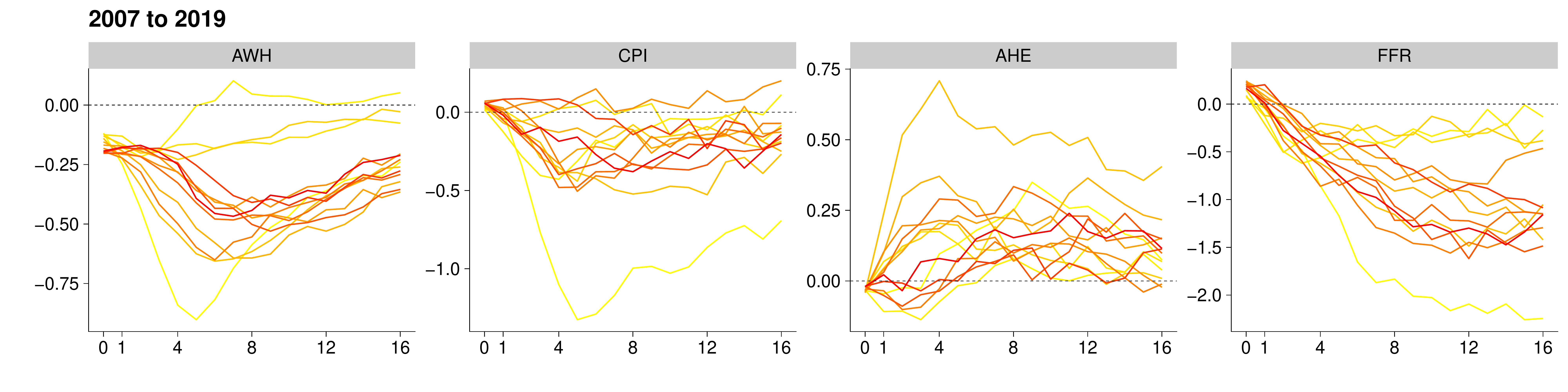}
\end{minipage}
\begin{minipage}{\textwidth}
\centering
\hspace*{15pt}\includegraphics[scale=0.3]{key_subperiods_median_legend.pdf}
\end{minipage}
\begin{minipage}{\textwidth}
\scriptsize \textit{Notes:} Impulse response functions to a positive one standard deviation shock in macroeconomic uncertainty in the GP-VAR-8 across sub-sample periods. Solid lines denote the yearly averaged posterior medians, with colors ranging from yellow (start of the sample) to red (end of the sample).
\end{minipage}
\label{fig:girfovert_app}
\end{figure}

\clearpage

\subsection{Additional results on model size and asymmetries}

\begin{figure}[!ht]
\caption{Impulse responses of non-focus variables across different information sets. \label{fig:compsize2}}
\centering
\begin{minipage}{\textwidth}
\centering
\vspace*{-10pt}
\includegraphics[width=\textwidth]{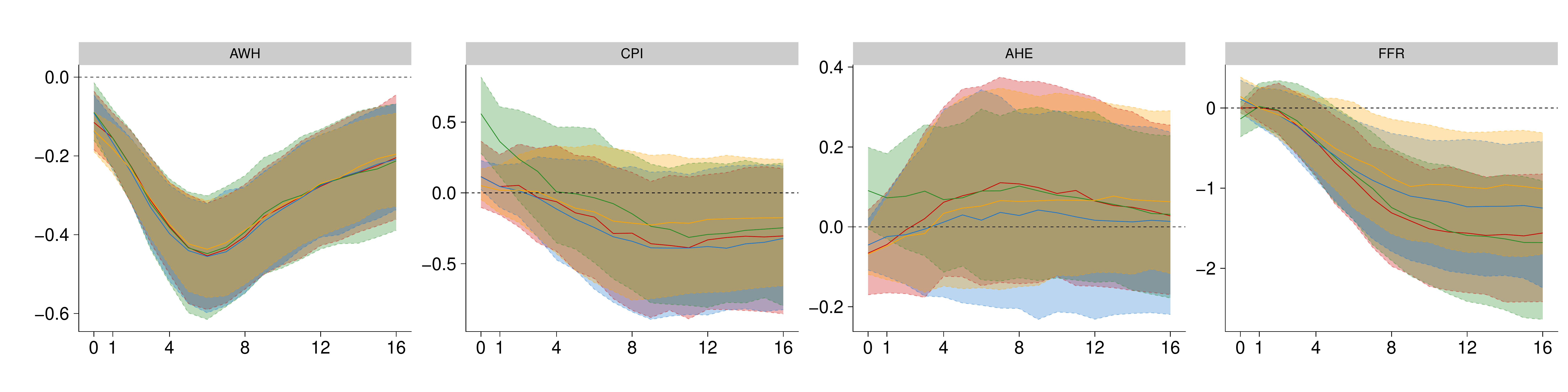}
\end{minipage}
\begin{minipage}{\textwidth}
\centering
\hspace*{50pt}
\includegraphics[scale=0.30]{key_avg_GP_all_legend.pdf}
\end{minipage}
\begin{minipage}{\textwidth}
\vspace*{-5pt}
\scriptsize \textbf{Notes:} Average generalized impulse responses (GIRFs, outlined in Sub-section \ref{sec:GIRFs}) to a positive one standard deviation shock in macroeconomic uncertainty. Solid lines denote the posterior medians, while shaded areas correspond to the $68\%$ posterior credible sets. 
\end{minipage}
\end{figure}

\begin{figure}[!ht]
\caption{Shock sign asymmetries in responses of non-focus variables for the GP-VAR-8.\label{fig:asym}}
\begin{minipage}{\textwidth}
\centering
\vspace*{-10pt}
\includegraphics[width = \textwidth]{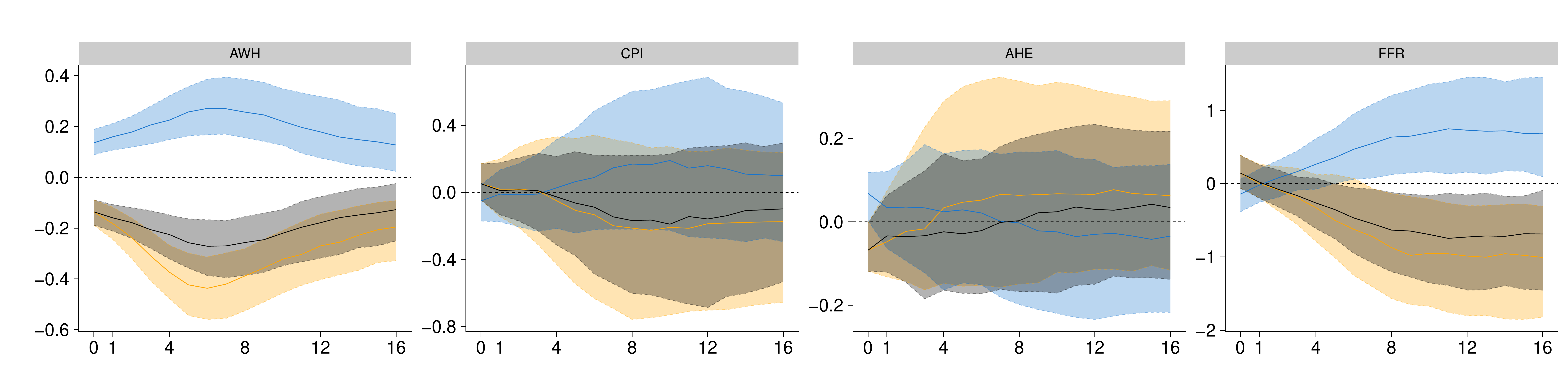}
\end{minipage}
\begin{minipage}{\textwidth}
\centering
\hspace*{50pt}
\includegraphics[scale=0.30]{key_avg_GPS_sign_legend.pdf}
\end{minipage}
\begin{minipage}{\textwidth}
\vspace*{-5pt}
\scriptsize \textbf{Notes:} Average generalized impulse responses (GIRFs, outlined in Sub-section \ref{sec:GIRFs}) to a negative (positive) one standard deviation shock in macroeconomic uncertainty. Solid lines denote the posterior medians, while shaded areas correspond to the $68\%$ posterior credible sets.  Here, negative  $\times (-1)$ denotes a negative one standard standard deviation shock with the respective responses being mirrored across the x-axis.
\end{minipage}
\end{figure}

\begin{figure}[!ht]
\caption{Shock size asymmetries in responses of non-focus variables for the GP-VAR-8. \label{fig:asym_shocks_app}}
\begin{minipage}{\textwidth}
\centering
\vspace*{-25pt}
\includegraphics[width = \textwidth]{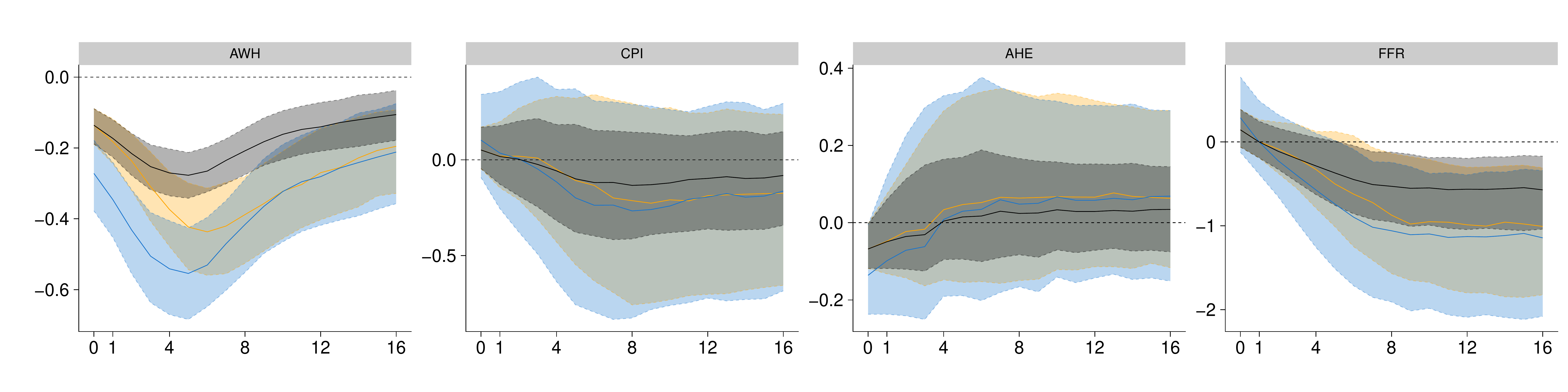}
\end{minipage}
\begin{minipage}{\textwidth}
\centering
\hspace*{50pt}
\includegraphics[scale=0.30]{key_avg_GPS_size_legend.pdf}
\end{minipage}
\begin{minipage}{\textwidth}
\vspace*{-5pt}
\scriptsize \textbf{Notes:} Average generalized impulse responses (GIRFs, outlined in Sub-section \ref{sec:GIRFs}) to a positive two (one) standard deviation shock in macroeconomic uncertainty. Solid lines denote the posterior medians, while shaded areas correspond to the $68\%$ posterior credible sets. Here, $1$ sd refers to a one standard deviation shock, $2$ sd indicates a two standard deviation shock, and $2$ sd  $\times (1/2)$ denotes a two standard deviation shock with the respective responses divided by two. 
 \end{minipage}
\end{figure}

\begin{figure}[!ht]
\caption{Shock sign asymmetries in impulse responses of focus variables for the GP-VAR-64.\label{fig:asymL}}
\begin{minipage}{\textwidth}
\centering
\includegraphics[scale=0.2]{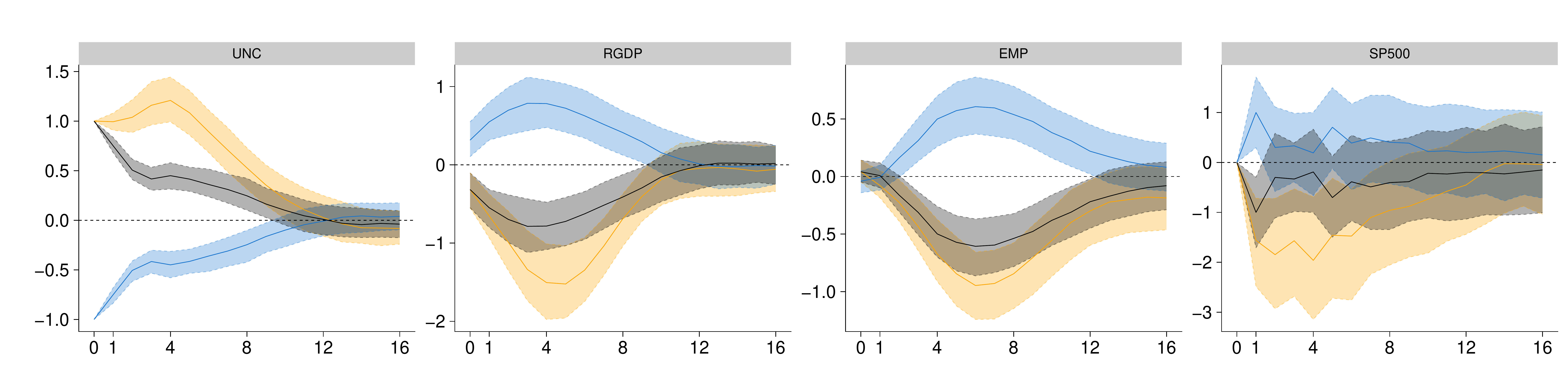}
\end{minipage}
\begin{minipage}{\textwidth}
\centering
\hspace*{50pt}
\includegraphics[scale=0.30]{key_avg_GPS_sign_legend.pdf}
\end{minipage}
\begin{minipage}{\textwidth}
\vspace*{-5pt}
\scriptsize \textbf{Notes:} Average generalized impulse responses (GIRFs, outlined in Sub-section \ref{sec:GIRFs}) to a negative (positive) one standard deviation shock in macroeconomic uncertainty. Solid lines denote the posterior medians, while shaded areas correspond to the $68\%$ posterior credible sets.  Here, negative x $(-1)$ denotes a negative one standard standard deviation shock with the respective responses being mirrored across the x-axis.
\end{minipage}
\end{figure}

\begin{figure}[!ht]
\caption{Shock size asymmetries in impulse responses of focus variables for the GP-VAR-64. \label{fig:asym_shocksL_app}}
\begin{minipage}{\textwidth}
\centering
\includegraphics[scale=0.2]{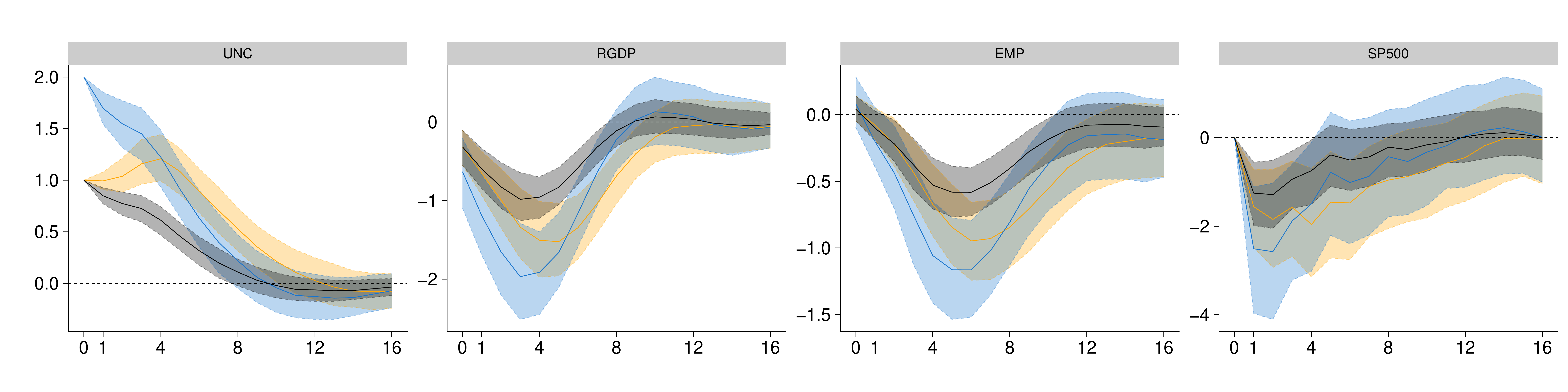}
\end{minipage}
\begin{minipage}{\textwidth}
\centering
\hspace*{50pt}
\includegraphics[scale=0.30]{key_avg_GPS_size_legend.pdf}
\end{minipage}
\begin{minipage}{\textwidth}
\vspace*{-5pt}
\scriptsize \textbf{Notes:} Average generalized impulse responses (GIRFs, outlined in Sub-section \ref{sec:GIRFs}) to a positive two (one) standard deviation shock in macroeconomic uncertainty. Solid lines denote the posterior medians, while shaded areas correspond to the $68\%$ posterior credible sets. Here, $1$ sd refers to a one standard deviation shock, $2$ sd indicates a two standard deviation shock, and $2$ sd x $(1/2)$ denotes a two standard deviation shock with the respective responses divided by two.
\end{minipage}
\end{figure}

\clearpage


\end{appendices}
\end{document}

%% file: LPSfinal_SVhom.tex
\begin{table}[!tbp]
{\tiny
\begin{center}
\caption{Log predictive Bayes factors (LPBFs) relative to a small-scale BVAR with stochastic volatility (SV). \label{tab:lpl3vars}}
\begin{tabular}{llcrrrrcrrrr}
\toprule
\multicolumn{2}{c}{\bfseries Specification}&\multicolumn{1}{c}{\bfseries }&\multicolumn{4}{c}{\bfseries One-step-ahead}&\multicolumn{1}{c}{\bfseries }&\multicolumn{4}{c}{\bfseries Four-step-ahead}\tabularnewline
\cmidrule{1-2} \cmidrule{4-7} \cmidrule{9-12} 
\multicolumn{1}{c}{M}&\multicolumn{1}{c}{Grid for kernel hyperparameters}&\multicolumn{1}{c}{}&\multicolumn{1}{c}{Joint}&\multicolumn{1}{c}{RGDP}&\multicolumn{1}{c}{CPI}&\multicolumn{1}{c}{FFR}&\multicolumn{1}{c}{}&\multicolumn{1}{c}{Joint}&\multicolumn{1}{c}{RGDP}&\multicolumn{1}{c}{CPI}&\multicolumn{1}{c}{FFR}\tabularnewline
\midrule
\multicolumn{12}{c}{GP-VAR homoskedastic}\tabularnewline
&&&&&&&&&&&\tabularnewline
64& semi-automatic && --90.310& --7.769& --12.312&--59.702&& --52.285&  1.021&  --9.443&--31.713\tabularnewline
& semi-automatic w/o linear scaling && --78.323& --1.832& --14.770&--49.583&& --45.304&\textbf{5.940}& --10.795&--22.169\tabularnewline
& naive && --84.216& --4.953& --23.869&--53.582&& --38.794&  3.785&  --4.312&--24.104\tabularnewline
&&&&&&&&&&&\tabularnewline
32&semi-automatic&& --82.036& --8.549& --11.667&--58.767&& --45.526&  1.816&  --6.588&--30.567\tabularnewline
&semi-automatic w/o linear scaling&& --72.908& --2.449& --16.077&--48.757&& --53.405&  4.485& --22.694&--17.920\tabularnewline
&naive&& --71.102& --3.640& --16.168&--51.738&& --46.112&  2.980& --14.723&--20.679\tabularnewline
&&&&&&&&&&&\tabularnewline
16&semi-automatic&& --80.595& --7.388& --11.332&--58.063&& --45.852&  1.234&  --6.506&--30.582\tabularnewline
&semi-automatic w/o linear scaling&& --83.747& --0.300& --18.696&--45.319&& --51.862&  2.849& --20.447&--18.794\tabularnewline
&naive&& --81.852&  0.088& --15.936&--47.053&& --42.958&  2.051& --10.318&--20.849\tabularnewline
&&&&&&&&&&&\tabularnewline
8&semi-automatic&& --75.047& --7.007& --11.688&--52.163&& --43.450&  3.441&  --6.034&--27.388\tabularnewline
&semi-automatic w/o linear scaling&& --78.854& --1.254& --21.087&--46.306&& --46.022&  4.743& --15.341&--19.580\tabularnewline
&naive&& --73.726& --2.155& --16.128&--46.877&& --43.949&  4.520& --12.480&--22.536\tabularnewline
\midrule
\multicolumn{12}{c}{GP-VAR SV}\tabularnewline
&&&&&&&&&&&\tabularnewline
64&semi-automatic&&  --7.995& --5.423&  --0.426& --2.106&&   6.370& --6.907&  11.058&  9.543\tabularnewline
&semi-automatic w/o linear scaling&& --15.400& --6.154&  --4.971& --1.117&&  --0.077&--14.521&   7.616&  8.992\tabularnewline
&naive&&  --4.490& --5.829&   3.954&  0.186&&  --4.247& --9.763&   9.384&  8.585\tabularnewline
&&&&&&&&&&&\tabularnewline
32&semi-automatic&&  --3.673& --4.881&  --0.672&  3.699&&  10.102& --4.945&  12.764&  7.434\tabularnewline
&semi-automatic w/o linear scaling&&  --5.219& --5.661&  --0.629&  2.134&&  --0.016&--12.096&  11.029&  4.948\tabularnewline
&naive&&  --5.751& --5.225&  --0.902&  3.905&&   3.513&--12.650&  11.346&  5.793\tabularnewline
&&&&&&&&&&&\tabularnewline
16&semi-automatic&&\textbf{10.206}& --4.387&   7.387&  7.222&&  \textbf{17.899} & --7.682&  15.316&\textbf{10.967}\tabularnewline
&semi-automatic w/o linear scaling&&   4.986& --4.640&   6.485&  6.010&&   7.336&--14.249&\textbf{18.058}&  9.332\tabularnewline
&naive&&   7.444& --2.179&   6.436&  9.980&&   6.735&--13.909&  15.195&  8.764\tabularnewline
&&&&&&&&&&&\tabularnewline
8&semi-automatic&&   4.459& --7.142&  10.887&  3.285&&   7.989& --9.383&  16.009&  8.377\tabularnewline
&semi-automatic w/o linear scaling&&   7.496& --7.652&\textbf{12.122}&  7.648&&  13.661&--10.402&  14.084&  9.753\tabularnewline
&naive&&   8.994& --7.158&  11.004&\textbf{11.154}&&   8.808&--11.680&  14.594&  9.055\tabularnewline
\midrule
\multicolumn{12}{c}{BART-VAR SV}\tabularnewline
&&&&&&&&&&&\tabularnewline
64&&& --41.432& --5.128& --12.146&--18.671&& --10.984&  2.714&   1.216&--11.179\tabularnewline
32&&& --39.967& --5.513&  --4.017&--26.650&& --17.564& --3.351&   5.918&--15.872\tabularnewline
16&&& --32.691& --2.987&  --2.448&--22.042&& --17.153&  0.580&   1.398&--17.512\tabularnewline
8&&& --33.579& --5.345&  --3.765&--20.196&& --13.522&  5.349&   0.769&--21.598\tabularnewline
\midrule
\multicolumn{12}{c}{Minnesota TVP-VAR SV}\tabularnewline
&&&&&&&&&&&\tabularnewline
8&&& 1.648& --1.531&   2.220& --2.268&&  --3.543& --1.877&   0.948& --0.359\tabularnewline
\midrule
\multicolumn{12}{c}{Minnesota BVAR SV}\tabularnewline
&&&&&&&&&&&\tabularnewline
64&&&  --0.220& --0.347&  --8.011&  8.268&& --23.563& --3.944& --18.503& --2.696\tabularnewline
32&&&  --7.983&\textbf{2.592}&  --9.862&  0.911&& --12.071& --4.846&  --6.127&  2.137\tabularnewline
16&&&   3.426&  2.296&   1.746& --0.003&&  --5.153& --1.373&  --2.152&  0.137\tabularnewline
&&&&&&&&&&&\tabularnewline
\shadeBench 8& Benchmark&&--223.987&--73.554&--138.633&--11.961&&--278.520&--91.359&--151.355&--41.429\tabularnewline
\bottomrule
\end{tabular}
\begin{minipage}{1.03\textwidth}
\scriptsize
\vspace*{5pt}
\noindent \textit{Notes:} The table shows joint and marginal LPBFs relative to a small-scale BVAR with SV for the one- and four-step-ahead horizons. The first column indicates model size, the second column provides information on how the hyperparameters are treated. We consider three cases: ``semi-automatic" refers to a two-dimensional grid for both the (inverse) length scale and the linear scaling parameter which is scaled using the median heuristic, ``semi-automatic w/o linear scaling" to a grid only for the (inverse) length scale  and sets $\xi_{jk} = 1$ while ``naive" refers to a grid only for the (inverse) length scale without scaling the grid using the median heuristic and $\xi_{jk} = 1$ ($j = 1, \dots, M$; $k = \{1,2\}$). Positive values imply that a particular model improves upon the small-scale BVAR with SV, while negative values suggest that the benchmark is preferred. The best specification for each variable and horizon combination is shown in bold. The red shaded entries associated with the benchmark give the actual log predictive likelihoods (LPLs).
\end{minipage}
\end{center}}
\end{table}

%% file: data_desc.tex
\begin{table}[!htbp]
{\tiny
\begin{center}
\caption{Data description.\label{tab:data}}
\scalebox{0.8}{
\begin{tabular}{lllccccc}
\toprule
\multicolumn{1}{l}{\ }&\multicolumn{1}{c}{\bfseries Mnemonic}&\multicolumn{1}{c}{\bfseries Description}&\multicolumn{1}{c}{\bfseries Trans.}&\multicolumn{1}{c}{\bfseries VAR-8}&\multicolumn{1}{c}{\bfseries VAR-16}&\multicolumn{1}{c}{\bfseries VAR-32}&\multicolumn{1}{c}{\bfseries VAR-64}\tabularnewline
\midrule
&UNC& Macroeconomic Uncertainty Index of \cite{jurado2015measuring} &1&x&x&x&x\tabularnewline
\midrule
~~&GDPC1 (RGDP) &Real Gross Domestic Product&$2$&x&x&x&x\tabularnewline
~~&CE16OV (EMP) &Civilian Employment (Thousands of Persons)&$2$&x&x&x&x\tabularnewline
~~&AWHMAN (AWH) &Average Weekly Hours of Production and Nonsupervisory Employees:  Manufacturing&$1$&x&x&x&x\tabularnewline
~~&CPIAUCSL (CPI) &Consumer Price Index for All Urban Consumers:  All Items&$2$&x&x&x&x\tabularnewline
~~&CES3000000008x (AHE) &Real Average Hourly Earnings of Production and Nonsupervisory Employees: Manufacturing&$2$&x&x&x&x\tabularnewline
~~&FEDFUNDS (FFR) &Effective Federal Funds Rate (Percent)&$1$&x&x&x&x\tabularnewline
~~&S.P.500 (SP500) &S\&P's Common Stock Price Index:  Composite&$3$&x&x&x&x\tabularnewline
\midrule
~~&PCECC96&Real Personal Consumption Expenditures&$2$&&x&x&x\tabularnewline
~~&FPIx&Real private fixed investment &$2$&&x&x&x\tabularnewline
~~&UNRATE&Civilian Unemployment Rate (Percent)&$1$&&x&x&x\tabularnewline
~~&CES0600000007&Average Weekly Hours of Production and Nonsupervisory Employees:  Goods-Producing&$1$&&x&x&x\tabularnewline
~~&CLAIMSx&Initial Claims&$2$&&x&x&x\tabularnewline
~~&HOUST&Housing Starts: Total: New Privately Owned Housing Units Started&$2$&&x&x&x\tabularnewline
~~&CES0600000008&Average Hourly Earnings of Production and Nonsupervisory Employees:&$2$&&x&x&x\tabularnewline
~~&M2REAL&Real M2 Money Stock&$2$&&x&x&x\tabularnewline
\midrule
~~&GCEC1&Real Government Consumption Expenditures and Gross Investment&$2$&&&x&x\tabularnewline
~~&INDPRO&IP:Total index Industrial Production Index (Index 2012=100)&$2$&&&x&x\tabularnewline
~~&CUMFNS&Capacity Utilization:  Manufacturing (SIC) (Percent of Capacity)&$1$&&&x&x\tabularnewline
~~&PAYEMS& Emp:Nonfarm All Employees: Total nonfarm (Thousands of Persons)&$2$&&&x&x\tabularnewline
~~&PERMIT&New Private Housing Units Authorized by Building Permits&$2$&&&x&x\tabularnewline
~~&PCECTPI&Personal Consumption Expenditures: Chain-type Price Index &$2$&&&x&x\tabularnewline
~~&GDPCTPI&Gross Domestic Product: Chain-type Price Index&$2$&&&x&x\tabularnewline
~~&CES2000000008x&Real Average Hourly Earnings of Production and Nonsupervisory Employees: Construction&$2$&&&x&x\tabularnewline
~~&BAA10YM&Moody's Seasoned Baa Corporate Bond Yield Relative to Yield on 10-Year Treasury&$1$&&&x&x\tabularnewline
~~&GS10TB3Mx&10-Year Treasury Constant Maturity Minus 3-Month Treasury Bill, secondary market&$1$&&&x&x\tabularnewline
~~&TB3SMFFM&3-Month Treasury Constant Maturity Minus Federal Funds Rate&$1$&&&x&x\tabularnewline
~~&AAAFFM&Moody's Seasoned Aaa Corporate Bond Minus Federal Funds Rate&$1$&&&x&x\tabularnewline
~~&BUSLOANSx&Real Commercial and Industrial Loans, All Commercial Banks&$2$&&&x&x\tabularnewline
~~&CONSUMERx&Real Consumer Loans at All Commercial Banks &$2$&&&x&x\tabularnewline
~~&NONREVSLx&Total Real Nonrevolving Credit Owned and Securitized, Outstanding&$2$&&&x&x\tabularnewline
~~&NONBORRES&Reserves Of Depository Institutions, Nonborrowed&$4$&&&x&x\tabularnewline
\midrule
~~&GPDIC1&Real Gross Private Domestic Investment&$2$&&&&x\tabularnewline
~~&PNFIx&Real private fixed investment:  Nonresidential &$2$&&&&x\tabularnewline
~~&PRFIx&Real private fixed investment:  Residential &$2$&&&&x\tabularnewline
~~&EXPGSC1&Real Exports of Goods and Services&$2$&&&&x\tabularnewline
~~&IMPGSC1&Real Imports of Goods and Services&$2$&&&&x\tabularnewline
~~&IPCONGD&IP:Consumer goods Industrial Production: Consumer Goods (Index 2012=100)&$2$&&&&x\tabularnewline
~~&UNRATELTx&Unemployment Rate for more than 27 weeks (Percent)&$1$&&&&x\tabularnewline
~~&AWOTMAN&Average Weekly Overtime Hours of Production and Nonsupervisory Employees: Manufacturing&$1$&&&&x\tabularnewline
~~&AMDMNOx&Real Manufacturers' New Orders:  Durable Goods (Millions of 2012 Dollars)&$2$&&&&x\tabularnewline
~~&GPDICTPI&Gross Private Domestic Investment: Chain-type Price Index &$2$&&&&x\tabularnewline
~~&DGDSRG3Q086SBEA&Personal consumption expenditures:  Goods &$2$&&&&x\tabularnewline
~~&DDURRG3Q086SBEA&Personal consumption expenditures:  Durable goods &$2$&&&&x\tabularnewline
~~&DSERRG3Q086SBEA&Personal consumption expenditures:  Services &$2$&&&&x\tabularnewline
~~&DNDGRG3Q086SBEA&Personal consumption expenditures:  Nondurable goods&$2$&&&&x\tabularnewline
~~&CPILFESL&Consumer Price Index for All Urban Consumers:  All Items Less Food \& Energy&$2$&&&&x\tabularnewline
~~&OILPRICEx&Real Crude Oil Prices:  West Texas Intermediate (WTI) - Cushing, Oklahoma&$2$&&&&x\tabularnewline
~~&COMPRNFB&Nonfarm Business Sector:  Real Compensation Per Hour (Index 2012=100)&$2$&&&&x\tabularnewline
~~&RCPHBS&Business Sector:  Real Compensation Per Hour (Index 2012=100)&$2$&&&&x\tabularnewline
~~&TB3MS&3-Month Treasury Bill: Secondary Market Rate (Percent)&$1$&&&&x\tabularnewline
~~&TB6MS&6-Month Treasury Bill: Secondary Market Rate (Percent)&$1$&&&&x\tabularnewline
~~&GS1&1-Year Treasury Constant Maturity Rate (Percent)&$1$&&&&x\tabularnewline
~~&GS10&10-Year Treasury Constant Maturity Rate (Percent)&$1$&&&&x\tabularnewline
~~&AAA&Moody's Seasoned Aaa Corporate Bond Yield (Percent)&$1$&&&&x\tabularnewline
~~&BAA&Moody's Seasoned Baa Corporate Bond Yield (Percent)&$1$&&&&x\tabularnewline
~~&TB6M3Mx&6-Month Treasury Bill Minus 3-Month Treasury Bill, secondary market (Percent)&$1$&&&&x\tabularnewline
~~&GS1TB3Mx&1-Year Treasury Constant Maturity Minus 3-Month Treasury Bill, secondary market&$1$&&&&x\tabularnewline
~~&CPF3MTB3Mx&3-Month Commercial Paper Minus 3-Month Treasury Bill, secondary market&$1$&&&&x\tabularnewline
~~&M1REAL& Real M1 Money Stock&$2$&&&&x\tabularnewline
~~&REALLNx&Real Real Estate Loans, All Commercial Banks&$2$&&&&x\tabularnewline
~~&EXUSUKx&U.S. / U.K. Foreign Exchange Rate&$2$&&&&x\tabularnewline
~~&S.P..indust&S\&P's Common Stock Price Index:  Industrials&$3$&&&&x\tabularnewline
~~&S.P.div.yield&S\&P's Composite Common Stock:  Dividend Yield&$1$&&&&x\tabularnewline
\bottomrule
\end{tabular}}
\begin{minipage}{1.09\textwidth}
\vspace*{5pt}
\scriptsize 
\noindent \textit{Notes:} We use the macroeconomic uncertainty measure of \cite{jurado2015measuring} provided (and regularly updated) on the web page of Sydney C. Ludvigson (available online via \href{https://www.sydneyludvigson.com/macro-and-financial-uncertainty-indexes}{sydneyludvigson.com/macro-and-financial-uncertainty-indexes}). Otherwise, we rely on the quarterly version of the dataset proposed in \cite{mccracken2016fred}. \texttt{Trans} indicates the transformation applied to each variable with $(1)$ implying no transformation, $(2)$ denoting year-on-year growth rates, $(3)$ denoting quarter-on-quarter growth rates, and $(4)$ refers to quarter-on-quarter percentage changes.  
\end{minipage}
\end{center}}
\end{table}

%% file: lit.bib
@article{d2012century,
  title={A century of inflation forecasts},
  author={D'Agostino, Antonello and Surico, Paolo},
  journal={Review of Economics and Statistics},
  volume={94},
  number={4},
  pages={1097--1106},
  year={2012},
  publisher={The MIT Press}
}

@article{van2008partially,
  title={Partially collapsed {G}ibbs samplers: Theory and methods},
  author={Van Dyk, David A and Park, Taeyoung},
  journal={Journal of the American Statistical Association},
  volume={103},
  number={482},
  pages={790--796},
  year={2008},
  publisher={Taylor \& Francis}
}

@article{jin2022infinite,
  title={Infinite {M}arkov pooling of predictive distributions},
  author={Jin, Xin and Maheu, John M and Yang, Qiao},
  journal={Journal of Econometrics},
  volume={228},
  number={2},
  pages={302--321},
  year={2022},
  publisher={Elsevier}
}

@article{bassetti2014beta,
  title={Beta-product dependent {P}itman--{Y}or processes for {B}ayesian inference},
  author={Bassetti, Federico and Casarin, Roberto and Leisen, Fabrizio},
  journal={Journal of Econometrics},
  volume={180},
  number={1},
  pages={49--72},
  year={2014},
  publisher={Elsevier}
}

@article{castelnuovo2022jes,
  title={Uncertainty before and during {COVID}19: A survey},
  author={Castelnuovo, Efrem},
  journal={Journal of Economic Surveys},
  volume={forthcoming},
  number={},
  pages={},
  year={2022},
  publisher={Wiley}
}

@article{hirano2002semiparametric,
  title={Semiparametric {B}ayesian inference in autoregressive panel data models},
  author={Hirano, Keisuke},
  journal={Econometrica},
  volume={70},
  number={2},
  pages={781--799},
  year={2002},
  publisher={JSTOR}
}

@article{billio2019bayesian,
  title={Bayesian nonparametric sparse {VAR} models},
  author={Billio, Monica and Casarin, Roberto and Rossini, Luca},
  journal={Journal of Econometrics},
  volume={212},
  number={1},
  pages={97--115},
  year={2019},
  publisher={Elsevier}
}

@article{kalli2018bayesian,
  title={Bayesian nonparametric vector autoregressive models},
  author={Kalli, Maria and Griffin, Jim E},
  journal={Journal of Econometrics},
  volume={203},
  number={2},
  pages={267--282},
  year={2018},
  publisher={Elsevier}
}

@article{clark2021tail,
  title={Tail forecasting with multivariate {B}ayesian additive regression trees},
  author={Clark, Todd E and Huber, Florian and Koop, Gary and Marcellino, Massimiliano and Pfarrhofer, Michael},
  year={2021},
  volume = {No. 21-08}, 
  journal={FRB of Cleveland Working Paper}
}

@article{huber2022inference,
  title={Inference in {B}ayesian additive vector autoregressive tree models},
  author={Huber, Florian and Rossini, Luca},
  journal={The Annals of Applied Statistics},
  volume={16},
  number={1},
  pages={104--123},
  year={2022},
  publisher={Institute of Mathematical Statistics}
}

@article{geweke2010comparing,
  title={Comparing and evaluating {B}ayesian predictive distributions of asset returns},
  author={Geweke, John and Amisano, Gianni},
  journal={International Journal of Forecasting},
  volume={26},
  number={2},
  pages={216--230},
  year={2010},
  publisher={Elsevier}
}

@article{chan2017stochastic,
  title={The stochastic volatility in mean model with time-varying parameters: An application to inflation modeling},
  author={Chan, Joshua C.C.},
  journal={Journal of Business \& Economic Statistics},
  volume={35},
  number={1},
  pages={17--28},
  year={2017},
  publisher={Taylor \& Francis}
}

@article{makalic2015simple,
  title={A simple sampler for the horseshoe estimator},
  author={Makalic, Enes and Schmidt, Daniel F},
  journal={IEEE Signal Processing Letters},
  volume={23},
  number={1},
  pages={179--182},
  year={2015},
  publisher={IEEE}
}

@article{kim1998stochastic,
  title={Stochastic volatility: likelihood inference and comparison with {ARCH} models},
  author={Kim, Sangjoon and Shephard, Neil and Chib, Siddhartha},
  journal={The Review of Economic Studies},
  volume={65},
  number={3},
  pages={361--393},
  year={1998},
  publisher={Wiley-Blackwell}
}

@article{cong2017fast,
  title={Fast simulation of hyperplane-truncated multivariate {N}ormal distributions},
  author={Cong, Yulai and Chen, Bo and Zhou, Mingyuan},
  journal={Bayesian Analysis},
  volume={12},
  number={4},
  pages={1017--1037},
  year={2017},
  publisher={International Society for Bayesian Analysis}
}

@article{coulombe2022neural,
  title={A Neural {P}hillips Curve and a Deep Output Gap},
  author={Coulombe, Philippe Goulet},
  journal={arXiv},
  volume = {2202.04146},
  year={2022}
}

@article{hornik1989multilayer,
  title={Multilayer feedforward networks are universal approximators},
  author={Hornik, Kurt and Stinchcombe, Maxwell and White, Halbert},
  journal={Neural networks},
  volume={2},
  number={5},
  pages={359--366},
  year={1989},
  publisher={Elsevier}
}

@article{novak2018bayesian,
  title={Bayesian deep convolutional networks with many channels are {G}aussian processes},
  author={Novak, Roman and Xiao, Lechao and Lee, Jaehoon and Bahri, Yasaman and Yang, Greg and Hron, Jiri and Abolafia, Daniel A and Pennington, Jeffrey and Sohl-Dickstein, Jascha},
  volume = {1810.05148},
  journal={arXiv},
  year={2018}
}

@article{carriero2022specification,
  title={Specification Choices in Quantile Regression for Empirical Macroeconomics},
  author={Carriero, Andrea and Clark, Todd E and Marcellino, Massimiliano Giuseppe},
  year={2022},
  volume = {No. 22-25},
  journal = {FRB of Cleveland Working Paper}
}

@article{jylanki2011robust,
  title={Robust {G}aussian Process Regression with a {S}tudent-t Likelihood.},
  author={Jyl{\"a}nki, Pasi and Vanhatalo, Jarno and Vehtari, Aki},
  journal={Journal of Machine Learning Research},
  volume={12},
  number={11},
  year={2011}
}

@article{stock2007has,
  title={Why has {US} inflation become harder to forecast?},
  author={Stock, James H and Watson, Mark W},
  journal={Journal of Money, Credit and Banking},
  volume={39},
  pages={3--33},
  year={2007},
  publisher={Wiley Online Library}
}

@article{stone1982optimal,
  title={Optimal global rates of convergence for nonparametric regression},
  author={Stone, Charles J},
  journal={The Annals of Statistics},
  pages={1040--1053},
  year={1982},
  publisher={JSTOR}
}

@article{van2008rates,
  title={Rates of contraction of posterior distributions based on {G}aussian process priors},
  author={van der Vaart, Aad W and van Zanten, J Harry},
  journal={The Annals of Statistics},
  volume={36},
  number={3},
  pages={1435--1463},
  year={2008},
  publisher={Institute of Mathematical Statistics}
}

@article{yang2017frequentist,
  title={Frequentist coverage and sup-norm convergence rate in {G}aussian process regression},
  author={Yang, Yun and Bhattacharya, Anirban and Pati, Debdeep},
  journal={arXiv},
  volume = {1708.04753},
  year={2017}
}

@article{teckentrup2020convergence,
  title={Convergence of Gaussian process regression with estimated hyper-parameters and applications in Bayesian inverse problems},
  author={Teckentrup, Aretha L},
  journal={SIAM/ASA Journal on Uncertainty Quantification},
  volume={8},
  number={4},
  pages={1310--1337},
  year={2020},
  publisher={SIAM}
}

@article{mercer1909xvi,
  title={Functions of positive and negative type, and their connection the theory of integral equations},
  author={Mercer, James},
  journal={Philosophical transactions of the Royal Society of London. Series A, containing papers of a mathematical or physical character},
  volume={209},
  number={441-458},
  pages={415--446},
  year={1909},
  publisher={The Royal Society London}
}

@article{fruhwirth2019here,
  title={From here to infinity: sparse finite versus {D}irichlet process mixtures in model-based clustering},
  author={Fr{\"u}hwirth-Schnatter, Sylvia and Malsiner-Walli, Gertraud},
  journal={Advances in Data Analysis and Classification},
  volume={13},
  number={1},
  pages={33--64},
  year={2019},
  publisher={Springer}
}

@article{neal2000markov,
  title={Markov chain sampling methods for {D}irichlet process mixture models},
  author={Neal, Radford M},
  journal={Journal of Computational and Graphical Statistics},
  volume={9},
  number={2},
  pages={249--265},
  year={2000},
  publisher={Taylor \& Francis}
}

@article{escobar1995bayesian,
  title={Bayesian density estimation and inference using mixtures},
  author={Escobar, Michael D and West, Mike},
  journal={Journal of the American Statistical Association},
  volume={90},
  number={430},
  pages={577--588},
  year={1995},
  publisher={Taylor \& Francis}
}

@article{carvalho2010horseshoe,
  title={The horseshoe estimator for sparse signals},
  author={Carvalho, Carlos M and Polson, Nicholas G and Scott, James G},
  journal={Biometrika},
  volume={97},
  number={2},
  pages={465--480},
  year={2010},
  publisher={Oxford University Press}
}

@article{kowal2017bayesian,
  title={A {B}ayesian multivariate functional dynamic linear model},
  author={Kowal, Daniel R and Matteson, David S and Ruppert, David},
  journal={Journal of the American Statistical Association},
  volume={112},
  number={518},
  pages={733--744},
  year={2017},
  publisher={Taylor \& Francis}
}

@article{gu2021autoencoder,
  title={Autoencoder asset pricing models},
  author={Gu, Shihao and Kelly, Bryan and Xiu, Dacheng},
  journal={Journal of Econometrics},
  volume={222},
  number={1},
  pages={429--450},
  year={2021},
  publisher={Elsevier}
}

@article{coulombe2020macroeconomy,
  title={The Macroeconomy as a Random Forest},
  author={Coulombe, Philippe Goulet},
  journal={arXiv},
  volume = {2006.12724},
  year={2020}
}

@article{chipman2010bart,
  title={{BART}: {B}ayesian additive regression trees},
  author={Chipman, Hugh A and George, Edward I and McCulloch, Robert E},
  journal={The Annals of Applied Statistics},
  volume={4},
  number={1},
  pages={266--298},
  year={2010},
  publisher={Institute of Mathematical Statistics}
}

@article{huber2020nowcasting,
  title={Nowcasting in a pandemic using non-parametric mixed frequency {VAR}s},
  author={Huber, Florian and Koop, Gary and Onorante, Luca and Pfarrhofer, Michael and Schreiner, Josef},
  journal={Journal of Econometrics},
  volume = {forthcoming},
  year={2020},
  publisher={Elsevier}
}

@article{cai2000functional,
  title={Functional-coefficient regression models for nonlinear time series},
  author={Cai, Zongwu and Fan, Jianqing and Yao, Qiwei},
  journal={Journal of the American Statistical Association},
  volume={95},
  number={451},
  pages={941--956},
  year={2000},
  publisher={Taylor \& Francis}
}

@article{caggiano2017estimating,
  title={Estimating the real effects of uncertainty shocks at the zero lower bound},
  author={Caggiano, Giovanni and Castelnuovo, Efrem and Pellegrino, Giovanni},
  journal={European Economic Review},
  volume={100},
  pages={257--272},
  year={2017},
  publisher={Elsevier}
}

@article{caggiano2021uncertainty,
  title={Uncertainty shocks and the great recession: Nonlinearities matter},
  author={Caggiano, Giovanni and Castelnuovo, Efrem and Pellegrino, Giovanni},
  journal={Economics Letters},
  volume={198, 109669},
  year={2021},
  publisher={Elsevier}
}

@article{friedman2001greedy,
  title={Greedy function approximation: a gradient boosting machine},
  author={Friedman, Jerome H},
  journal={Annals of Statistics},
  pages={1189--1232},
  year={2001},
  publisher={JSTOR}
}

@article{mumtaz2018changing,
  title={The changing transmission of uncertainty shocks in the {US}},
  author={Mumtaz, Haroon and Theodoridis, Konstantinos},
  journal={Journal of Business \& Economic Statistics},
  volume={36},
  number={2},
  pages={239--252},
  year={2018},
  publisher={Taylor \& Francis}
}

@article{accm2017changing,
  title={Have standard {VAR}s remained stable since the crisis?},
  author={Aastveit, Knut Are and Carriero, Andrea and Clark, Todd E. and Marcellino, Massimiliano},
  journal={Journal of Applied Econometrics},
  volume={32},
  number={5},
  pages={931-951},
  year={2017},
  publisher={Wiley}
}

@article{koop2013forecasting,
  title={Forecasting with medium and large {B}ayesian {VAR}s},
  author={Koop, Gary},
  journal={Journal of Applied Econometrics},
  volume={28},
  number={2},
  pages={177--203},
  year={2013},
  publisher={Wiley Online Library}
}

@book{tong1990non,
  title={Non-linear Time Series: {A} Dynamical System Approach},
  author={Tong, Howell},
  year={1990},
  publisher={Oxford University Press}
}

@book{chan_koop_poirier_tobias_2019, 
place={Cambridge}, 
edition={2}, 
series={Econometric Exercises}, 
title={Bayesian Econometric Methods},  
publisher={Cambridge University Press}, 
author={Chan, Joshua C.C. and Koop, Gary and Poirier, Dale J. and Tobias, Justin L.}, 
year={2019}, 
collection={Econometric Exercises}
}

@article{nyblom1989testing,
  title={Testing for the constancy of parameters over time},
  author={Nyblom, Jukka},
  journal={Journal of the American Statistical Association},
  volume={84},
  number={405},
  pages={223--230},
  year={1989},
  publisher={Taylor \& Francis}
}

@article{hamilton1989new,
  title={A new approach to the economic analysis of nonstationary time series and the business cycle},
  author={Hamilton, James D.},
  journal={Econometrica},
  pages={357--384},
  year={1989},
  publisher={JSTOR}
}

@article{stock1996evidence,
  title={Evidence on structural instability in macroeconomic time series relations},
  author={Stock, James H. and Watson, Mark W.},
  journal={Journal of Business \& Economic Statistics},
  volume={14},
  number={1},
  pages={11--30},
  year={1996},
  publisher={Taylor \& Francis}
}

@article{terasvirta1994specification,
  title={Specification, estimation, and evaluation of smooth transition autoregressive models},
  author={Ter{\"a}svirta, Timo},
  journal={Journal of the American Statistical Association},
  volume={89},
  number={425},
  pages={208--218},
  year={1994},
  publisher={Taylor \& Francis}
}

@InProceedings{robinson1991time,
author="Robinson, Peter M.",
editor="Hackl, Peter
and Westlund, Anders Holger",
title="Time-Varying Nonlinear Regression",
booktitle="Economic Structural Change",
year="1991",
publisher="Springer Berlin Heidelberg",
address="Berlin, Heidelberg",
pages="179--190",
isbn="978-3-662-06824-3"
}

@article{kapetanios2019large,
  title={Large time-varying parameter {VAR}s: {A} nonparametric approach},
  author={Kapetanios, George and Marcellino, Massimiliano and Venditti, Fabrizio},
  journal={Journal of Applied Econometrics},
  volume={34},
  number={7},
  pages={1027--1049},
  year={2019},
  publisher={Wiley Online Library}
}

@article{giraitis2018inference,
  title={Inference on multivariate heteroscedastic time varying random coefficient models},
  author={Giraitis, Liudas and Kapetanios, George and Yates, Tony},
  journal={Journal of Time Series Analysis},
  volume={39},
  number={2},
  pages={129--149},
  year={2018},
  publisher={Wiley Online Library}
}

@article{giraitis2014inference,
  title={Inference on stochastic time-varying coefficient models},
  author={Giraitis, Liudas and Kapetanios, George and Yates, Tony},
  journal={Journal of Econometrics},
  volume={179},
  number={1},
  pages={46--65},
  year={2014},
  publisher={Elsevier}
}

@article{bloom2014fluctuations,
  title={Fluctuations in uncertainty},
  author={Bloom, Nicholas},
  journal={Journal of Economic Perspectives},
  volume={28},
  number={2},
  pages={153--76},
  year={2014}
}

@article{chen2012testing,
  title={Testing for smooth structural changes in time series models via nonparametric regression},
  author={Chen, Bin and Hong, Yongmiao},
  journal={Econometrica},
  volume={80},
  number={3},
  pages={1157--1183},
  year={2012},
  publisher={Wiley Online Library}
}

@article{mccracken2016fred,
  title={{FRED-MD}: {A} monthly database for macroeconomic research},
  author={McCracken, Michael W. and Ng, Serena},
  journal={Journal of Business \& Economic Statistics},
  volume={34},
  number={4},
  pages={574--589},
  year={2016},
  publisher={Taylor \& Francis}
}

@article{jurado2015measuring,
  title={Measuring uncertainty},
  author={Jurado, Kyle and Ludvigson, Sydney C. and Ng, Serena},
  journal={American Economic Review},
  volume={105},
  number={3},
  pages={1177--1216},
  year={2015}
}

@article{banbura2010large,
  title={Large {B}ayesian vector auto regressions},
  author={Ba{\'n}bura, Marta and Giannone, Domenico and Reichlin, Lucrezia},
  journal={Journal of Applied Econometrics},
  volume={25},
  number={1},
  pages={71--92},
  year={2010},
  publisher={Wiley Online Library}
}

@article{chan2021ijof,
title = {Minnesota-type adaptive hierarchical priors for large {B}ayesian {VAR}s},
author = {Chan, Joshua C.C.},
journal = {International Journal of Forecasting},
volume = {37},
number = {3},
pages = {1212-1226},
year = {2021}
}

@article{huber2019adaptive,
  title={Adaptive shrinkage in {B}ayesian vector autoregressive models},
  author={Huber, Florian and Feldkircher, Martin},
  journal={Journal of Business \& Economic Statistics},
  volume={37},
  number={1},
  pages={27--39},
  year={2019},
  publisher={Taylor \& Francis}
}

@article{koop1996impulse,
  title={Impulse response analysis in nonlinear multivariate models},
  author={Koop, Gary and Pesaran, M. Hashem and Potter, Simon M.},
  journal={Journal of Econometrics},
  volume={74},
  number={1},
  pages={119--147},
  year={1996},
  publisher={Elsevier}
}

@inproceedings{chaudhuri2017mean,
  title={The mean and median criteria for kernel bandwidth selection for support vector data description},
  author={Chaudhuri, Arin and Kakde, Deovrat and Sadek, Carol and Gonzalez, Laura and Kong, Seunghyun},
  booktitle={2017 IEEE International Conference on Data Mining Workshops (ICDMW)},
  pages={842--849},
  year={2017},
  organization={IEEE}
}

@book{koop2003bayesian,
  title={Bayesian Econometrics},
  author={Koop, Gary},
  year={2003},
  publisher = {Wiley, Chichester}
}

@book{williams2006gaussian,
  title={Gaussian processes for machine learning},
  author={Williams, Christopher K. and Rasmussen, Carl Edward},
  volume={2},
  number={3},
  year={2006},
  publisher={MIT press Cambridge, MA}
}

@article{alessandri2019financial,
  title={Financial regimes and uncertainty shocks},
  author={Alessandri, Piergiorgio and Mumtaz, Haroon},
  journal={Journal of Monetary Economics},
  volume={101},
  pages={31--46},
  year={2019},
  publisher={Elsevier}
}

@article{crawford2019variable,
  title={Variable prioritization in nonlinear black box methods: {A} genetic association case study},
  author={Crawford, Lorin and Flaxman, Seth R. and Runcie, Daniel E. and West, Mike},
  journal={The Annals of Applied Statistics},
  volume={13},
  number={2},
  pages={958--989},
  year={2019},
  publisher={NIH Public Access}
}

@article{aastveit2017economic,
  title={Economic uncertainty and the influence of monetary policy},
  author={Aastveit, Knut Are and Natvik, Gisle James and Sola, Sergio},
  journal={Journal of International Money and Finance},
  volume={76},
  pages={50--67},
  year={2017},
  publisher={Elsevier}
}

@article{primiceri2005time,
  title={Time varying structural vector autoregressions and monetary policy},
  author={Primiceri, Giorgio E.},
  journal={The Review of Economic Studies},
  volume={72},
  number={3},
  pages={821--852},
  year={2005},
  publisher={Wiley-Blackwell}
}

@article{koop2013large,
  title={Large time-varying parameter {VAR}s},
  author={Koop, Gary and Korobilis, Dimitris},
  journal={Journal of Econometrics},
  volume={177},
  number={2},
  pages={185--198},
  year={2013},
  publisher={Elsevier}
}
